\documentclass[10pt,twocolumn,superscriptaddress,amssymb,amsmath,aps,pra]{revtex4-2}
\usepackage{graphicx}

\usepackage{slashed}

\usepackage{bbm}

\usepackage[normalem]{ulem}

\usepackage{color}

\newcommand{\cev}[1]{\reflectbox{\ensuremath{\vec{\reflectbox{\ensuremath{#1}}}}}}

\begin{document}


\title{QED based on an eight-dimensional spinorial wave equation of the\\ electromagnetic field and the emergence of quantum gravity}
\date{March 27, 2024}
\author{Mikko Partanen}
\affiliation{Photonics Group, Department of Electronics and Nanoengineering, 
Aalto University, P.O. Box 13500, 00076 Aalto, Finland}
\author{Jukka Tulkki}
\affiliation{Engineered Nanosystems Group, School of Science, Aalto University, 
P.O. Box 12200, 00076 Aalto, Finland}

\begin{abstract}
Quantum electrodynamics (QED) is the most accurate of all experimentally verified physical theories. How QED and other theories of fundamental interactions couple to gravity through special unitary symmetries, on which the standard model of particle physics is based, is, however, still unknown. Here we develop a coupling between the electromagnetic field, Dirac electron-positron field, and the gravitational field based on an eight-component spinorial representation of the electromagnetic field. Our spinorial representation is analogous to the well-known representation of particles in the Dirac theory but it is given in terms of $8\times8$ bosonic gamma matrices. In distinction from earlier works on the spinorial representations of the electromagnetic field, we reformulate QED using eight-component spinors. This enables us to introduce the generating Lagrangian density of gravity based on the special unitary symmetry of the eight-dimensional spinor space. The generating Lagrangian density of gravity plays, in the definition of the gauge theory of gravity and its symmetric stress-energy-momentum tensor source term, a similar role as the conventional Lagrangian density of the free Dirac field plays in the definition of the gauge theory of QED and its electric four-current density source term. The fundamental consequence, the Yang-Mills gauge theory of unified gravity, is studied in a separate work [arXiv:2310.01460], where the theory is also extended to cover the other fundamental interactions of the standard model. We devote ample space for details of the eight-spinor QED to provide solid mathematical basis for the present work and the related work on the Yang-Mills gauge theory of unified gravity.
\end{abstract}

\maketitle


\section{Introduction}

Quantum electrodynamics (QED) agrees with experiments to an exceedingly high accuracy \cite{Peskin2018,Fan2023,Parker2018}. In this regard, it can be considered as the most successful theory of physics ever developed. The present formulation of QED has, however, not enabled coupling of the electromagnetic and Dirac electron-positron fields to the gravitational field through the special unitary symmetries on which the standard model of particle physics is based \cite{Schwartz2014,Peskin2018}. Therefore, alternative approaches, such as string theory \cite{Becker2007,Green1987} and loop quantum gravity \cite{Ashtekar1986,Jacobson1988,Rovelli1990}, are being developed. We approach the problem from a different point of view: It is proposed that identifying a special-unitary-symmetry-based coupling of the present standard model to the gravitational field leads to the Yang-Mills gauge theory of unified gravity, the theory of all known fundamental interactions of nature. Thus, it is our interest to study whether QED, the simplest quantum field theory of the standard model interactions \cite{Schwartz2014,Peskin2018}, can be reformulated in alternative ways using novel mathematical structures that may help us to identify the special-unitary-symmetry-based coupling to gravity. This is the ultimate goal of the present work. Due to the very specific goal of the present work, the review of the relation of the present spinorial representation of the electromagnetic field to earlier spinorial representations \cite{Sachs1961,Sachs1964,Perkins1978,Kulyabov2017,Kiessling2018,Hong2019} will be left as a topic of further work.

Advances in optics and photonics technologies enable detailed experimental investigations of the spin and orbital angular momenta of light \cite{Shen2019,Padgett2017,Devlin2017,Li2017} and the related physics of structured light fields \cite{Forbes2021,Forbes2019,He2022,RubinszteinDunlop2017,Shen2022} and chiral quantum optics \cite{Aiello2022,Chen2023,Lininger2023,Lodahl2017}. Classical Maxwell's equations and QED provide solid basis for theoretical understanding of the related physical phenomena \cite{Jackson1999,Landau1982,Bliokh2015b,Bliokh2015c,Bliokh2017a,WangM2018,Partanen2018a,Partanen2022b}. However, in spite of overall success, theories of light are also associated with well-known enigma that exist independently of the question of the gravitational coupling.

In QED, the usual approach to quantize the electromagnetic field \cite{Glauber2006,Klauder1968} is based on the transverse-vector-potential eigenstates describing the transverse photons \cite{Loudon2000,Landau1982,Sakurai1967}. In optical spectroscopy \cite{Parson2015}, only radiative processes, which can be associated with transverse photons, are discovered. One enigmatic feature of the transverse-vector-potential eigenstates is that their Lorentz transformation between inertial frames leads to a nonzero scalar potential and a nonzero longitudinal component in the transformed inertial frame \cite{Gupta1950}. Thus, the transformed state does not belong to the set of transverse-vector-potential eigenstates, which excludes scalar and longitudinal photon states \cite{Feynman1961}. This is associated with the vector potential being gauge dependent \cite{Landau1982,Feynman1961}. The related enigma of the definition of the photon wave function has been a subject of a continuing dialogue \cite{BialynickiBirula1994,BialynickiBirula1996,Sipe1995,Newton1949,Kiessling2018,Dressel2015,Landau1982,Kidd1989,Hawton2019,Hawton1999,SmithB2007,Sebens2019}. The same applies to the definition of the relativistic quantum spin structure of light \cite{Yang2022,YangL2021}. The enigma is conventionally avoided only in the formalism of second quantization.

For comparison, the wave functions of spin-$\frac{1}{2}$ particles and their antiparticles are described by the Dirac equation whose solutions are four-component spinors \cite{Feynman1961,Landau1982}. The Dirac spinors transform between inertial frames by their own spinorial Lorentz transformation, which is not equivalent to the Lorentz transformation of four-vectors. The Lorentz-transformed spinors belong to the set of eigenstates in contrast with the case of the transverse-four-potential eigenstates of photons as discussed above.

Furthermore, the spin emerges from the Dirac theory through the generators of Lorentz transformations on the Dirac spinors \cite{Peskin2018}. The origin of the spin in the Dirac theory suggests that a similar Lorentz-transformation-based origin may be found for the spin of photons. A strict relativistic description of the spin of photons does not emerge within the principles of classical physics from Maxwell's equations, from the electromagnetic field tensor, or from the electromagnetic four-potential. The conventional formulations of the quantum field theory of photons neither lead to a natural emergence of the covariant and gauge-invariant spin structure of light \cite{Barnett2016,Andrews2013}. The previous two-component spinorial representations of the electromagnetic field \cite{Sachs1961,Sachs1964,Perkins1978} are covariant but they have been formulated so that the electromagnetic spinors in vacuum satisfy the Weyl equation. Regarding the spin properties of the theory, the Weyl equation using Pauli spin matrices is more natural in the description of massless spin-$\frac{1}{2}$ particles, known as Weyl fermions, as discussed by Perkins \cite{Perkins1978}. The rank-two bi-spinor formulation of Kiessling \emph{et al.}~\cite{Kiessling2018} satisfies the massless Dirac equation, which is likewise typically used to describe spin-$\frac{1}{2}$ particles.

The initial goal of the present work was to shed light on the wave function and spin properties of light by constructing a covariant and gauge-invariant spinorial wave function and its Lorentz transformation for photons with a natural emergence of the spin structure of light. However, we discovered later that the theory of light based on the spinorial electromagnetic field leads to much more far-reaching consequences on the gravitational coupling as discussed below. First, we introduce the eight-component \emph{spinorial Maxwell equation}, the solutions of which are called \emph{electromagnetic spinors}. This approach leads to the appearance of $8\times8$ matrices in the spinorial Maxwell equation and to the electromagnetic spinors having eight components. The relativistic description of the quantum spin of light emerges from this theory through the generators of Lorentz transformations on electromagnetic spinors in analogy to the emergence of the spin in the Dirac theory. Since the electromagnetic spinors are gauge independent, they also avoid the gauge dependence problem of the four-potential eigenstates.

The spinorial representation of the electromagnetic field enables rewriting the conventional QED and its Lagrangian density without changes in the physical predictions of the theory. However, as a fundamental consequence, the eight-spinor formulation of the theory enables the description of \emph{the generating Lagrangian density of gravity} based on the special unitary symmetry of the eight-dimensional spinor space. The generating Lagrangian density of gravity leads to an elegant derivation of the \emph{symmetric} stress-energy-momentum (SEM) tensors of the electromagnetic and Dirac electron-positron fields in accordance with Noether's theorem \cite{Noether1918,Zee2010}. Previously, the SEM tensors have been derived only by utilizing the external space-time symmetry. Furthermore, the canonical SEM tensors of the conventional QED are asymmetric, being as such incompatible with \emph{general relativity}. This suggests that the generating Lagrangian density of gravity provides the basis for the derivation of the quantum field theory of gravity. Accordingly, the fundamental consequence of the present eight-spinor theory, the Yang-Mills gauge theory of unified gravity, is investigated in a separate work \cite{Partanen2023c}. In Ref.~\cite{Partanen2023c}, we also extend the theory to cover, not just QED, but all fundamental interactions of the standard model. The present work provides a detailed discussion of many physical and mathematical concepts that are extensively used in Ref.~\cite{Partanen2023c}.

We present a self-contained description of the theory and reserve plenty of space for the detailed study of the foundations so that the reader can become convinced that the mathematical and physical formulations are technically sound in every detail. This work is organized as follows: Section \ref{sec:Dirac} introduces the electromagnetic spinor and formulates the spinorial Maxwell equation, which is the basis of the present work. Section \ref{sec:Lorentz} is devoted to the description of Lorentz and Poincar\'e transformations and their generators, which provide a natural way to introduce a relativistically consistent spin structure to the theory of light. The quantum operators in the first quantization of the field are presented in Sec.~\ref{sec:quantumoptics}. The spinorial photon eigenstates and their density expectation values and CPT symmetry are described in Sec.~\ref{sec:spinorialwavefunctions}. The conventional QED Lagrangian density is reformulated using eight-spinors, and Euler-Lagrange equations are investigated in Sec.~\ref{sec:Lagrangian}. Section \ref{sec:secondquantization} briefly presents the foundations of the second quantization using eight-spinors. This section is provided as a technical tool for interested readers to outline how the key properties of QED emerge from the conventional quantization procedure when eight-spinors are used. Section \ref{sec:Lagrangian2} shows how the eight-spinor formalism of QED enables the description of the generating Lagrangian density of gravity that acts as the basis for the Yang-Mills gauge theory of unified gravity \cite{Partanen2023c}. Related elegant special-unitary-symmetry-based derivations of the electromagnetic and Dirac field SEM tensors are also presented. Finally, conclusions are drawn in Sec.~\ref{sec:conclusions}.

\section{\label{sec:Dirac}Spinorial Maxwell equation}
\subsection{\label{sec:fields}Electromagnetic and charge-current spinors}

We start the formulation of the theory by introducing an \emph{electromagnetic spinor} $\Psi$ and a \emph{charge-current spinor} $\Phi$, which both have eight components. As we show below, the advantage of using an eight-dimensional electromagnetic spinor is that it is \emph{gauge-independent}. The eight-component spinors describe the physics of \emph{all four Maxwell's equations} including the charge and current densities: one component is related to Gauss's law of magnetism, three components to the Amp\`ere-Maxwell law, one component to Gauss's law of electricity, and three components to Faraday's law. In any inertial frame, the components of $\Psi(t,\mathbf{r})$ are composed of the electric and magnetic fields, $\mathbf{E}(t,\mathbf{r})$ and $\mathbf{B}(t,\mathbf{r})$, and the components of $\Phi(t,\mathbf{r})$ are composed of the total electric charge and current densities $\rho_\mathrm{e}(t,\mathbf{r})$ and $\mathbf{J}_\mathrm{e}(t,\mathbf{r})$. Omitting the time and position arguments, we write
\begin{equation}
 \Psi=\sqrt{\frac{\varepsilon_0}{2}}\left[
 \begin{array}{c}
  0\\
  \mathbf{E}\\
  0\\
  ic\mathbf{B}
 \end{array}\right],
 \hspace{0.5cm}\Phi=\sqrt{\frac{\varepsilon_0}{2}}\left[
 \begin{array}{c}
  0\\
  \mu_0c\mathbf{J}_\mathrm{e}\\
  \rho_\mathrm{e}/\varepsilon_0\\
  \mathbf{0}
 \end{array}\right].
 \label{eq:wavefunctions}
\end{equation}
Here $\varepsilon_0$, $\mu_0$, and $c=1/\sqrt{\varepsilon_0\mu_0}$ are the permittivity, permeability, and the speed of light in vacuum, respectively. The zero elements of the electromagnetic spinor $\Psi$ are related to the property of all electromagnetic spinors to be eigenstates of the spin operator squared, having a well-defined spin $S=1$. The zero elements of the charge-current spinor $\Phi$ are related to the nonexistence of magnetic monopoles as elementary particles \cite{Jackson1999}. The definition of the covariant spin operator will be given in Secs.~\ref{sec:operators3D} and \ref{sec:operators4D}, and the eigenstates are discussed in Sec.~\ref{sec:eigenvalues}.

The fields appearing in the electromagnetic spinor in Eq.~\eqref{eq:wavefunctions} can be either real- or complex-valued solutions of the complete set of Maxwell's equations. They will then automatically satisfy the eight-component spinorial Maxwell equation, given in Sec.~\ref{sec:equation}. The spinors made of real-valued fields $\mathbf{E}_\Re$ and $\mathbf{B}_\Re$, charge density $\rho_\mathrm{e\Re}$, and current density $\mathbf{J}_\mathrm{e\Re}$ are denoted by the subscript $\Re$ as $\Psi_\Re$ and $\Phi_\Re$. The spinors are normalized so that $\Psi_\Re^\dag\Psi_\Re=\frac{1}{2}(\varepsilon_0\mathbf{E}_\Re^2+\mathbf{B}_\Re^2/\mu_0)$ corresponds the classical expression of the energy density of the electromagnetic field. Using complex-valued amplitudes of time-harmonic fields, the energy density expectation value is $\frac{1}{2}\Psi^\dag\Psi=\frac{1}{4}(\varepsilon_0|\mathbf{E}|^2+|\mathbf{B}|^2/\mu_0)$, which corresponds to the expression of the energy density of a classical time-harmonic field averaged over the harmonic cycle. When we consider the electromagnetic spinor in Eq.~\eqref{eq:wavefunctions} as a quantum-mechanical concept, we end up to the second quantization and the related description of Fock states. Thus, the quantum-mechanical information content is not equal to that of a classical time-harmonic field, but limited by complete lack of the knowledge of the phase of single-photon states in accordance with the \emph{uncertainty principle} of the photon number and the phase \cite{Smithey1993}. The density expectation values and the classical and quantum-mechanical information content of the electromagnetic spinor in Eq.~\eqref{eq:wavefunctions} will be addressed in more detail in Sec.~\ref{sec:spinorialwavefunctions}.

The electromagnetic spinor in Eq.~\eqref{eq:wavefunctions} can be compared with previous approaches to construct a wave-function-like concept for photons. One conventional approach is to use the electromagnetic vector potential or four-potential in the role of the photon wave function \cite{Landau1982}. The Lorentz transformation of the electromagnetic four-potential predicts, for a radiation field in a general inertial frame, unobserved longitudinal and scalar photons \cite{Gupta1950,Yang2022}. The present eight-component electromagnetic spinor $\Psi$ is seen to be a generalization of the six-component quantity $\sqrt{\frac{\varepsilon_0}{2}}[\mathbf{E},c\mathbf{B}]^T$ used in some previous literature \cite{Sommerfeld1910a,Sommerfeld1910b,Darwin1932,Berry1990,Berry2009,Mohr2010,Barnett2014,Bliokh2014a,Bliokh2015a,Bliokh2017a,Bliokh2017b}, where $T$ indicates the transpose. Depending on the work, the imaginary unit can be present as a factor of the electric or magnetic-field component. In the case of real-valued fields, the sum of the electric and magnetic components of $\Psi_\Re$ is also seen to correspond to the three-component Weber vector $\sqrt{\frac{\varepsilon_0}{2}}(\mathbf{E}_\Re+ic\mathbf{B}_\Re)$ \cite{Weber1900,Kiessling2018}, also known as the Riemann-Silberstein vector \cite{Silberstein1907a,Silberstein1907b,Bateman1915,Stratton1941,BialynickiBirula1996,BialynickiBirula2003,BialynickiBirula2011,Good1957}. An expression that reminds the Weber vector, but generalizes to complex-valued fields, is the representation of the electromagnetic field in the space-time algebra \cite{Hestenes1966,Hestenes1987,Hestenes2003,Hestenes2005,Dressel2015} through a quantity $\sqrt{\frac{\varepsilon_0}{2}}(\mathbf{E}+Ic\mathbf{B})$, where the algebraic element $I$ satisfing $I^2=-1$ is not the imaginary unit, but a geometric object. In the previous two-component spinorial representations \cite{Sachs1961,Sachs1964,Perkins1978}, the spinors are formed by mixing different components of the electric and magnetic fields, e.g., $\varphi_1=\sqrt{\frac{\varepsilon_0}{2}}[(cB_{\Re z}-iE_{\Re z}),(cB_{\Re x}-iE_{\Re x})+i(cB_{\Re y}-iE_{\Re y})]^T$ and $\varphi_2=\sqrt{\frac{\varepsilon_0}{2}}[-(cB_{\Re x}-iE_{\Re x})+i(cB_{\Re y}-iE_{\Re y}),(cB_{\Re z}-iE_{\Re z})]^T$ corresponding to the representation by Sachs and Schwebel \cite{Sachs1961}. This representation does not generalize to complex-valued field amplitudes.

\subsection{\label{sec:equation}Spinorial Maxwell equation and the associated gamma and sigma matrices}

Our formulation of electrodynamics equations below is closely analogous to the approach of Dirac starting from the second-order Klein-Gordon equation to derive a first-order equation, the Dirac equation \cite{Dirac1928,Dirac1958}. Dirac's clever idea was to take a square root of the wave operator, which is equivalent to finding matrices $\boldsymbol{\gamma}^a$, which satisfy $\mathbf{I}_n\partial^a\partial_a=(\boldsymbol{\gamma}^a\partial_a)^2$, where the Einstein summation convention is used and $\mathbf{I}_n$ is the $n\times n$ identity matrix. In this work, the Latin indices $a,b,c,d\in\{0,x,y,z\}$ range over the four dimensions of the Minkowski space-time, the Latin indices $i,j,k\in\{x,y,z\}$ range over the three spatial dimensions, and the Greek indices $\mu,\nu,\rho,\sigma\in\{x^0,x^1,x^2,x^3\}$ range over the four general space-time dimensions. Dirac found that $\mathbf{I}_4\partial^a\partial_a=(\boldsymbol{\gamma}_\mathrm{F}^a\partial_a)^2$ is satisfied with $4\times4$ matrices $\boldsymbol{\gamma}_\mathrm{F}^a$, which became known as the Dirac gamma matrices. Here we use the subscript F to indicate that these gamma matrices are associated with fermionic fields in distinction to the gamma matrices of bosonic fields introduced below.

Our goal is to define gamma matrices, which enable writing the well-known second-order wave equation of the four-potential, given by $\partial^a\partial_a A^b=\mu_0J_\mathrm{e}^b$, as a first-order equation for the electromagnetic spinor $\Psi$ in Eq.~\eqref{eq:wavefunctions}, which, unlike the Dirac equation, also includes the source term expressed by the charge-current spinor $\Phi$. We next show that it is possible to construct $8\times8$ gamma matrices, which operate on the electromagnetic spinor $\Psi$ in Eq.~\eqref{eq:wavefunctions} leading to Maxwell's equations and to $\Psi$ having desired properties. In analogy to the Dirac theory, our gamma matrices $\boldsymbol{\gamma}_\mathrm{B}^a$ satisfy $\mathbf{I}_8\partial^a\partial_a=(\boldsymbol{\gamma}_\mathrm{B}^a\partial_a)^2$. Here we use the subscript B to indicate that these gamma matrices are associated with bosonic fields in distinction to the gamma matrices of fermionic fields in the Dirac theory.

We write the spinorial Maxwell equation in a form that closely reminds the Dirac equation describing spin-$\frac{1}{2}$ particles in the case when the mass of the particle is set to zero and when an external source term is added. The spinorial Maxwell equation, which, as shown in Sec.~\ref{sec:Maxwell}, describes all of classical electromagnetism, is written as
\begin{equation}
 \boldsymbol{\gamma}_\mathrm{B}^a\partial_a\Psi=-\Phi.
 \label{eq:photonDirac}
\end{equation}
In analogy to the Dirac equation for electrons, the spinorial Maxwell equation is a first-order differential equation that is covariant in Lorentz transformations between inertial frames. The $8\times8$ bosonic gamma matrices $\boldsymbol{\gamma}_\mathrm{B}^a$ in Eq.~\eqref{eq:photonDirac} are defined as
\begin{equation}
 \boldsymbol{\gamma}_\mathrm{B}^0=\left[
 \begin{array}{cc}
  \mathbf{I}_4 & \mathbf{0}\\
  \mathbf{0} & -\mathbf{I}_4
 \end{array}\right],\hspace{0.5cm}
 \boldsymbol{\gamma}_\mathrm{B}^i=\left[
 \begin{array}{cc}
  \mathbf{0} & \boldsymbol{\sigma}_\mathrm{B}^i\\
  -\boldsymbol{\sigma}_\mathrm{B}^i & \mathbf{0}
 \end{array}\right].
 \label{eq:gamma}
\end{equation}
The $4\times4$ bosonic sigma matrices $\boldsymbol{\sigma}_\mathrm{B}^x$, $\boldsymbol{\sigma}_\mathrm{B}^y$, and $\boldsymbol{\sigma}_\mathrm{B}^z$ in Eq.~\eqref{eq:gamma} are given in the Cartesian basis by
\begin{equation}
 \boldsymbol{\sigma}_\mathrm{B}^i=\mathbf{K}_\mathrm{boost}^i+i\mathbf{K}_\mathrm{rot}^i.
 \label{eq:sigma}
\end{equation}
Here $\mathbf{K}_\mathrm{boost}^i$ and $\mathbf{K}_\mathrm{rot}^i$ are generators of the proper orthochronous Lorentz group SO$^+$(1,3).
The Lorentz boost generators $\mathbf{K}_\mathrm{boost}^i$ are given by \cite{Baskal2015}
\begin{align}
 &\mathbf{K}_\mathrm{boost}^x=\left[
 \begin{array}{cccc}
  0 & 1 & 0 & 0\\
  1 & 0 & 0 & 0\\
  0 & 0 & 0 & 0\\
  0 & 0 & 0 & 0
 \end{array}\right]\!\!,\hspace{0.1cm}
 \mathbf{K}_\mathrm{boost}^y=\left[
 \begin{array}{cccc}
  0 & 0 & 1 & 0\\
  0 & 0 & 0 & 0\\
  1 & 0 & 0 & 0\\
  0 & 0 & 0 & 0
 \end{array}\right]\!\!,\nonumber\\
 &\hspace{1.5cm}\mathbf{K}_\mathrm{boost}^z=\left[
 \begin{array}{cccc}
  0 & 0 & 0 & 1\\
  0 & 0 & 0 & 0\\
  0 & 0 & 0 & 0\\
  1 & 0 & 0 & 0
 \end{array}\right]\!\!,
 \label{eq:boostgenerators}
\end{align}
and the SO(3) rotation generators $\mathbf{K}_\mathrm{rot}^i$ are given by
\begin{align}
 &\mathbf{K}_\mathrm{rot}^x=\left[
 \begin{array}{cccc}
  0 & 0 & 0 & 0\\
  0 & 0 & 0 & 0\\
  0 & 0 & 0 & -1\\
  0 & 0 & 1 & 0
 \end{array}\right]\!\!,\hspace{0.1cm}
 \mathbf{K}_\mathrm{rot}^y=\left[
 \begin{array}{cccc}
  0 & 0 & 0 & 0\\
  0 & 0 & 0 & 1\\
  0 & 0 & 0 & 0\\
  0 & -1 & 0 & 0
 \end{array}\right]\!\!,\nonumber\\
 &\hspace{1.5cm}\mathbf{K}_\mathrm{rot}^z=\left[
 \begin{array}{cccc}
  0 & 0 & 0 & 0\\
  0 & 0 & -1 & 0\\
  0 & 1 & 0 & 0\\
  0 & 0 & 0 & 0
 \end{array}\right]\!\!.
 \label{eq:rotationgenerators}
\end{align}
Apart from not having the imaginary unit and the reduced Planck constant $\hbar$ as a factor, the lower right $3\times3$ space components of the rotation generators in Eq.~\eqref{eq:rotationgenerators} are equal to the well-known spin matrices of the three-dimensional space \cite{deShalit1963,Bliokh2015a,Berry1998,Akhiezer1965,BialynickiBirula1996}.

The form of the bosonic gamma matrices $\boldsymbol{\gamma}_\mathrm{B}^a$ in Eq.~\eqref{eq:gamma} reminds the conventional definition of the $4\times4$ fermionic gamma matrices $\boldsymbol{\gamma}_\mathrm{F}^a$ in the Dirac theory \cite{Zee2010}, but the dimensions and the corresponding representations of the sigma matrices are different. In the Dirac theory, the $2\times2$ sigma matrices $\boldsymbol{\sigma}_\mathrm{F}^i$ are known as Pauli matrices.
The $8\times8$ gamma matrices of the present theory can also be seen as the generalization of the $6\times6$ gamma matrices that have been used together with the six-component wave-function-like concept in previous literature \cite{Mohr2010,Barnett2014}. In comparison to the $6\times6$ gamma matrices, one particular advantage of our $8\times8$ gamma matrices is that all Maxwell's equations can be presented as a single equation allowing us to rewrite the QED in the eight-spinor notation as presented in Sec.~\ref{sec:Lagrangian}. The bosonic gamma matrices in Eq.~\eqref{eq:gamma} can be used to define the spinorial Lorentz transformations as will be shown in Sec.~\ref{sec:Lorentz}. This makes the electromagnetic spinor in Eq.~\eqref{eq:wavefunctions} a true spinor with spin 1 in the full physical meaning of the term \cite{Penrose1987,Cartan1981,Corson1953}.

\subsection{\label{sec:Maxwell}Dynamical equations in terms of the fields, charges, and currents}

When the expressions of the electromagnetic spinor $\Psi$ and the charge-current spinor $\Phi$ from Eq.~\eqref{eq:wavefunctions} are substituted into the spinorial Maxwell equation in Eq.~\eqref{eq:photonDirac}, we obtain the full set of four Maxwell's equations in the conventional form as
\begin{equation}
 \nabla\cdot\mathbf{B}=0,
 \label{eq:maxwell1}
\end{equation}
\begin{equation}
 \nabla\times\mathbf{B}=\mu_0\mathbf{J}_\mathrm{e}+\frac{1}{c^2}\frac{\partial}{\partial t}\mathbf{E},
\end{equation}
\begin{equation}
 \nabla\cdot\mathbf{E}=\frac{\rho_\mathrm{e}}{\varepsilon_0},
\end{equation}
\begin{equation}
 \nabla\times\mathbf{E}=-\frac{\partial}{\partial t}\mathbf{B}.
 \label{eq:maxwell4}
\end{equation}
Thus, the very compact representation of the spinorial Maxwell equation in Eq.~\eqref{eq:photonDirac} together with the electromagnetic and charge-current spinors in Eq.~\eqref{eq:wavefunctions} is \emph{equivalent} to the full theory of classical electromagnetism \cite{Jackson1999}. Again, the fields and charge and current densities in Eqs.~\eqref{eq:maxwell1}--\eqref{eq:maxwell4} can be either real or complex valued.

Operating on Eq.~\eqref{eq:photonDirac} side by side with an operator $\boldsymbol{\gamma}_\mathrm{B}^a\partial_a$ and using $(\boldsymbol{\gamma}_\mathrm{B}^a\partial_a)^2=\mathbf{I}_8\partial^a\partial_a$ on the left-hand side, we obtain $\partial^a\partial_a\Psi=-\boldsymbol{\gamma}_\mathrm{B}^a\partial_a\Phi$. This equation is equivalent to the \emph{continuity equation of the electric four-current density} $J_\mathrm{e}^a=(c\rho_\mathrm{e},\mathbf{J}_\mathrm{e})$ and the \emph{inhomogeneous wave equations of the electric and magnetic fields} as
\begin{equation}
 \frac{\partial\rho_\mathrm{e}}{\partial t}+\nabla\cdot\mathbf{J}_\mathrm{e}=0,
\end{equation}
\begin{equation}
 \nabla^2\mathbf{E}-\frac{1}{c^2}\frac{\partial^2}{\partial t^2}\mathbf{E}=\mu_0\frac{\partial}{\partial t}\mathbf{J}_\mathrm{e}+\frac{1}{\varepsilon_0}\nabla\rho_\mathrm{e},
\end{equation}
\begin{equation}
 \nabla^2\mathbf{B}-\frac{1}{c^2}\frac{\partial^2}{\partial t^2}\mathbf{B}=-\mu_0\nabla\times\mathbf{J}_\mathrm{e}.
\end{equation}
Therefore, in addition to the conventional Maxwell's equations, the spinorial Maxwell equation also compactly describes the physics related to the \emph{conservation of the electric four-current density} \cite{Jackson1999}. If the hypothetical magnetic charge and current densities were added in the theory through substituting them in the zero elements of the charge-current spinor in Eq.~\eqref{eq:wavefunctions}, the conservation law of the magnetic four-current density would also follow.

\subsection{\label{sec:potentialspinor}Potential spinor}

For the Lagrangian formulation of QED using the present spinorial notation, to be presented in Sec.~\ref{sec:Lagrangian}, we next provide the spinorial representation of the four-potential. The electromagnetic four-potential is a four-vector $A^a=(\phi_\mathrm{e}/c,\mathbf{A})$, where $\phi_\mathrm{e}$ is the electric scalar potential and $\mathbf{A}$ is the vector potential. Apart from constant prefactors, the \emph{potential spinor} $\Theta$ is constructed from the components of $A^a=(\phi_\mathrm{e}/c,\mathbf{A})$ in the same way as the charge-current spinor is made of the components of the electric four-current density $J_\mathrm{e}^a=(c\rho_\mathrm{e},\mathbf{J}_\mathrm{e})$ as presented in Eq.~\eqref{eq:wavefunctions}. Consequently, we write $\Theta$ as
\begin{equation}
 \Theta=\sqrt{\frac{\varepsilon_0}{2}}\left[
 \begin{array}{c}
  0\\
  c\mathbf{A}\\
  \phi_\mathrm{e}\\
  \mathbf{0}
 \end{array}\right].
 \label{eq:Theta}
\end{equation}
Then, operating on this spinor by the operator $-\boldsymbol{\gamma}_\mathrm{B}^a\partial_a$ we obtain
\begin{align}
 -\boldsymbol{\gamma}_\mathrm{B}^a\partial_a\Theta &=\sqrt{\frac{\varepsilon_0}{2}}\left[
 \begin{array}{c}
  0\\
  -\nabla\phi_\mathrm{e}-\frac{\partial}{\partial t}\mathbf{A}\\
  c\partial_a A^a\\
  ic\nabla\times\mathbf{A}
 \end{array}\right]\nonumber\\
 &\hspace{0.4cm}\xrightarrow{\partial_a A^a\rightarrow0}\sqrt{\frac{\varepsilon_0}{2}}\left[
 \begin{array}{c}
  0\\
  \mathbf{E}\\
  0\\
  ic\mathbf{B}
 \end{array}\right]=\Psi.
 \label{eq:fourpotentialspinorrelation}
\end{align}
Here we have used the conventional expressions of the fields in terms of the vector and scalar potentials, given by $\mathbf{E}=-\nabla\phi_\mathrm{e}-\frac{\partial}{\partial t}\mathbf{A}$ and $\mathbf{B}=\nabla\times\mathbf{A}$ \cite{Jackson1999}. Equation \eqref{eq:fourpotentialspinorrelation} shows that the four-potential can be used to construct the electromagnetic spinor $\Psi$ by the operator $-\boldsymbol{\gamma}_\mathrm{B}^a\partial_a$ when the Lorenz gauge condition $\partial_a A^a=0$ is satisfied. Otherwise, the fifth component of the electromagnetic spinor would be nonzero as seen from Eq.~\eqref{eq:fourpotentialspinorrelation}. This would violate the property of all electromagnetic spinors to be eigenstates of the spin operator squared, having a well-defined spin of $S=1$, as presented in Sec.~\ref{sec:eigenvalues}. Note the fact that the Lorenz gauge condition does not determine the four-potential uniquely since there are remaining degrees of freedom corresponding to gauge functions which satisfy the wave equation. This is called residual gauge symmetry \cite{Swanson2022}. If the Coulomb gauge condition $\nabla\cdot\mathbf{A}=0$ is used, then the Lorenz gauge condition must be simultaneously satisfied. This means that we must separately have the time derivative of the scalar potential equal to zero.  If the electric and magnetic fields are transverse, we can set the scalar potential to zero, and with this further condition, the Coulomb gauge is called the radiation gauge \cite{Jackson1999}.

From the perspective of the four-potential, the formulation of the spinorial photon equation can be seen as follows: The wave equation of the four-potential in the Lorenz gauge is given by $\partial^a\partial_a A^b=\mu_0J_\mathrm{e}^b$. To transform this equation to an equation for the potential spinor, we use $A^b$ and $J_\mathrm{e}^b$ to form spinors $\Theta$ and $\Phi$ according to Eqs.~\eqref{eq:wavefunctions} and \eqref{eq:Theta}. Thus, the wave equation of the four-potential is rewritten as
\begin{equation}
 \partial^a\partial_a\Theta=\Phi.
 \label{eq:photonDiracA1}
\end{equation}
To transform this equation into a first-order partial differential equation for the electromagnetic spinor $\Psi$, we write the wave operator as $\mathbf{I}_8\partial^a\partial_a=(-\boldsymbol{\gamma}_\mathrm{B}^a\partial_a)^2$. Thus, the left-hand-side of the wave equation of the potential spinor in Eq.~\eqref{eq:photonDiracA1} is rewritten as $\partial^a\partial_a\Theta=(-\boldsymbol{\gamma}_\mathrm{B}^a\partial_a)^2\Theta=-\boldsymbol{\gamma}_\mathrm{B}^a\partial_a\Psi$, where we have used Eq.~\eqref{eq:fourpotentialspinorrelation}. Thus, from Eq.~\eqref{eq:photonDiracA1}, we obtain
\begin{equation}
 -\partial^a\partial_a\Theta=\boldsymbol{\gamma}_\mathrm{B}^a\partial_a\Psi=-\Phi.
 \label{eq:photonDiracA2}
\end{equation}
This equation is equivalent to the spinorial Maxwell equation in Eq.~\eqref{eq:photonDirac}.

\subsection{\label{sec:fifthgamma}Left-handed and right-handed spinor components and the fifth gamma matrix}

In analogy to the Dirac theory \cite{Peskin2018}, the electromagnetic spinor can be projected onto its left-handed and right-handed chiral components $\Psi_\mathrm{L}$ and $\Psi_\mathrm{R}$ as
\begin{equation}
 \Psi_\mathrm{L}=\frac{\mathbf{I}_8-\boldsymbol{\gamma}_\mathrm{B}^5}{2}\Psi,
 \hspace{0.5cm}\Psi_\mathrm{R}=\frac{\mathbf{I}_8+\boldsymbol{\gamma}_\mathrm{B}^5}{2}\Psi.
 \label{eq:projections}
\end{equation}
The components $\Psi_\mathrm{L}$ and $\Psi_\mathrm{R}$ are eigenstates of $\boldsymbol{\gamma}_\mathrm{B}^5$ with eigenvalues $\pm 1$ as $\boldsymbol{\gamma}_\mathrm{B}^5\Psi_\mathrm{L}=-\Psi_\mathrm{L}$ and $\boldsymbol{\gamma}_\mathrm{B}^5\Psi_\mathrm{R}=\Psi_\mathrm{R}$. The fifth gamma matrix $\boldsymbol{\gamma}_\mathrm{B}^5$ is the chirality operator, and it is defined in terms of the other gamma matrices by the relation $\boldsymbol{\gamma}_\mathrm{B}^5=\frac{i}{4!}\varepsilon_{abcd}\boldsymbol{\gamma}_\mathrm{B}^a\boldsymbol{\gamma}_\mathrm{B}^b\boldsymbol{\gamma}_\mathrm{B}^c\boldsymbol{\gamma}_\mathrm{B}^d=i\boldsymbol{\gamma}_\mathrm{B}^0\boldsymbol{\gamma}_\mathrm{B}^x\boldsymbol{\gamma}_\mathrm{B}^y\boldsymbol{\gamma}_\mathrm{B}^z$, where $\varepsilon_{abcd}$ is the four-dimensional Levi-Civita symbol. This relation is of the same form as the corresponding relation for conventional Dirac gamma matrices \cite{Peskin2018}. The matrix $\boldsymbol{\gamma}_\mathrm{B}^5$ anticommutes with the other gamma matrices as $\{\boldsymbol{\gamma}_\mathrm{B}^5,\boldsymbol{\gamma}_\mathrm{B}^a\}=\mathbf{0}$ and commutes with the Lorentz transformation matrices that will be presented in Sec.~\ref{sec:Lorentz}. For massless particles, such as for photons in the present case, the chirality is equivalent to the helicity divided by $\hbar$ \cite{Peskin2018}. For the definition of the helicity operator, see Sec.~\ref{sec:operators3D}. For further discussion of the helicity-chirality equivalence, see Appendix \ref{apx:spinorialwavefunctions}. As will be shown in Sec.~\ref{sec:Lagrangian2}, $\boldsymbol{\gamma}_\mathrm{B}^5$ plays a fundamental role in the definition of the generating Lagrangian density of gravity and its special unitary symmetry.

\subsection{Eight-spinor adjoint}

Next, we define the eight-spinor adjoints of the electromagnetic and charge-current spinors and any othe eight-spinors. In analogy to the Dirac adjoint $\bar{\psi}=\psi^\dag\boldsymbol{\gamma}_\mathrm{F}^0$ of Dirac spinors $\psi$, we define the eight-spinor adjoints in terms of the timelike gamma matrix and the Hermitian conjugates of the spinors. For the electromagnetic spinor, we thus write
\begin{equation}
 \bar{\Psi}=\Psi^\dag\boldsymbol{\gamma}_\mathrm{B}^0.
\end{equation}
The eight-spinor adjoints of other eight-spinor quantities are obtained by similar relations. For a generic $8\times8$ matrix $\mathbf{H}$, the corresponding adjoint is defined as $\bar{\mathbf{H}}=\boldsymbol{\gamma}_\mathrm{B}^0\mathbf{H}^\dag\boldsymbol{\gamma}_\mathrm{B}^0$. The eight-spinor adjoint can be used to write the local scalar product for two arbitrary eight-spinors. Each of the spinors can be an electromagnetic spinor, charge-current spinor, or any other eight-spinor. As an example, using the electromagnetic spinor symbols $\Psi_1$ and $\Psi_2$, the local scalar product is written as
\begin{equation}
 \bar{\Psi}_1\Psi_2=\Psi_1^\dag\boldsymbol{\gamma}_\mathrm{B}^0\Psi_2.
 \label{eq:scalarproductDirac}
\end{equation}
A fundamental property of the local scalar product in Eq.~\eqref{eq:scalarproductDirac} is that, for two electromagnetic spinors, $\bar{\Psi}_1\Psi_2$ transforms as a Lorentz scalar between inertial frames, which is not the case for $\Psi_1^\dag\Psi_2$. Thus, the local scalar product of eight-spinors in Eq.~\eqref{eq:scalarproductDirac} highlights the fundamental analogy between the present eight-spinor theory of the electromagnetic field and the Dirac theory of the electron-positron field. In previous constructions of wave-function-like concepts for photons or representations of the electromagnetic field in general \cite{BialynickiBirula1996,Landau1982,Kiessling2018,Dressel2015}, such a transparent analogy is not typically observable. However, the previous six-component representation of the wave-function-like quantity for the electromagnetic field \cite{Darwin1932,Mohr2010,Barnett2014} can be seen to be analogous to the Dirac theory, but it neither possess all properties of the present theory, such as the presentation of all Maxwell's equations by a single equation.

\subsection{\label{sec:algebra}Algebraic properties of the Lorentz generators and sigma and gamma matrices}

The Lorentz boost and rotation generators $\mathbf{K}_\mathrm{boost}^i$ and $\mathbf{K}_\mathrm{rot}^i$ in Eqs.~\eqref{eq:boostgenerators} and \eqref{eq:rotationgenerators} satisfy the commutation relations $[\mathbf{K}_\mathrm{boost}^i,\mathbf{K}_\mathrm{boost}^j]=-\varepsilon_{ijk}\mathbf{K}_\mathrm{rot}^k$,
$[\mathbf{K}_\mathrm{rot}^i,\mathbf{K}_\mathrm{rot}^j]=\varepsilon_{ijk}\mathbf{K}_\mathrm{rot}^k$, and $[\mathbf{K}_\mathrm{rot}^i,\mathbf{K}_\mathrm{boost}^j]=\varepsilon_{ijk}\mathbf{K}_\mathrm{boost}^k$, where $\varepsilon_{ijk}$ is the three-dimensional Levi-Civita symbol. The Lorentz Lie algebra relations and the presentation of the Lorentz transformations in terms of the Lorentz generators are discussed in Sec.~\ref{sec:Lorentz}.

The sigma matrices $\boldsymbol{\sigma}_\mathrm{B}^i$ in Eq.~\eqref{eq:sigma} are Hermitian, involutory, and unitary satisfying $\boldsymbol{\sigma}_\mathrm{B}^x\boldsymbol{\sigma}_\mathrm{B}^x=\boldsymbol{\sigma}_\mathrm{B}^y\boldsymbol{\sigma}_\mathrm{B}^y=\boldsymbol{\sigma}_\mathrm{B}^z\boldsymbol{\sigma}_\mathrm{B}^z=-i\boldsymbol{\sigma}_\mathrm{B}^x\boldsymbol{\sigma}_\mathrm{B}^y\boldsymbol{\sigma}_\mathrm{B}^z=\mathbf{I}_4$. Their determinants are equal to unity as $\mathrm{Det}(\boldsymbol{\sigma}_\mathrm{B}^i)=1$ and their traces are zero as $\mathrm{Tr}(\boldsymbol{\sigma}_\mathrm{B}^i)=0$. The eigenvalues of each $\boldsymbol{\sigma}_\mathrm{B}^i$ are $\pm1$. The commutation and anticommutation relations of the sigma matrices are given by $[\boldsymbol{\sigma}_\mathrm{B}^i,\boldsymbol{\sigma}_\mathrm{B}^j]=2i\varepsilon_{ijk}\boldsymbol{\sigma}_\mathrm{B}^k$ and $\{\boldsymbol{\sigma}_\mathrm{B}^i,\boldsymbol{\sigma}_\mathrm{B}^j\}=2\delta^{ij}\mathbf{I}_4$, where $\delta^{ij}$ is the Kronecker delta. Apart from the sign of the determinants of $\boldsymbol{\sigma}_\mathrm{B}^i$, these relations correspond to the relations of the Pauli matrices in the conventional $2\times 2$-dimensional case.

In analogy to the Dirac theory, the timelike gamma matrix $\boldsymbol{\gamma}_\mathrm{B}^0$ is Hermitian and the spacelike gamma matrices $\boldsymbol{\gamma}_\mathrm{B}^x$, $\boldsymbol{\gamma}_\mathrm{B}^y$, and $\boldsymbol{\gamma}_\mathrm{B}^z$ are anti-Hermitian. The gamma matrices are generators of the gamma group $G_{1,3}$ \cite{Petitjean2020}. Furthermore, the gamma matrices satisfy the anticommutation relation $\{\boldsymbol{\gamma}_\mathrm{B}^a,\boldsymbol{\gamma}_\mathrm{B}^b\}=2\eta^{ab}\mathbf{I}_8$, where $\eta^{ab}$ is the Minkowski metric tensor with signature $(+,-,-,-)$. The conventional Dirac gamma matrices are well-known to satisfy the corresponding relation $\{\boldsymbol{\gamma}_\mathrm{F}^a,\boldsymbol{\gamma}_\mathrm{F}^b\}=2\eta^{ab}\mathbf{I}_4$ \cite{Peskin2018}. This is the defining relation of a Dirac algebra, which is a Clifford algebra $\mathcal{C}\ell_{1,3}(\mathbb{C})$ over the four-dimensional Minkowski space-time. Therefore, in spite of being $8\times8$ matrices, the present bosonic gamma matrices share the same group-theoretical and algebraic properties with the conventional $4\times4$ gamma matrices of the Dirac theory. One particularly useful relation to note is $\boldsymbol{\gamma}_\mathrm{B}^{a\dag}=\boldsymbol{\gamma}_\mathrm{B}^0\boldsymbol{\gamma}_\mathrm{B}^a\boldsymbol{\gamma}_\mathrm{B}^0$. To preserve the forms of the electromagnetic and charge-current spinors in Eq.~\eqref{eq:wavefunctions} in terms of their component fields in all inertial frames, the present theory follows the convention that the spinors transform under Lorentz transformations and the gamma matrices are constant having the same form in all inertial frames \cite{Feynman1961}.

\section{\label{sec:Lorentz}Lorentz and Poincar\'e transformations and invariants}

In the present theory, there are two types of Lorentz transformations for eight-spinors: the Lorentz transformation of spinors generated by four-vectors, called four-vector spinors, such as the charge-current spinor and the potential spinor; and the Lorentz transformation of spin-1 field spinors, such as the electromagnetic spinor. These transformations preserve the forms of the electromagnetic and charge-current spinors in Eq.~\eqref{eq:wavefunctions} and the potential spinor in Eq.~\eqref{eq:Theta} in all inertial frames. The transformations are analogous to the Lorentz transformation of Dirac spinors in their construction through the Lorentz generators in a way that they differ from the Lorentz transformation of four-vectors. For completeness, the Lorentz transformation of the conventional spin-$\frac{1}{2}$ Dirac field spinors is also presented. In addition, we construct the Lorentz transformation of spin-2 fields, which can be presented using $8\times8$ matrices in the present theory \cite{Partanen2023c}. The Lorentz transformations are described in detail in the sections below. In the last three sections, we present the complete Lorentz transformations of fields, which involve the transformations of the coordinate arguments; the Poincar\'e transformations, which also involve space-time translations; and the construction of Lorentz invariants.

\subsection{\label{sec:Lorentzvector}Lorentz transformation of four-vectors}

We start with the conventional Lorentz transformation of four-vectors. In terms of the generators of the proper orthochronous Lorentz group SO$^+$(1,3) in Eqs.~\eqref{eq:boostgenerators} and \eqref{eq:rotationgenerators}, the Lorentz boost of four-vectors is given by $\boldsymbol{\Lambda}_\mathrm{boost}(\boldsymbol{\zeta})=\exp(-\boldsymbol{\zeta}\cdot\mathbf{K}_\mathrm{boost})$ and the spatial rotation of four-vectors is given by $\boldsymbol{\Lambda}_\mathrm{rot}(\boldsymbol{\theta})=\exp(\boldsymbol{\theta}\cdot\mathbf{K}_\mathrm{rot})$, where $\boldsymbol{\zeta}=(\zeta_x,\zeta_y,\zeta_z)=\arctan(v/c)\mathbf{v}/|\mathbf{v}|$ is the rapidity vector, in which $\mathbf{v}$ is the boost velocity, and $\boldsymbol{\theta}=(\theta_x,\theta_y,\theta_z)$ is the axis-angle vector \cite{Jackson1999}. The total combined Lorentz transformation of four-vectors is given by $\boldsymbol{\Lambda}(\boldsymbol{\boldsymbol{\zeta},\theta})=\exp(-\boldsymbol{\zeta}\cdot\mathbf{K}_\mathrm{boost}+\boldsymbol{\theta}\cdot\mathbf{K}_\mathrm{rot})$, where $\mathbf{K}_\mathrm{boost}=(\mathbf{K}_\mathrm{boost}^x,\mathbf{K}_\mathrm{boost}^y,\mathbf{K}_\mathrm{boost}^z)$ and $\mathbf{K}_\mathrm{rot}=(\mathbf{K}_\mathrm{rot}^x,\mathbf{K}_\mathrm{rot}^y,\mathbf{K}_\mathrm{rot}^z)$.

A more compact presentation of the full Lorentz transformation involving both a boost and rotation is obtained by substituting the components of the rapidity and axis-angle vectors in an antisymmetric Lorentz parameter matrix $\Omega_{ab}$, given by
\begin{equation}
 \Omega_{ab}=\left[
 \begin{array}{cccc}
  0 & -\zeta_x & -\zeta_y & -\zeta_z\\
  \zeta_x & 0 & \theta_z & -\theta_y\\
  \zeta_y & -\theta_z & 0 & \theta_x\\
  \zeta_z & \theta_y & -\theta_x & 0
 \end{array}\right],
 \label{eq:Omega}
\end{equation}
and combining the representation of the Lorentz generators in Eqs.~\eqref{eq:boostgenerators} and \eqref{eq:rotationgenerators} by defining \cite{Maggiore2005} 
\begin{equation}
 (K^{ab})^c_{\;\,d}=\eta^{ac}\delta^b_{\;\,d}-\eta^{bc}\delta^a_{\;\,d}.
 \label{eq:K}
\end{equation}
For different values of indices $c$ and $d$, this expression gives matrices corresponding to the Lorentz generators in Eqs.~\eqref{eq:boostgenerators} and \eqref{eq:rotationgenerators}. Using Eq.~\eqref{eq:K}, one can show that the generators $K^{ab}$ obey the Lorentz Lie algebra relations, $[K^{ab},K^{cd}]=\eta^{bc}K^{ad}-\eta^{ac}K^{bd}+\eta^{ad}K^{bc}-\eta^{bd}K^{ac}$, and the Lorentz transformation of four-vectors is given by \cite{Maggiore2005} 
\begin{equation}
 \textstyle\boldsymbol{\Lambda}=\exp(\frac{1}{2}\Omega_{ab}K^{ab}).
 \label{eq:lambda}
\end{equation}
Using this Lorentz transformation matrix, a four-vector $X$ transforms between inertial frames as $X'=\boldsymbol{\Lambda}X$. For boosts, the real-valued Lorentz transformation matrix $\boldsymbol{\Lambda}$ in Eq.~\eqref{eq:lambda} is symmetric, i.e., $\boldsymbol{\Lambda}=\boldsymbol{\Lambda}^T$. For SO(3) rotations, it is orthogonal, i.e., $\boldsymbol{\Lambda}\boldsymbol{\Lambda}^T=\mathbf{I}_4$.

\subsection{\label{sec:Lorentzchargecurrentspinor}Lorentz transformation of four-vector spinors}

The commutator of the boost parts of the bosonic gamma matrices determines the Lorentz transformation of eight-component four-vector spinors, such as the charge-current spinor in Eq.~\eqref{eq:wavefunctions} and the potential spinor in Eq.~\eqref{eq:Theta}. With this commutator, we define $8\times8$ antisymmetric matrices $\Sigma^{ab}_\mathrm{L}$, the generators of Lorentz transformations on four-vector spinors, as
\begin{equation}
 \Sigma_\mathrm{L}^{ab}=[\boldsymbol{\gamma}_\mathrm{B,boost}^a,\boldsymbol{\gamma}_\mathrm{B,boost}^b].
 \label{eq:Lmat}
\end{equation}
The boost parts $\boldsymbol{\gamma}_\mathrm{B,boost}^a$ of the gamma matrices are determined through Eq.~\eqref{eq:gamma} with $\boldsymbol{\sigma}_\mathrm{B}^i$ replaced by $\mathbf{K}_\mathrm{boost}^i$ and $\mathbf{I}_4$ replaced by $\frac{1}{2}\mathbf{I}_4$. The generators $\Sigma^{ab}_\mathrm{L}$ obey the Lorentz Lie algebra relations,
$[\Sigma^{ab}_\mathrm{L},\Sigma^{cd}_\mathrm{L}]=\eta^{bc}\Sigma^{ad}_\mathrm{L}-\eta^{ac}\Sigma^{bd}_\mathrm{L}+\eta^{ad}\Sigma^{bc}_\mathrm{L}-\eta^{bd}\Sigma^{ac}_\mathrm{L}$.

Under the Lorentz transformation, four-vector spinors transform as $\Phi'=\boldsymbol{\Lambda}_\mathrm{L}\Phi$, where the $8\times8$ transformation matrix $\boldsymbol{\Lambda}_\mathrm{L}$ is in analogy to Eq.~\eqref{eq:lambda} given by 
\begin{equation}
 \textstyle\boldsymbol{\Lambda}_\mathrm{L}=\exp(\frac{1}{2}\Omega_{ab}\Sigma_\mathrm{L}^{ab}).
 \label{eq:UL1}
\end{equation}
The appearance of the Lorentz parameter matrix $\Omega_{ab}$ in Eq.~\eqref{eq:UL1} ensures that this transformation is a manifestation of the same Lorentz transformation as presented for four-vectors in Eq.~\eqref{eq:lambda}. For boosts, the Lorentz transformation matrix $\boldsymbol{\Lambda}_\mathrm{L}$ in Eq.~\eqref{eq:UL1} is Hermitian, i.e., $\boldsymbol{\Lambda}_\mathrm{L}=\boldsymbol{\Lambda}_\mathrm{L}^\dag$. For SO(3) rotations, it is unitary, i.e., $\boldsymbol{\Lambda}_\mathrm{L}\boldsymbol{\Lambda}_\mathrm{L}^\dag=\mathbf{I}_8$. In analogy to the Dirac theory \cite{Peskin2018}, using the identity $\boldsymbol{\gamma}_\mathrm{B}^0\boldsymbol{\Lambda}_\mathrm{L}^\dag\boldsymbol{\gamma}_\mathrm{B}^0=\boldsymbol{\Lambda}_\mathrm{L}^{-1}$, for the Lorentz transformation of $\bar{\Phi}=\Phi^\dag\boldsymbol{\gamma}_\mathrm{B}^0$, we obtain $\bar{\Phi}'=\bar{\Phi}\boldsymbol{\Lambda}_\mathrm{L}^{-1}$. Thus, $\boldsymbol{\Lambda}_\mathrm{L}$ preserves the scalar product in Eq.~\eqref{eq:scalarproductDirac} for arbitrary four-vector spinors $\Phi_1$ and $\Phi_2$ as $\overline{\boldsymbol{\Lambda}_\mathrm{L}\Phi_1}\boldsymbol{\Lambda}_\mathrm{L}\Phi_2=\bar{\Phi}_1\Phi_2$.

The consistency of the Lorentz transformation of our eight-component four-vector spinors in Eq.~\eqref{eq:UL1} can be straightforwardly verified by observing that it reproduces the conventional Lorentz transformation of the electric four-current density, when it is used to transform a generic charge-current spinor of the form in Eq.~\eqref{eq:wavefunctions}.

\subsection{Lorentz transformation of spin-$\frac{1}{2}$ field spinors}

The commutator of the conventional fermionic gamma matrices determines the Lorentz transformation of spin-$\frac{1}{2}$ field spinors, i.e., the Dirac spinors. With this commutator, one defines antisymmetric spin matrices $\Sigma^{ab}_\mathrm{F}$, the generators of Lorentz transformations on spin-$\frac{1}{2}$ field spinors as \cite{Peskin2018}
\begin{equation}
 \Sigma^{ab}_\mathrm{F}=\frac{1}{4}[\boldsymbol{\gamma}_\mathrm{F}^a,\boldsymbol{\gamma}_\mathrm{F}^b].
 \label{eq:Fmat}
\end{equation}
The generators $\Sigma^{ab}_\mathrm{F}$ obey the Lorentz Lie algebra relations,
$[\Sigma^{ab}_\mathrm{F},\Sigma^{cd}_\mathrm{F}]=\eta^{bc}\Sigma^{ad}_\mathrm{F}-\eta^{ac}\Sigma^{bd}_\mathrm{F}+\eta^{ad}\Sigma^{bc}_\mathrm{F}-\eta^{bd}\Sigma^{ac}_\mathrm{F}$.

Under the Lorentz transformation, spin-$\frac{1}{2}$ field spinors $\psi$ transform as $\psi'=\boldsymbol{\Lambda}_\mathrm{F}\psi$, where the $4\times4$ transformation matrix $\boldsymbol{\Lambda}_\mathrm{F}$ is in analogy to Eq.~\eqref{eq:lambda} given by \cite{Peskin2018}
\begin{equation}
 \textstyle\boldsymbol{\Lambda}_\mathrm{F}=\exp(\frac{1}{2}\Omega_{ab}\Sigma_\mathrm{F}^{ab}).
 \label{eq:UF}
\end{equation}
Again, the appearance of the Lorentz parameter matrix $\Omega_{ab}$ in the transformation in Eq.~\eqref{eq:UF} ensures that this transformation manifests the same Lorentz transformation for spin-$\frac{1}{2}$ field spinors as was defined for four-vectors in Eq.~\eqref{eq:lambda}. For boosts, the Lorentz transformation matrix $\boldsymbol{\Lambda}_\mathrm{F}$ in Eq.~\eqref{eq:UF} is Hermitian, i.e., $\boldsymbol{\Lambda}_\mathrm{F}=\boldsymbol{\Lambda}_\mathrm{F}^\dag$. For SO(3) rotations, it is unitary, i.e., $\boldsymbol{\Lambda}_\mathrm{F}\boldsymbol{\Lambda}_\mathrm{F}^\dag=\mathbf{I}_4$. Using the identity $\boldsymbol{\gamma}_\mathrm{F}^0\boldsymbol{\Lambda}_\mathrm{F}^\dag\boldsymbol{\gamma}_\mathrm{F}^0=\boldsymbol{\Lambda}_\mathrm{F}^{-1}$, for the Lorentz transformation of $\bar{\psi}=\psi^\dag\boldsymbol{\gamma}_\mathrm{F}^0$, we obtain $\bar{\psi}'=\bar{\psi}\boldsymbol{\Lambda}_\mathrm{F}^{-1}$ \cite{Peskin2018}. Thus, $\boldsymbol{\Lambda}_\mathrm{F}$ preserves the scalar product $\bar{\psi}_1\psi_2$ for arbitrary Dirac spinors $\psi_1$ and $\psi_2$ as $\overline{\boldsymbol{\Lambda}_\mathrm{F}\psi_1}\boldsymbol{\Lambda}_\mathrm{F}\psi_2=\bar{\psi}_1\psi_2$.

\subsection{\label{sec:Lorentzphotonspinor}Lorentz transformation of spin-1 field spinors}

The commutator of the rotational parts of the bosonic gamma matrices determines the Lorentz transformation of electromagnetic spinors or any other spinors corresponding to a spin-1 field. With this commutator, we define $8\times8$ antisymmetric spin matrices $\Sigma^{ab}_\mathrm{S}$, the generators of Lorentz transformations on spin-1 field spinors as
\begin{equation}
 \Sigma^{ab}_\mathrm{S}=[\boldsymbol{\gamma}_\mathrm{B,rot}^a,\boldsymbol{\gamma}_\mathrm{B,rot}^b].
 \label{eq:Smat}
\end{equation}
The rotational parts $\boldsymbol{\gamma}_\mathrm{B,rot}^a$ of the gamma matrices are determined through Eq.~\eqref{eq:gamma} with $\boldsymbol{\sigma}_\mathrm{B}^i$ replaced by $i\mathbf{K}_\mathrm{rot}^i$ and $\mathbf{I}_4$ replaced by $\frac{1}{2}\mathbf{I}_4$. The generators $\Sigma^{ab}_\mathrm{S}$ obey the Lorentz Lie algebra relations,
$[\Sigma^{ab}_\mathrm{S},\Sigma^{cd}_\mathrm{S}]=\eta^{bc}\Sigma^{ad}_\mathrm{S}-\eta^{ac}\Sigma^{bd}_\mathrm{S}+\eta^{ad}\Sigma^{bc}_\mathrm{S}-\eta^{bd}\Sigma^{ac}_\mathrm{S}$.

Under the Lorentz transformation, spin-1 field spinors transform as $\Psi'=\boldsymbol{\Lambda}_\mathrm{S}\Psi$, where the $8\times8$ transformation matrix $\boldsymbol{\Lambda}_\mathrm{S}$ is in analogy to Eqs.~\eqref{eq:lambda}, \eqref{eq:UL1}, and \eqref{eq:UF} given by 
\begin{equation}
 \textstyle\boldsymbol{\Lambda}_\mathrm{S}=\exp(\frac{1}{2}\Omega_{ab}\Sigma_\mathrm{S}^{ab}).
 \label{eq:US1}
\end{equation}
Again, the appearance of the Lorentz parameter matrix $\Omega_{ab}$ in the transformation in Eq.~\eqref{eq:US1} ensures that this transformation is a manifestation of the same Lorentz transformation for spin-1 field spinors as was defined for four-vectors in Eq.~\eqref{eq:lambda}. For boosts, the Lorentz transformation matrix $\boldsymbol{\Lambda}_\mathrm{S}$ in Eq.~\eqref{eq:US1} is Hermitian, i.e., $\boldsymbol{\Lambda}_\mathrm{S}=\boldsymbol{\Lambda}_\mathrm{S}^\dag$. For SO(3) rotations, it is unitary, i.e., $\boldsymbol{\Lambda}_\mathrm{S}\boldsymbol{\Lambda}_\mathrm{S}^\dag=\mathbf{I}_8$. In analogy to the Dirac theory \cite{Peskin2018}, using the identity $\boldsymbol{\gamma}_\mathrm{B}^0\boldsymbol{\Lambda}_\mathrm{S}^\dag\boldsymbol{\gamma}_\mathrm{B}^0=\boldsymbol{\Lambda}_\mathrm{S}^{-1}$, for the Lorentz transformation of $\bar{\Psi}=\Psi^\dag\boldsymbol{\gamma}_\mathrm{B}^0$, we obtain $\bar{\Psi}'=\bar{\Psi}\boldsymbol{\Lambda}_\mathrm{S}^{-1}$. Thus, $\boldsymbol{\Lambda}_\mathrm{S}$ preserves the local scalar product in Eq.~\eqref{eq:scalarproductDirac} for arbitrary electromagnetic spinors $\Psi_1$ and $\Psi_2$ as $\overline{\boldsymbol{\Lambda}_\mathrm{S}\Psi_1}\boldsymbol{\Lambda}_\mathrm{S}\Psi_2=\bar{\Psi}_1\Psi_2$.

The consistency of the Lorentz transformation in Eq.~\eqref{eq:US1} for our eight-component spin-1 field spinors can be straightforwardly verified by observing that it reproduces the conventional Lorentz transformation of the electric and magnetic fields, when it is used to transform a generic electromagnetic spinor of the form in Eq.~\eqref{eq:wavefunctions}.

\subsection{\label{sec:Lorentztensor}Lorentz transformation of spin-2 fields}

The commutator of the full bosonic gamma matrices determines the Lorentz transformation of spin-2 fields, which can be presented using $8\times8$ matrices in the present theory \cite{Partanen2023c}. With this commutator, we define $8\times8$ antisymmetric matrices $\Sigma^{ab}_\mathrm{J}$, the generators of Lorentz transformations on spin-2 fields as
\begin{equation}
 \Sigma^{ab}_\mathrm{J}=\frac{1}{4}[\boldsymbol{\gamma}_\mathrm{B}^a,\boldsymbol{\gamma}_\mathrm{B}^b].
 \label{eq:LambdaJmat}
\end{equation}
The generators $\Sigma^{ab}_\mathrm{J}$ obey the Lorentz Lie algebra relations, $[\Sigma^{ab}_\mathrm{J},\Sigma^{cd}_\mathrm{J}]=\eta^{bc}\Sigma^{ad}_\mathrm{J}-\eta^{ac}\Sigma^{bd}_\mathrm{J}+\eta^{ad}\Sigma^{bc}_\mathrm{J}-\eta^{bd}\Sigma^{ac}_\mathrm{J}$.

Under the Lorentz transformation, a tensor field $\mathbf{H}$ transforms as $\mathbf{H}'=\boldsymbol{\Lambda}_\mathrm{J}\mathbf{H}\boldsymbol{\Lambda}_\mathrm{J}^{-1}$, where the $8\times8$ transformation matrix $\boldsymbol{\Lambda}_\mathrm{J}$ is in analogy to Eqs.~\eqref{eq:lambda}, \eqref{eq:UL1}, \eqref{eq:UF}, and \eqref{eq:US1} given by 
\begin{equation}
 \textstyle\boldsymbol{\Lambda}_\mathrm{J}=\exp(\frac{1}{2}\Omega_{ab}\Sigma_\mathrm{J}^{ab}).
 \label{eq:UJ1}
\end{equation}
The appearance of the Lorentz parameter matrix $\Omega_{\rho\sigma}$ in the transformation in Eq.~\eqref{eq:UJ1} ensures that this transformation is a manifestation of the same Lorentz transformation for gamma matrices as was defined for four-vectors and other spinors above. For boosts, the Lorentz transformation matrix $\boldsymbol{\Lambda}_\mathrm{J}$ in Eq.~\eqref{eq:UJ1} is Hermitian, i.e., $\boldsymbol{\Lambda}_\mathrm{J}=\boldsymbol{\Lambda}_\mathrm{J}^\dag$. For SO(3) rotations, it is unitary, i.e., $\boldsymbol{\Lambda}_\mathrm{J}\boldsymbol{\Lambda}_\mathrm{J}^\dag=\mathbf{I}_8$. Using the identity $\boldsymbol{\gamma}_\mathrm{B}^0\boldsymbol{\Lambda}_\mathrm{J}^\dag\boldsymbol{\gamma}_\mathrm{B}^0=\boldsymbol{\Lambda}_\mathrm{J}^{-1}$, for the Lorentz transformation of $\bar{\mathbf{H}}=\boldsymbol{\gamma}_\mathrm{B}^0\mathbf{H}^\dag\boldsymbol{\gamma}_\mathrm{B}^0$, we obtain $\bar{\mathbf{H}}'=\boldsymbol{\Lambda}_\mathrm{J}\bar{\mathbf{H}}\boldsymbol{\Lambda}_\mathrm{J}^{-1}$.

For a general Lorentz transformation, using the gamma matrix algebra, it can be verified that the operation of $\boldsymbol{\Lambda}_\mathrm{J}$ on the matrix product of two arbitrary complex-conjugated gamma matrices is invariant as $\boldsymbol{\Lambda}_\mathrm{J}\boldsymbol{\gamma}_\mathrm{B}^{a*}\boldsymbol{\gamma}_\mathrm{B}^{b*}\boldsymbol{\Lambda}_\mathrm{J}^{-1}=\boldsymbol{\gamma}_\mathrm{B}^{a*}\boldsymbol{\gamma}_\mathrm{B}^{b*}$. Similarly, we have $\boldsymbol{\Lambda}_\mathrm{J}\boldsymbol{\gamma}_\mathrm{B}^{5*}\boldsymbol{\gamma}_\mathrm{B}^{a*}\boldsymbol{\gamma}_\mathrm{B}^{b*}\boldsymbol{\Lambda}_\mathrm{J}^{-1}=\boldsymbol{\gamma}_\mathrm{B}^{5*}\boldsymbol{\gamma}_\mathrm{B}^{a*}\boldsymbol{\gamma}_\mathrm{B}^{b*}$. From this relation, it follows that the representation of a spin-2 tensor field as a linear combination of the symmetry transform generators in the special unitary symmetry studied in Sec.~\ref{sec:unitary} is invariant. This observation is utilized in the Yang-Mills gauge theory of unified gravity that we elaborate in a separate work in Ref.~\cite{Partanen2023c}. It is known that a massless spin-2 field can be associated with gravity \cite{Misner1973}. Therefore, due to the connection to gravity, we call the Lorentz transformation $\boldsymbol{\Lambda}_\mathrm{J}$ the Lorentz transformation of spin-2 fields. However, since spin-2 fields are invariant with respect to $\boldsymbol{\Lambda}_\mathrm{J}$ as briefly mentioned above, $\boldsymbol{\Lambda}_\mathrm{J}$ does not represent an actual transformation of such fields.

\subsection{Complete Lorentz transformations of space-time-dependent fields}
In the case of fields that are functions of space-time, when performing the Lorentz transformations, it is necessary to transform the vector or spinor components of the fields, as discussed above, and also to reexpress the space-time-coordinate arguments of the functions via the coordinates of the transformed frame using $\boldsymbol{\Lambda}^{-1}$. In the cases of the space-time-dependent four-vector field $X(\mathrm{x})$, four-vector-spinor field $\Phi(\mathrm{x})$, spin-$\frac{1}{2}$ field spinor $\psi(\mathrm{x})$, spin-1 field spinor $\Psi(\mathrm{x})$, and the spin-2 field $\mathbf{H}(\mathrm{x})$, in the Minkowski space-time with $\mathrm{x}=(ct,x,y,z)^T$, we then have
\begin{equation}
 X'(\mathrm{x})=\boldsymbol{\Lambda}X(\boldsymbol{\Lambda}^{-1}\mathrm{\mathrm{x}}),
\end{equation}
\begin{equation}
 \Phi'(\mathrm{x})=\boldsymbol{\Lambda}_\mathrm{L}\Phi(\boldsymbol{\Lambda}^{-1}\mathrm{x}),
\end{equation}
\begin{equation}
 \psi'(\mathrm{x})=\boldsymbol{\Lambda}_\mathrm{F}\psi(\boldsymbol{\Lambda}^{-1}\mathrm{x}),
\end{equation}
\begin{equation}
 \Psi'(\mathrm{x})=\boldsymbol{\Lambda}_\mathrm{S}\Psi(\boldsymbol{\Lambda}^{-1}\mathrm{x}),
\end{equation}
\begin{equation}
 \mathbf{H}'(\mathrm{x})=\boldsymbol{\Lambda}_\mathrm{J}\mathbf{H}(\boldsymbol{\Lambda}^{-1}\mathrm{x})\boldsymbol{\Lambda}_\mathrm{J}^{-1}.
\end{equation}
For infinitesimal Lorentz transformations, the relations above give
\begin{equation}
 X'(\mathrm{x})=\Big(\mathbf{I}_4+\frac{1}{2i\hbar}\Omega_{ab}\hat{J}_\mathrm{K}^{ab}\Big)X(\mathrm{x}),
 \label{eq:LX}
\end{equation}
\begin{equation}
 \Phi'(\mathrm{x})=\Big(\mathbf{I}_8+\frac{1}{2i\hbar}\Omega_{ab}\hat{J}_\mathrm{L}^{ab}\Big)\Phi(\mathrm{x}),
\end{equation}
\begin{equation}
 \psi'(\mathrm{x})=\Big(\mathbf{I}_4+\frac{1}{2i\hbar}\Omega_{ab}\hat{J}_\mathrm{F}^{ab}\Big)\psi(\mathrm{x}),
\end{equation}
\begin{equation}
 \Psi'(\mathrm{x})=\Big(\mathbf{I}_8+\frac{1}{2i\hbar}\Omega_{ab}\hat{J}_\mathrm{S}^{ab}\Big)\Psi(\mathrm{x}).
 \label{eq:LPsi}
\end{equation}
\begin{equation}
 \mathbf{H}'(\mathrm{x})=\Big(\mathbf{I}_8+\frac{1}{2i\hbar}\Omega_{ab}\hat{J}_\mathrm{J}^{ab}\Big)\mathbf{H}(\mathrm{x}).
 \label{eq:LH}
\end{equation}
Here we have defined the total angular-momentum tensor operators, the generators of the complete infinitesimal Lorentz transformations of the pertinent fields, as
\begin{equation}
 \hat{J}_\mathrm{K}^{ab}=\hat{L}_4^{ab}+i\hbar K^{ab},
 \label{eq:JK}
\end{equation}
\begin{equation}
 \hat{J}_\mathrm{L}^{ab}=\hat{L}_8^{ab}+i\hbar\Sigma_\mathrm{L}^{ab},
\end{equation}
\begin{equation}
 \hat{J}_\mathrm{F}^{ab}=\hat{L}_4^{ab}+i\hbar\Sigma_\mathrm{F}^{ab},
\end{equation}
\begin{equation}
 \hat{J}_\mathrm{S}^{ab}=\hat{L}_8^{ab}+i\hbar\Sigma_\mathrm{S}^{ab},
 \label{eq:JS}
\end{equation}
\begin{equation}
 \hat{J}_\mathrm{J}^{ab}=\hat{L}_8^{ab}.
 \label{eq:JJ}
\end{equation}
The operators $\hat{L}_n^{ab}=-i\hbar\mathbf{I}_n(K^{ab})_{\;\,d}^c\mathrm{x}^d\partial_c$ are the orbital angular-momentum tensor operators that arise from the Lorentz transformations of the function arguments. The latter terms are the spin angular-momentum tensor operators, given in terms of the vector-space Lorentz generator matrices of the pertinent fields discussed above. The total angular-momentum tensor operators in Eqs.~\eqref{eq:JK}--\eqref{eq:JJ} satisfy the Lorentz Lie algebra relations \cite{Maggiore2005}
\begin{equation}
 [\hat{J}^{ab},\hat{J}^{cd}]=i\hbar(\eta^{bc}\hat{J}^{ad}-\eta^{ac}\hat{J}^{bd}+\eta^{ad}\hat{J}^{bc}-\eta^{bd}\hat{J}^{ac}).
 \label{eq:LorentzLie}
\end{equation}
The orbital and spin angular-momentum tensor operators commute with each other and separately satisfy the Lorentz Lie algebra relations similar to Eq.~\eqref{eq:LorentzLie}. For a more detailed discussion of the angular-momentum operators of the electromagnetic spinor field, see Sec.~\ref{sec:quantumoptics} below. Equations \eqref{eq:LX}--\eqref{eq:LH} show that the total angular-momentum tensor operators are the generators of the infinitesimal Lorentz transformations involving the transformations of the coordinate arguments of the fields.

\subsection{Poincar\'e transformations}

The Poincar\'e transformations, also called inhomogeneous Lorentz transformations, in the Minkowski space-time consist of the Lorentz transformation and a translation of coordinates with an arbitrary constant four-vector $\mathrm{a}=(\mathrm{a}_0,\mathrm{a}_x,\mathrm{a}_y,\mathrm{a}_z)^T$ and $b=\boldsymbol{\Lambda}\mathrm{a}$ as \cite{Misner1973,Maggiore2005}
\begin{equation}
 \mathrm{x}'=\boldsymbol{\Lambda}\mathrm{x}+\mathrm{b}
 \hspace{0.4cm}\Leftrightarrow\hspace{0.4cm}
 \mathrm{x}=\boldsymbol{\Lambda}^{-1}\mathrm{x}'-\mathrm{a}.
\end{equation}
For the Poincar\'e transformations of the space-time-dependent fields of different types, we then have
\begin{equation}
 X'(\mathrm{x})=\boldsymbol{\Lambda}X(\boldsymbol{\Lambda}^{-1}\mathrm{x}-\mathrm{a}),
\end{equation}
\begin{equation}
 \Phi'(\mathrm{x})=\boldsymbol{\Lambda}_\mathrm{L}\Phi(\boldsymbol{\Lambda}^{-1}\mathrm{x}-\mathrm{a}),
\end{equation}
\begin{equation}
 \psi'(\mathrm{x})=\boldsymbol{\Lambda}_\mathrm{F}\psi(\boldsymbol{\Lambda}^{-1}\mathrm{x}-\mathrm{a}),
\end{equation}
\begin{equation}
 \Psi'(\mathrm{x})=\boldsymbol{\Lambda}_\mathrm{S}\Psi(\boldsymbol{\Lambda}^{-1}\mathrm{x}-\mathrm{a}).
\end{equation}
\begin{equation}
 \mathbf{H}'(\mathrm{x})=\boldsymbol{\Lambda}_\mathrm{J}\mathbf{H}(\boldsymbol{\Lambda}^{-1}\mathrm{x}-\mathrm{a})\boldsymbol{\Lambda}_\mathrm{J}^{-1}.
\end{equation}
The infinitesimal Poincar\'e transformations are given by
\begin{equation}
 X'(\mathrm{x})=\Big(\mathbf{I}_4+\frac{i}{\hbar}\mathrm{a}_a\hat{P}_4^a+\frac{1}{2i\hbar}\Omega_{ab}\hat{J}_\mathrm{K}^{ab}\Big)X(\mathrm{x}),
 \label{eq:PoincareX}
\end{equation}
\begin{equation}
 \Phi'(\mathrm{x})=\Big(\mathbf{I}_8+\frac{i}{\hbar}\mathrm{a}_a\hat{P}_8^a+\frac{1}{2i\hbar}\Omega_{ab}\hat{J}_\mathrm{L}^{ab}\Big)\Phi(\mathrm{x}),
\end{equation}
\begin{equation}
 \psi'(\mathrm{x})=\Big(\mathbf{I}_4+\frac{i}{\hbar}\mathrm{a}_a\hat{P}_4^a+\frac{1}{2i\hbar}\Omega_{ab}\hat{J}_\mathrm{F}^{ab}\Big)\psi(\mathrm{x}),
\end{equation}
\begin{equation}
 \Psi'(\mathrm{x})=\Big(\mathbf{I}_8+\frac{i}{\hbar}\mathrm{a}_a\hat{P}_8^a+\frac{1}{2i\hbar}\Omega_{ab}\hat{J}_\mathrm{S}^{ab}\Big)\Psi(\mathrm{x}),
 \label{eq:PoincarePsi}
\end{equation}
\begin{equation}
 \mathbf{H}'(\mathrm{x})=\Big(\mathbf{I}_8+\frac{i}{\hbar}\mathrm{a}_a\hat{P}_8^a+\frac{1}{2i\hbar}\Omega_{ab}\hat{J}_\mathrm{J}^{ab}\Big)\mathbf{H}(\mathrm{x}).
 \label{eq:PoincareH}
\end{equation}
Here $\hat{P}_n^a=i\hbar\mathbf{I}_n\partial^a$ are the four-momentum operators that will be discussed in Sec.~\ref{sec:operators4D} below.
The terms dependent on $\hat{P}_n^a$ originate from the space-time translation. Thus, $\hat{P}_n^a$ can be called translation generators. The complete infinitesimal Poincar\'e transformations are, thus, generated by $\hat{P}_n^a$ and the total angular-momentum tensor operators of the pertinent fields. The Poincar\'e algebra is the Lie algebra of the Poincar\'e group. It is a Lie algebra extension of the Lorentz Lie algebra. The commutation relations of the Poincar\'e Lie algebra are given by \cite{Maggiore2005}
\begin{equation}
 [\hat{P}^a,\hat{P}^b]=\mathbf{0},
 \label{eq:PoincareLie1}
\end{equation}
\begin{equation}
 [\hat{J}^{ab},\hat{P}^{c}]=i\hbar(\eta^{bc}\hat{P}^{a}-\eta^{ac}\hat{P}^{b}),
\end{equation}
\begin{equation}
 [\hat{J}^{ab},\hat{J}^{cd}]=i\hbar(\eta^{bc}\hat{J}^{ad}-\eta^{ac}\hat{J}^{bd}+\eta^{ad}\hat{J}^{bc}-\eta^{bd}\hat{J}^{ac}).
 \label{eq:PoincareLie3}
\end{equation}
From Eq.~\eqref{eq:PoincareLie1}, it is seen that the momentum operators $\hat{P}^a$ represent a commutative subalgebra of the Poincar\'e algebra. Using the Lie theorem, this implies that the space-time translations form an Abelian subgroup of the Poincar\'e group. The commutation relation of the total angular-momentum tensor operators in Eq.~\eqref{eq:PoincareLie3} is equivalent to the Lorentz Lie algebra relation in Eq.~\eqref{eq:LorentzLie}.

\subsection{Lorentz invariants}

Next, we briefly describe how the Lorentz invariant scalars and pseudoscalars can be constructed from fields of different types. As a space-time-dependent quantity, a pseudoscalar behaves like a scalar, except that it changes sign under a parity transformation \cite{Peskin2018}. For the parity transformation, see Sec.~\ref{sec:CPTsymmetry}. A four-vector $X^a$ produces a scalar by a contraction with itself as $X^aX_a$, e.g., in the case of the electric four-current density we have $J_\mathrm{e\Re}^a J_{\mathrm{e\Re}a}=c^2\rho_\mathrm{e\Re}^2-\mathbf{J}_\mathrm{e\Re}^2$. A four-vector spinor $\Phi_\Re$ produces a scalar by a product with the eight-spinor adjoint as $\bar{\Phi}_\Re\Phi_\Re$, e.g., in the case of the charge-current spinor we have $\bar{\Phi}_\Re\Phi_\Re=-\frac{\mu_0}{2}J_\mathrm{e\Re}^a J_{\mathrm{e\Re}a}=\frac{\mu_0}{2}(\mathbf{J}_\mathrm{e\Re}^2-c^2\rho_\mathrm{e\Re}^2)$. A spin-$\frac{1}{2}$ field spinor $\psi$ produces a scalar by a product with the Dirac adjoint as $\bar{\psi}\psi$ and a pseudoscalar through the use of the fermionic fifth gamma matrix as $i\bar{\psi}\boldsymbol{\gamma}_\mathrm{F}^5\psi$ \cite{Peskin2018}. The imaginary unit factor has been chosen to make the expression Hermitian. Correspondingly, a spin-1 field spinor $\Psi_\Re$ produces a scalar by a product with the eight-spinor adjoint as $\bar{\Psi}_\Re\Psi_\Re$ and a pseudoscalar through the use of the bosonic fifth gamma matrix as $i\bar{\Psi}_\Re\boldsymbol{\gamma}_\mathrm{B}^5\Psi_\Re$. In the case of the electromagnetic spinor, we have
\begin{equation}
 \bar{\Psi}_\Re\Psi_\Re
 =-\frac{1}{4\mu_0}F_{ab}F^{ab}
 =\frac{1}{2}\Big(\varepsilon_0\mathbf{E}_\Re^2-\frac{1}{\mu_0}\mathbf{B}_\Re^2\Big),
 \label{eq:scalar}
\end{equation}
\begin{equation}
 i\bar{\Psi}_\Re\boldsymbol{\gamma}_\mathrm{B}^5\Psi_\Re
 =\frac{1}{4\mu_0}F_{ab}\widetilde{F}^{ab}
 =-\varepsilon_0c\mathbf{E}_\Re\cdot\mathbf{B}_\Re.
 \label{eq:pseudoscalar}
\end{equation}
In Eqs.~\eqref{eq:scalar} and \eqref{eq:pseudoscalar}, $F^{ab}$ is the electromagnetic tensor, which is given in terms of the four-potential and the corresponding electric and magnetic fields as \cite{Jackson1999,Landau1989}
\begin{align}
 F^{ab} &=\partial^a A_\Re^b-\partial^b A_\Re^a\nonumber\\
 &=\left[\begin{array}{cccc}
0 & -E_\Re^x/c & -E_\Re^y/c & -E_\Re^z/c\\
E_\Re^x/c & 0 & -B_\Re^z & B_\Re^y\\
E_\Re^y/c & B_\Re^z & 0 & -B_\Re^x\\
E_\Re^z/c & -B_\Re^y & B_\Re^x & 0
\end{array}\right].
\label{eq:Ftensor}
\end{align}
The dual electromagnetic field tensor $\widetilde{F}^{\mu\nu}$ in Eq.~\eqref{eq:pseudoscalar} is defined as \cite{Jackson1999}
\begin{equation}
 \widetilde{F}^{\mu\nu}=\frac{1}{2}\varepsilon^{\mu\nu\rho\sigma}F_{\rho\sigma}=\left[\begin{array}{cccc}
0 & -B_\Re^x & -B_\Re^y & -B_\Re^z\\
B_\Re^x & 0 & E_\Re^z/c & -E_\Re^y/c\\
B_\Re^y & -E_\Re^z/c & 0 & E_\Re^x/c\\
B_\Re^z & E_\Re^y/c & -E_\Re^x/c & 0
\end{array}\right].
\label{eq:chiralFtensor}
\end{equation}

The Lorentz scalar in Eq.~\eqref{eq:scalar} is equal to the well-known Lagrangian density of the electromagnetic field discussed in Sec.~\ref{sec:Lagrangian} below. The pseudoscalar in Eq.~\eqref{eq:pseudoscalar} emerges as an important quantity in the context of QED low-energy effective-field theories \cite{Heisenberg1936,Gies2017}. It is also used in the context of dual electromagnetism \cite{Bliokh2013b}.

\section{\label{sec:quantumoptics}Quantum operators in the first quantization}

In this section, we determine the key quantum-mechanical operators for electromagnetic spinors given in Eq.~\eqref{eq:wavefunctions}. These operators are in the literature called the operators in the first quantization. For the quantum field of many photons, one will need the corresponding second quantization operators described in Sec.~\ref{sec:secondquantization}.

\subsection{Hamiltonian operator}
By defining the $8\times8$ matrices $\boldsymbol{\alpha}_\mathrm{B}^i=\boldsymbol{\gamma}_\mathrm{B}^0\boldsymbol{\gamma}_\mathrm{B}^i$ and $\boldsymbol{\beta}_\mathrm{B}=\boldsymbol{\gamma}_\mathrm{B}^0$, the associated vector $\boldsymbol{\alpha}_\mathrm{B}=(\boldsymbol{\alpha}_\mathrm{B}^x,\boldsymbol{\alpha}_\mathrm{B}^y,\boldsymbol{\alpha}_\mathrm{B}^z)$, and the momentum operator $\hat{\mathbf{p}}=-i\hbar\nabla$, the spinorial Maxwell equation in Eq.~\eqref{eq:photonDirac}, corresponding to Maxwell's equations in Eqs.~\eqref{eq:maxwell1}--\eqref{eq:maxwell4}, can be written as
\begin{equation}
 c\boldsymbol{\alpha}_\mathrm{B}\cdot\hat{\mathbf{p}}\Psi-i\hbar c\boldsymbol{\beta}_\mathrm{B}\Phi=i\hbar\frac{\partial\Psi}{\partial t}.
 \label{eq:photonDiracalpha}
\end{equation}
Here the time derivative has been separated to the right-hand side. For nonzero charge-current spinors $\Phi$, the spinorial Maxwell equation in Eq.~\eqref{eq:photonDiracalpha} is not of the conventional form of a time-dependent wave equation of a particle. However, for external field components or for the total field in the absence of charges and currents, we can write the spinorial Maxwell equation in Eq.~\eqref{eq:photonDiracalpha} as an explicit time-dependent wave equation in the conventional form, given by
\begin{equation}
 \hat{H}\Psi=i\hbar\frac{\partial\Psi}{\partial t}.
 \label{eq:Hamiltonianeq}
\end{equation}
Here the electromagnetic field Hamiltonian operator $\hat{H}$ is simply obtained from the first term of Eq.~\eqref{eq:photonDiracalpha} as
\begin{equation}
 \hat{H}=c\boldsymbol{\alpha}_\mathrm{B}\cdot\hat{\mathbf{p}}.
 \label{eq:Hamiltonian}
\end{equation}
The form of the spinorial Maxwell equation in Eq.~\eqref{eq:Hamiltonianeq} corresponds to an electromagnetic field propagating in vacuum, and it is analogous to the corresponding equation for \emph{free} particles in the Dirac theory \cite{Dirac1958}. However, note that, in the present theory, the square of the electromagnetic spinor describes the electromagnetic energy density expectation value, while the square of the Dirac spinor describes the probability density of the Dirac fermions. For narrow-frequency-band spinorial electromagnetic wavepacket states, the probability density can be defined as discussed in Sec.~\ref{sec:probabilitycurrent}.

\subsection{\label{sec:operators3D}Scalar and three-vector quantum operators}

Next, we consider other relevant quantum-mechanical operators and their commutation relations. We start with the most trivial operator, which is the photon number operator that is, by definition, an identity operator for single-photon states as
\begin{equation}
 \hat{n}=1.
 \label{eq:singlephotonnumber}
\end{equation}
The single-photon number operator naturally commutes with all other operators defined for single-photon states.

Then, we write the conventional energy and momentum operators corresponding to the pertinent classical quantities. The energy operator $\hat{E}$ and the momentum operator $\hat{\mathbf{p}}$, also used in Eqs.~\eqref{eq:photonDiracalpha}--\eqref{eq:Hamiltonian} above, are given by \cite{Messiah1961,Sakurai1967}
\begin{equation}
 \hat{E}=i\hbar\frac{\partial}{\partial t},
 \label{eq:energyoperator}
\end{equation}
\begin{equation}
 \hat{\mathbf{p}}=-i\hbar\nabla.
 \label{eq:momentumoperator}
\end{equation}
With space and time coordinates, the energy and momentum operators satisfy the well-known commutation relations $[r_i,\hat{p}_j]=i\hbar\delta_{ij}$, $[t,\hat{E}]=-i\hbar$, $[r_i,\hat{E}]=0$, and $[t,\hat{p}_i]=0$. The energy operator $\hat{E}$ and the momentum operator components $\hat{p}_i$ commute with each other and with the Hamiltonian operator as $[\hat{E},\hat{p}_i]=0$, $[\hat{p}_i,\hat{p}_j]=0$, $[\hat{E},\hat{H}]=\mathbf{0}$, and $[\hat{p}_i,\hat{H}]=\mathbf{0}$. The commutation relation between the position and the Hamiltonian operator is given by $[\mathbf{r},\hat{H}]=i\hbar c\boldsymbol{\alpha}_\mathrm{B}$.

The angular momentum is the rotational analog of linear momentum. As conventional in quantum mechanics, the total angular-momentum operator $\hat{\mathbf{J}}$ is defined as a sum of the orbital-angular momentum operator $\hat{\mathbf{L}}$ and the spin angular-momentum operator $\hat{\mathbf{S}}$ as \cite{Messiah1961,Sakurai1967}
\begin{equation}
 \hat{\mathbf{J}}=\hat{\mathbf{L}}+\hat{\mathbf{S}}.
 \label{eq:totalangularmomentumoperator}
\end{equation}
The orbital angular momentum operator is the quantum-mechanical counterpart of the classical angular momentum. It follows from the classical definition of the angular momentum as a cross product between the position and momentum vectors of a particle by replacing the classical momentum with the momentum operator in Eq.~\eqref{eq:momentumoperator} \cite{Messiah1961,Sakurai1967}. With an additional multiplication by an identity matrix, the components of the orbital angular momentum operator are matrices, given by
\begin{equation}
 \hat{\mathbf{L}}_i=-i\hbar\mathbf{I}_8(\mathbf{r}\times\nabla)_i,
 \label{eq:orbitalangularmomentumoperator}
\end{equation}
For light beams, the orbital angular momentum is known to be associated with helical wavefronts following from a phase proportional to the azimuthal angle \cite{Allen1992b,Piccirillo2013,Andrews2013,Barnett2016,Ji2020,Devlin2017}.

As known for spin-$\frac{1}{2}$ particles, the spin angular-momentum operator has in general no classical counterpart. It is generally a purely quantum-mechanical property related to elementary particles. The spin angular-momentum operator depends on the type of the elementary particle as it operates in the internal vector space associated with the elementary particle. In the present formulation of electromagnetic spinors, the components of the spin angular-momentum operator are matrices, given by
\begin{equation}
 \hat{\mathbf{S}}_i=\left[
 \begin{array}{cc}
  i\hbar\mathbf{K}_\mathrm{rot}^i & \mathbf{0}\\
  \mathbf{0} & i\hbar\mathbf{K}_\mathrm{rot}^i
 \end{array}\right].
 \label{eq:spinoperator}
\end{equation}
Even though the spin has in general no classical counterpart, for the electromagnetic field, the spin is known to be related to the polarization of light \cite{Allen1992b,Andrews2013,Mechelen2018,CohenTannoudji1989}.

The orbital and spin angular-momentum operators satisfy the conventional commutation relations, given by $[\hat{\mathbf{L}}_i,\hat{\mathbf{L}}_j]=i\hbar\varepsilon_{ijk}\hat{\mathbf{L}}_k$,
$[\hat{\mathbf{S}}_i,\hat{\mathbf{S}}_j]=i\hbar\varepsilon_{ijk}\hat{\mathbf{S}}_k$, and $[\hat{\mathbf{L}}_i,\hat{\mathbf{S}}_j]=\mathbf{0}$. For the energy, momentum, and position operators, the commutation relations are given by $[\hat{\mathbf{L}}_i,\hat{E}]=\mathbf{0}$, $[\hat{\mathbf{S}}_i,\hat{E}]=\mathbf{0}$, $[\hat{\mathbf{L}}_i,\hat{p}_j]=i\hbar\mathbf{I}_8\varepsilon_{ijk}\hat{p}_k$, $[\hat{\mathbf{S}}_i,\hat{p}_j]=\mathbf{0}$, $[\hat{\mathbf{L}}_i,r_j]=i\hbar\mathbf{I}_8\varepsilon_{ijk}r_k$, and $[\hat{\mathbf{S}}_i,r_j]=\mathbf{0}$. In the same way as in the Dirac theory \cite{Bransden2000}, the spin and orbital angular-momentum operators do not commute with the Hamiltonian operator. Their commutation relations with the Hamiltonian operator in Eq.~\eqref{eq:Hamiltonian} are given by $[\hat{\mathbf{L}},\hat{H}]=i\hbar c\boldsymbol{\alpha}_\mathrm{B}\times\hat{\mathbf{p}}$ and $[\hat{\mathbf{S}},\hat{H}]=-i\hbar c\boldsymbol{\alpha}_\mathrm{B}\times\hat{\mathbf{p}}$. Thus, the total angular-momentum operator $\hat{\mathbf{J}}$ in Eq.~\eqref{eq:totalangularmomentumoperator} commutes with the Hamiltonian operator as $[\hat{\mathbf{J}},\hat{H}]=\mathbf{0}$. The total angular-momentum operator satisfies the same commutation relations as the orbital and spin angular-momentum operators, given by $[\hat{\mathbf{J}}_i,\hat{\mathbf{J}}_j]=i\hbar\varepsilon_{ijk}\hat{\mathbf{J}}_k$.

By their definition, it is seen that the spin and orbital angular-momentum operators commute with each other, but neither of them commutes with the Hamiltonian operator. This implies that they cannot be independently constants of motion simultaneously with energy. Unlike in the case of the Dirac electron-positron field in an external electromagnetic field \cite{Landau1982}, it is not possible to define a separate interaction energy for the coupling of the spin and orbital angular momenta of light. The interaction of magnetic momenta related to orbital and spin angular momenta of an electron is best understood in the nonrelativistic limit, which cannot be taken for the electromagnetic field.

In this section, we rely on the definition of the spin and orbital angular-momentum operators in the conventional first-quantized quantum mechanics of particles. Thus, our approach is analogous to the Dirac theory before the quantization of the field resulting in the many-particle picture. In previous literature, there are several classical and quantum field formulations of the spin and orbital angular momenta of light in terms of the electric, magnetic and vector-potential fields and the corresponding operators \cite{vanEnk1994a,vanEnk1994b,Yang2022,Andrews2013,Barnett2016,Piccirillo2013}. These approaches are not directly comparable to the first-quantized operators of this section since they are based on the many-photon picture obtained in the second quantization. Such a picture is obtained also in the present theory after the quantization of the field as presented in Sec.~\ref{sec:secondquantization}. Detailed comparison of the present theory with the extensive previous literature on the spin and orbital angular momenta of light is a topic of a separate work.

Related to the energy and momentum operators, it is convenient to define the orbital-boost-momentum operator $\hat{\mathbf{N}}$ \cite{Bliokh2013b,Bliokh2018a,Smirnova2018,Partanen2019a,Partanen2021b}. This operator is needed in the determination of the orbital angular-momentum tensor operator as presented in the next section. The components of $\hat{\mathbf{N}}$ are defined as matrix operators, given by
\begin{equation}
 \hat{\mathbf{N}}_i=\mathbf{I}_8\Big(\frac{\hat{E}}{c^2}r_i-\hat{p}_it\Big).
 \label{eq:boostmomentumoperator}
\end{equation}
From the commutation relations of the energy and momentum operators, it follows that the orbital-boost-momentum operator satisfies the commutation relations $[\hat{\mathbf{N}}_i,\hat{E}]=i\hbar\mathbf{I}_8\hat{p}_i$, $[\hat{\mathbf{N}}_i,\hat{p}_j]=i\hbar\mathbf{I}_8\delta_{ij}\hat{E}/c^2$, $[\hat{\mathbf{N}}_i,\hat{\mathbf{N}}_j]=-i\hbar\varepsilon_{ijk}\hat{\mathbf{L}}_k/c^2$, $[\hat{\mathbf{N}}_i,\hat{\mathbf{L}}_j]=i\hbar\varepsilon_{ijk}\hat{\mathbf{N}}_k$, $[\hat{\mathbf{N}}_i,r_j]=i\hbar\mathbf{I}_8\delta_{ij}t$, $[\hat{\mathbf{N}}_i,t]=i\hbar\mathbf{I}_8r_i/c^2$, and $[\hat{\mathbf{N}},\hat{H}]=i\hbar\boldsymbol{\alpha}_\mathrm{B}\hat{E}/c$.

For use in the determination of the spin angular-momentum tensor operator in the next section, we define the spin-boost-momentum operator $\hat{\boldsymbol{\mathcal{N}}}$. The components of $\hat{\boldsymbol{\mathcal{N}}}$ are defined as matrix operators, given by
\begin{equation}
 \hat{\boldsymbol{\mathcal{N}}}_i=-\boldsymbol{\gamma}_\mathrm{B}^5\hat{\mathbf{S}}_i/c.
 \label{eq:spinboostmomentumoperator}
\end{equation}
It follows that the spin-boost-momentum operator satisfies the commutation relations $[\hat{\boldsymbol{\mathcal{N}}}_i,\hat{\boldsymbol{\mathcal{N}}}_j]=i\hbar\varepsilon_{ijk}\hat{\mathbf{S}}_k/c^2$, $[\hat{\boldsymbol{\mathcal{N}}}_i,\hat{\mathbf{S}}_j]=i\hbar\varepsilon_{ijk}\hat{\boldsymbol{\mathcal{N}}}_k$, and $[\hat{\boldsymbol{\mathcal{N}}},\hat{H}]=i\hbar\boldsymbol{\gamma}_\mathrm{B}^5\boldsymbol{\alpha}_\mathrm{B}\times\hat{\mathbf{p}}$. These relations are analogous to the commutation relations of the orbital-boost-momentum operator above.

The helicity operator is defined as the projection of the spin angular-momentum operator along the direction of the momentum, and it is written as \cite{Landau1982,Sakurai1967,Messiah1961,Alpeggiani2018,Peskin2018}
\begin{equation}
 \hat{\mathfrak{h}}=\frac{\hat{\mathbf{S}}\cdot\hat{\mathbf{p}}}{|\mathbf{p}|}.
 \label{eq:helicityoperator}
\end{equation}
By studying the action of the helicity operator in Eq.~\eqref{eq:helicityoperator} on electromagnetic spinors, it is found that the helicity operator divided by $\hbar$ swaps the upper and lower four components of the electromagnetic spinors. This corresponds to the action of the chirality operator $\boldsymbol{\gamma}_\mathrm{B}^5$ discussed in Sec.~\ref{sec:fifthgamma}. The correspondence $\hat{\mathfrak{h}}/\hbar\leftrightarrow\boldsymbol{\gamma}_\mathrm{B}^5$ for electromagnetic spinors is discussed further, in the case of photon states, in Appendix \ref{apx:spinorialwavefunctions}. The helicity operator commutes with the Hamiltonian operator as $[\hat{\mathfrak{h}},\hat{H}]=\mathbf{0}$.

\subsection{\label{sec:operators4D}Four-vector and tensor quantum operators}

The energy and momentum operators in Eqs.~\eqref{eq:energyoperator} and \eqref{eq:momentumoperator} are components of the four-momentum operator $\hat{P}^a$ whose contravariant form is given by
\begin{equation}
 \hat{P}^a=i\hbar\mathbf{I}_8\partial^a=i\hbar\mathbf{I}_8\eta^{ab}\partial_b=(\mathbf{I}_8\hat{E}/c,\mathbf{I}_8\hat{\mathbf{p}}).
 \label{eq:fourmomentumoperator}
\end{equation}
The four-vector corresponding to an eigenvalue of the four-momentum operator in Eq.~\eqref{eq:fourmomentumoperator} can be transformed between inertial frames using the conventional Lorentz transformation of four-vectors discussed in Sec.~\ref{sec:Lorentzvector}. Together with the total angular-momentum tensor operator, discussed below, the four-momentum operator in Eq.~\eqref{eq:fourmomentumoperator} satisfies the Poincar\'e Lie algebra relations in Eqs.~\eqref{eq:PoincareLie1}--\eqref{eq:PoincareLie3}.

The components of the angular momentum do not form a four-vector \cite{Fayngold2008,Misner1973}. Therefore, it is useful to form a tensor operator, through which the components of the angular momentum can be transformed between inertial frames by using the Lorentz transformation of this tensor. Consequently, the relativistic orbital angular momentum is \emph{defined} as a tensor operator $\hat{L}^{ab}$ that is different from the three-vector operator $\hat{\mathbf{L}}$, whose components are given in Eq.~\eqref{eq:orbitalangularmomentumoperator}. This second-rank antisymmetric tensor operator, called the orbital angular-momentum tensor operator, also includes components of the boost-momentum operator $\hat{\mathbf{N}}$ in Eq.~\eqref{eq:boostmomentumoperator}, and, in analogy to the relativistic structure of the corresponding classical quantity \cite{Fayngold2008}, with $\mathrm{x}=(ct,x,y,z)^T$, it is given by
\begin{align}
 \hat{L}^{ab} &=-i\hbar\mathbf{I}_8(K^{ab})_{\;\,d}^c\mathrm{x}^d\partial_c=\mathrm{x}^a\hat{P}^b-\mathrm{x}^b\hat{P}^a\nonumber\\
 &=\left[
 \begin{array}{cccc}
  \mathbf{0} & -c\hat{\mathbf{N}}_x & -c\hat{\mathbf{N}}_y & -c\hat{\mathbf{N}}_z\\
  c\hat{\mathbf{N}}_x & \mathbf{0} & \hat{\mathbf{L}}_z & -\hat{\mathbf{L}}_y \\
  c\hat{\mathbf{N}}_y & -\hat{\mathbf{L}}_z & \mathbf{0} & \hat{\mathbf{L}}_x \\
  c\hat{\mathbf{N}}_z & \hat{\mathbf{L}}_y & -\hat{\mathbf{L}}_x & \mathbf{0}
 \end{array}\right].
 \label{eq:orbitaltensoroperator}
\end{align}
The corresponding expressions for the components of $\hat{L}^{ab}$ are given by $\hat{L}^{i0}=c\hat{\mathbf{N}}_i$, $\hat{L}^{0i}=-c\hat{\mathbf{N}}_i$, and $\hat{L}^{ij}=\varepsilon_{ijk}\hat{\mathbf{L}}_k$.

Similarly, the relativistic spin angular momentum is \emph{defined} as a tensor operator $\hat{S}^{ab}$, which is different from the three-vector operator $\hat{\mathbf{S}}$, whose components are given in Eq.~\eqref{eq:spinoperator}. This second-rank antisymmetric tensor operator is made of generators $\Sigma_\mathrm{S}^{ab}$ of the Lorentz transformation on electromagnetic spinors, given in Eq.~\eqref{eq:Smat}. This is analogous to the case of the spin angular momentum in the Dirac theory \cite{Peskin2018}. Accordingly, the spin tensor operator $\hat{S}^{ab}$ is given by
\begin{equation}
 \hat{S}^{ab}=i\hbar\Sigma_\mathrm{S}^{ab}
 =\left[
 \begin{array}{cccc}
  \mathbf{0} & -ic\hat{\boldsymbol{\mathcal{N}}}_x & -ic\hat{\boldsymbol{\mathcal{N}}}_y & -ic\hat{\boldsymbol{\mathcal{N}}}_z\\
  ic\hat{\boldsymbol{\mathcal{N}}}_x & \mathbf{0} & \hat{\mathbf{S}}_z & -\hat{\mathbf{S}}_y \\
  ic\hat{\boldsymbol{\mathcal{N}}}_y & -\hat{\mathbf{S}}_z & \mathbf{0} & \hat{\mathbf{S}}_x \\
  ic\hat{\boldsymbol{\mathcal{N}}}_z & \hat{\mathbf{S}}_y & -\hat{\mathbf{S}}_x & \mathbf{0}
 \end{array}\right].
 \label{eq:spintensoroperator}
\end{equation}
The operator $\hat{S}^{ab}$ is antisymmetric. All the component matrices can be expressed in terms of the components $\hat{\mathbf{S}}_i$ of the three-vector spin operator in Eq.~\eqref{eq:spinoperator} as $\hat{S}^{i0}=ic\hat{\boldsymbol{\mathcal{N}}}_i$, $\hat{S}^{0i}=-ic\hat{\boldsymbol{\mathcal{N}}}_i$, and $\hat{S}^{ij}=\varepsilon_{ijk}\hat{\mathbf{S}}_k$. These relations are of the same form as the corresponding relations in the Dirac theory \cite{Peskin2018}. The appearance of the spin-boost-momentum operators in the definition of the spin tensor operator in Eq.~\eqref{eq:spintensoroperator}, in analogy to the Dirac theory, strongly indicates that the present spinorial formulation of the theory is necessary for the consistent determination of the relativistic spin structure of light.

The total angular-momentum tensor operator is a sum of the orbital angular-momentum tensor operator in Eq.~\eqref{eq:orbitaltensoroperator} and the spin tensor operator in Eq.~\eqref{eq:spintensoroperator} as
\begin{equation}
 \hat{J}^{ab}=\hat{L}^{ab}+\hat{S}^{ab}
 =\mathrm{x}^a\hat{P}^b-\mathrm{x}^b\hat{P}^a+\hat{S}^{ab}.
 \label{eq:totalamtensoroperator}
\end{equation}
This relation is again analogous to the corresponding relation in the Dirac theory \cite{Peskin2018}. Equation \eqref{eq:totalamtensoroperator} is equivalent to Eq.~\eqref{eq:JS} with $\hat{J}^{ab}=\hat{J}_\mathrm{S}^{ab}$ and $\hat{L}^{ab}=\hat{L}_8^{ab}$. Again, note that, together with the four-momentum operator in Eq.~\eqref{eq:fourmomentumoperator}, the total angular-momentum tensor operator in Eq.~\eqref{eq:totalamtensoroperator} satisfies the Poincar\'e Lie algebra relations in Eqs.~\eqref{eq:PoincareLie1}--\eqref{eq:PoincareLie3}.

\subsection{Density operators}
For systematic calculation of density expectation values of quantum operators, we present an arbitrary first-quantized density operator $\hat{\rho}_{\hat{O}}$ corresponding to the pertinent first-quantized operator $\hat{O}$ as
\begin{equation}
 \hat{\rho}_{\hat{O}}(\mathbf{r},\mathbf{r}')=\frac{1}{2}\{\hat{O}(\mathbf{r}'),\delta(\mathbf{r}-\mathbf{r}')\}.
 \label{eq:densityoperator}
\end{equation}
The argument $\mathbf{r}$ indicates the position at which the density is calculated and $\mathbf{r}'$ is the coordinate of the field. The anticommutator is necessary in Eq.~\eqref{eq:densityoperator} for obtaining real-valued density expectation values in the case of operators that do not commute with the position operator. Equation \eqref{eq:densityoperator} is justified by the physically meaningful results obtained for the density expectation values in Sec.~\ref{sec:densityexpectationvalues} and for the second-quantized density operators in Sec.~\ref{sec:secondquantizedoperators}.

As an example of the use of the first-quantized density operator in Eq.~\eqref{eq:densityoperator}, we study the number density. Substituting the trivial single-photon number operator from Eq.~\eqref{eq:singlephotonnumber} into Eq.~\eqref{eq:densityoperator} gives the single-photon number density operator as $\hat{\rho}_{\hat{n}}(\mathbf{r},\mathbf{r}')=\delta(\mathbf{r}-\mathbf{r}')$ \cite{Fetter1971}. Thus, for the first-quantized number-density expectation value, we obtain
\begin{align}
 \rho_{\hat{n}}^\mathrm{(M)}(\mathbf{r}) &=\frac{\int\Psi^\dag(\mathbf{r}')\hat{\rho}_{\hat{n}}(\mathbf{r},\mathbf{r}')\Psi(\mathbf{r}')d^3r'}{\int\Psi^\dag(\mathbf{r}')\Psi(\mathbf{r}')d^3r'}\nonumber\\
 &=\frac{\Psi^\dag(\mathbf{r})\Psi(\mathbf{r})}{\int\Psi^\dag(\mathbf{r}')\Psi(\mathbf{r}')d^3r'}.
 \label{eq:rhon}
\end{align}
The superscript $\mathrm{(M)}$ is used to indicate that the expectation value density is calculated for the Maxwell electromagnetic spinor field. Equation \eqref{eq:rhon} shows that the number density of first-quantized electromagnetic field is simply equal to the squared norm of the electromagnetic spinor normalized by the volume integral of the same quantity. The state must be normalizable for Eq.~\eqref{eq:rhon} to be well-defined. Equation \eqref{eq:rhon} is in accordance with Born's probability interpretation of the wave function \cite{Liboff1980}. Since the photon number of single-photon states is fixed to one, according to the \emph{uncertainty principle} of the photon number and the phase \cite{Smithey1993}, there is complete lack of the knowledge of the \emph{phase of the photon state}. For the number density and other density expectation values in the second-quantization picture, see Sec.~\ref{sec:secondquantization}.

\section{\label{sec:spinorialwavefunctions}Spinorial photon eigenstates}

In this section, we present selected well-known special cases of photon eigenstates in the spinorial formulation of the electromagnetic field. To be able to define the second-quantized electromagnetic field, we need a complete set of solutions of the free-field spinorial Maxwell equation in Eq.~\eqref{eq:Hamiltonianeq}, which we call photon spinors. Eigenvalue equations of quantum operators, presented below, cannot be satisfied for the electromagnetic spinor in Eq.~\eqref{eq:wavefunctions} if it is made of classical real-valued fields. Therefore, the eigenvalue equations of quantum operators require the determination of complex-valued photon spinors. The real-valued physical quantities are recovered in this formalism as expectation values.

Complete sets of complex-valued photon spinors are conveniently given in the plane-wave and spherical state bases. Their mathematical forms are presented in Appendix \ref{apx:spinorialwavefunctions}. These sets describe a transverse electromagnetic radiation field. In contrast with the conventional vector-potential eigenstates, the photon spinors remain transverse in the Lorentz transformation defined in Sec.~\ref{sec:Lorentz}. In other words, these sets are closed under the action of the Lorentz group. Therefore, photon spinors avoid the gauge dependence problem of the conventional vector-potential eigenstates. The radiation gauge is not Lorentz invariant, so each inertial coordinate system requires its own gauge condition.

\subsection{\label{sec:eigenvalues}Eigenvalues of physical observables in the plane-wave and spherical state bases}

When the field is an eigenstate of the energy, momentum, and helicity operators, we obtain the following set of eigenvalue equations of commuting observables, written as
\begin{equation}
 \left\{\begin{array}{l}
 \hat{\mathbf{S}}^2\Psi_{\mathbf{k},q}=\hbar^2S(S+1)\Psi_{\mathbf{k},q},\\
 \hat{H}\Psi_{\mathbf{k},q}=\hat{E}\Psi_{\mathbf{k},q}=\hbar\omega_\mathbf{k}\Psi_{\mathbf{k},q},\\
 \hat{\mathbf{p}}\Psi_{\mathbf{k},q}=\hbar\mathbf{k}\Psi_{\mathbf{k},q},\\
 \hat{\mathfrak{h}}\Psi_{\mathbf{k},q}=\hbar q\Psi_{\mathbf{k},q}.
 \end{array}\right.
 \label{eq:planewavestatesset}
\end{equation}
Here $S=1$ is the spin quantum number, $\mathbf{k}$ is the wave vector, $\omega_\mathbf{k}=c|\mathbf{k}|$, and $q=\pm1$ is the helicity quantum number. We call the photon spinors $\Psi_{\mathbf{k},q}$ plane-wave states. One can produce identical photons corresponding to plane-wave states given in Eq.~\eqref{eq:planewavestatesset} and measure any of the quantum numbers $\omega_\mathbf{k}$, $\mathbf{k}$, and $q$. In repeated measurements, one then always obtains the same results. Thus, these quantum numbers refer to physical quantities, which are constants of motion.

As presented in Appendix \ref{apx:spinorialwavefunctions}, the spinor $\Psi_{\mathbf{k},q}$ is formed from the complex-valued normalized plane-wave electric and magnetic-field amplitudes $\boldsymbol{\mathcal{E}}_{\mathbf{k},q}$ and $\boldsymbol{\mathcal{B}}_{\mathbf{k},q}$ \cite{Loudon2000} corresponding to Eq.~\eqref{eq:wavefunctions} as $\Psi_{\mathbf{k},q}=[0,\boldsymbol{\mathcal{E}}_{\mathbf{k},q},0,i\boldsymbol{\mathcal{B}}_{\mathbf{k},q}]^T$. The operator $\hat{\mathbf{S}}^2=\hat{\mathbf{S}}_x^2+\hat{\mathbf{S}}_y^2+\hat{\mathbf{S}}_z^2$, formed using the operators for the components of the spin in Eq.~\eqref{eq:spinoperator}, is a diagonal matrix whose first and fifth diagonal elements are zero and its eigenvalue corresponds to $S=1$. As briefly mentioned in the context of the definition of the electromagnetic spinor in Eq.~\eqref{eq:wavefunctions}, this corresponds to the property of all electromagnetic and photon spinors to be eigenstates of $\hat{\mathbf{S}}^2$ with the first and fifth spinor components being zero. Together with this condition, from the free-field spinorial Maxwell equation in Eq.~\eqref{eq:Hamiltonianeq}, it follows that the field must be transverse and the helicity eigenvalues correspond to the helicity quantum number $q=1$ for right-handed circular polarization and $q=-1$ for the left-handed circular polarization.

Correspondingly, when the field is an eigenstate of the energy operator, and the square and $z$ component of the total angular-momentum operator, we obtain
\begin{equation}
 \left\{\begin{array}{l}
 \hat{\mathbf{S}}^2\Psi_{\omega,J,M}=\hbar^2S(S+1)\Psi_{\omega,J,M},\\
 \hat{H}\Psi_{\omega,J,M}=\hat{E}\Psi_{\omega,J,M}=\hbar\omega\Psi_{\omega,J,M},\\
 \hat{\mathbf{J}}^2\Psi_{\omega,J,M}=\hbar^2 J(J+1)\Psi_{\omega,J,M},\\
 \hat{\mathbf{J}}_z\Psi_{\omega,J,M}=\hbar M\Psi_{\omega,J,M}.
 \end{array}\right.
 \label{eq:sphericalstatesset}
\end{equation}
We call these photon spinors spherical states. As presented in Appendix \ref{apx:spinorialwavefunctions}, using the definition of the electromagnetic spinor in Eq.~\eqref{eq:wavefunctions}, we can write $\Psi_{\omega,J,M}$ using the well-known normalized complex-valued spherical electric and magnetic-field amplitudes $\boldsymbol{\mathcal{E}}_{\omega,J,M}$ and $\boldsymbol{\mathcal{B}}_{\omega,J,M}$ \cite{Landau1982} as $\Psi_{\omega,J,M}=[0,\boldsymbol{\mathcal{E}}_{\omega,J,M},0,i\boldsymbol{\mathcal{B}}_{\omega,J,M}]^T$.

Both sets of quantum numbers, $(\mathbf{k},q)$ and $(\omega,J,M)$, considered above, unambiguously specify the quantum state of the photon. Accordingly, both $(\hat{\mathbf{S}}^2,\hat{H},\hat{E},\hat{\mathbf{p}},\hat{\mathfrak{h}})$ and $(\hat{\mathbf{S}}^2,\hat{H},\hat{E},\hat{\mathbf{J}}^2,\hat{\mathbf{J}}_z)$ are complete sets of mutually commuting operators \cite{Sakurai1994}. Furthermore, the photon spinors corresponding to the latter set can be split into two linearly independent parts associated with the electric and magnetic multipoles of the multipole expansion of the electromagnetic field \cite{Landau1982,Jackson1999}. This will be discussed in Appendix \ref{apx:spinorialwavefunctions}.

\subsection{\label{sec:densityexpectationvalues}Density expectation values}

Next, we consider the density expectation values for eigenstates of the Hamiltonian operator, which are complex-valued spinors as discussed above. We assume that the state is normalizable as $\frac{1}{2}\int\Psi^\dag\Psi d^3r=\hbar\omega$, where $\omega$ is the angular frequency of the field. To fulfill this normalization condition, we must take a narrow-frequency-band superposition of the spinorial photon states in Eq.~\eqref{eq:planewavestatesset} or \eqref{eq:sphericalstatesset}. This is because the eigenstates of the Hamiltonian operator in Eq.~\eqref{eq:Hamiltonian} do not vanish sufficiently rapidly at infinity and are, thus, only normalizable with respect to the Dirac delta function as presented in Appendix \ref{apx:spinorialwavefunctions}. Accordingly, we use the narrow-frequency-band plane-wave wavepacket state, defined as
\begin{align}
 \Psi_{\mathbf{k}_0,q} &=\int u(\mathbf{k},\mathbf{k}_0)\Psi_{\mathbf{k},q}d^3k,\nonumber\\
 u(\mathbf{k},\mathbf{k}_0) &=\frac{\exp\Big(-\frac{(k_x-k_{0x})^2}{2(\Delta k_{0x})^2}-\frac{(k_y-k_{0y})^2}{2(\Delta k_{0y})^2}-\frac{(k_z-k_{0z})^2}{2(\Delta k_{0z})^2}\Big)}{(\sqrt{2\pi})^3\Delta k_{0x}\Delta k_{0y}\Delta k_{0z}}.
 \label{eq:narrow}
\end{align}
Here $\mathbf{k}=(k_x,k_y,k_z)^T$ is the component representation of the wave vector, $\mathbf{k}_0=(k_{0x},k_{0y},k_{0z})^T$ is the central wave vector of the wavepacket, and $\Delta k_{0i}$ are the standard deviations of the wave vector components, which satisfy $\Delta k_{0i}/|\mathbf{k}_0|\ll1$, with $|\mathbf{k}_0|=\omega/c$. Without this narrow-frequency-band wavepacket approximation, the probability density cannot be defined for photons using a wave-function-like picture. This is discussed in previous literature \cite{Sipe1995,Landau1982,Newton1949,BialynickiBirula1996,Hawton2019,Hawton1999,SmithB2007}.

Since the narrow-frequency-band superposition in Eq.~\eqref{eq:narrow} is a solution of the free-field spinorial Maxwell equation in Eq.~\eqref{eq:Hamiltonianeq}, we can understand the first-principles quantum-mechanics-based spinor in Eq.~\eqref{eq:narrow} also as a complex-valued electromagnetic spinor of Eq.~\eqref{eq:wavefunctions}, given by $\Psi_{\mathbf{k}_0,q}=\sqrt{\varepsilon_0/2}[0,\mathbf{E}_{\mathbf{k}_0,q},0,ic\mathbf{B}_{\mathbf{k}_0,q}]^T$. Omitting the subscripts $\mathbf{k}_0$ and $q$ for brevity, we obtain the density expectation values for selected key quantum operators, given by
\begin{align}
 \rho_{\hat{n}}^\mathrm{(M)} &=\frac{\int\Psi^\dag\hat{\rho}_{\hat{n}}\Psi d^3r'}{\int\Psi^\dag\Psi d^3r'}
 =\frac{\Psi^\dag\Psi}{\int\Psi^\dag\Psi d^3r'}
 =\frac{\Psi^\dag\Psi}{2\hbar\omega}
 =\frac{\rho_{\hat{H}}^\mathrm{(M)}}{\hbar\omega}\nonumber\\
 &=\frac{1}{4\hbar\omega}\Big(\varepsilon_0|\mathbf{E}|^2+\frac{1}{\mu_0}|\mathbf{B}|^2\Big),
 \label{eq:rhon2}
\end{align}
\begin{align}
 \rho_{\hat{H}}^\mathrm{(M)} &=\frac{\int\Psi^\dag\hat{\rho}_{\hat{H}}\Psi d^3r'}{\int\Psi^\dag\Psi d^3r'}=\frac{\Psi^\dag\hat{H}\Psi}{\int\Psi^\dag\Psi d^3r'}
 =\frac{\Psi^\dag\Psi}{2}
 =\hbar\omega\rho_{\hat{n}}^\mathrm{(M)}\nonumber\\
 &=\frac{1}{4}\Big(\varepsilon_0|\mathbf{E}|^2+\frac{1}{\mu_0}|\mathbf{B}|^2\Big),
 \label{eq:rhoE}
\end{align}
\begin{align}
 \boldsymbol{\rho}_{\hat{\mathbf{p}}}^\mathrm{(M)}
 &=\frac{\int\Psi^\dag\hat{\boldsymbol{\rho}}_{\hat{\mathbf{p}}}\Psi d^3r'}{\int\Psi^\dag\Psi d^3r'}
 =\frac{1}{2}\frac{\Psi^\dag\hat{\mathbf{p}}\Psi-(\hat{\mathbf{p}}\Psi^\dag)\Psi}{\int\Psi^\dag\Psi d^3r'}\nonumber\\
 &=\frac{1}{4\omega}\mathrm{Im}\Big[\varepsilon_0\mathbf{E}^*\cdot(\nabla)\mathbf{E}+\frac{1}{\mu_0}\mathbf{B}^*\cdot(\nabla)\mathbf{B}\Big],
 \label{eq:rhop}
\end{align}
\begin{align}
 \boldsymbol{\rho}_{\hat{\mathbf{L}}}^\mathrm{(M)}
 &=\frac{\int\Psi^\dag\hat{\boldsymbol{\rho}}_{\hat{\mathbf{L}}}\Psi d^3r'}{\int\Psi^\dag\Psi d^3r'}
 =\frac{1}{2}\frac{\Psi^\dag\hat{\mathbf{L}}\Psi-(\hat{\mathbf{L}}\Psi^\dag)\Psi}{\int\Psi^\dag\Psi d^3r'}
 =\mathbf{r}\times\boldsymbol{\rho}_{\hat{\mathbf{p}}}^\mathrm{(M)}\nonumber\\
 &=\frac{1}{4\omega}\mathbf{r}\times\mathrm{Im}\Big[\varepsilon_0\mathbf{E}^*\cdot(\nabla)\mathbf{E}+\frac{1}{\mu_0}\mathbf{B}^*\cdot(\nabla)\mathbf{B}\Big],
 \label{eq:rhoL}
 \end{align}
\begin{align}
 \boldsymbol{\rho}_{\hat{\mathbf{S}}}^\mathrm{(M)} &=\frac{\int\Psi^\dag\hat{\boldsymbol{\rho}}_{\hat{\mathbf{S}}}\Psi d^3r'}{\int\Psi^\dag\Psi d^3r'}
 =\frac{\Psi^\dag\hat{\mathbf{S}}\Psi}{\int\Psi^\dag\Psi d^3r'}\nonumber\\
 &=\frac{1}{4\omega}\mathrm{Im}\Big[\varepsilon_0\mathbf{E}^*\times\mathbf{E}+\frac{1}{\mu_0}\mathbf{B}^*\times\mathbf{B}\Big],
 \label{eq:rhoS}
 \end{align}
\begin{align}
 \rho_{\hat{\mathfrak{h}}}^\mathrm{(M)} &=\frac{\int\Psi^\dag\hat{\rho}_{\hat{\mathfrak{h}}}\Psi d^3r'}{\int\Psi^\dag\Psi d^3r'}
 =\frac{\Psi^\dag\hat{\mathfrak{h}}\Psi}{\int\Psi^\dag\Psi d^3r'}\nonumber\\
 &=\frac{c}{4\omega^2}\mathrm{Re}\Big[\varepsilon_0\mathbf{E}^*\cdot(\nabla\times\mathbf{E})+\frac{1}{\mu_0}\mathbf{B}^*\cdot(\nabla\times\mathbf{B})\Big]\nonumber\\
 &=\frac{\varepsilon_0c}{2\omega}\mathrm{Im}(\mathbf{B}^*\cdot\mathbf{E}).
 \label{eq:rhoh}
\end{align}
Boldface is used for the density operators and density expectation values of vector quantities. Similar relations apply to the narrow-frequency-band wavepacket states based on the spherical photon states. All density expectation values in Eqs.~\eqref{eq:rhon2}--\eqref{eq:rhoh} are approximately independent of time since they are given for an approximate eigenstate of the energy operator. If the energy eigenstate is also an approximate eigenstate of the momentum operator, which is the case with the narrow-frequency-band plane-wave wavepacket state in Eq.~\eqref{eq:narrow}, the density expectation values are additionally approximately independent of the position. The results in Eqs.~\eqref{eq:rhon2}--\eqref{eq:rhoh} are seen to be consistent with previous literature \cite{Bliokh2015a,Bliokh2017b}, as expected.

When comparing the energy density expectation value in Eq.~\eqref{eq:rhoE} to the well-known energy density of classical field, one must account for the complete lack of knowledge of the phase of photon spinors and their narrow-frequency-band approximations. In Eq.~\eqref{eq:rhoE}, the phase information is lost in the squares of the absolute values of the complex-valued fields. Thus, the density expectation value of energy in Eq.~\eqref{eq:rhoE} corresponds to the expression of the energy density of a classical time-harmonic field averaged over the harmonic cycle as discussed in Sec.~\ref{sec:fields}. The lack of phase information applies similarly to the other density expectation values in Eqs.~\eqref{eq:rhon2}--\eqref{eq:rhoh}. Applying the eight-spinor formalism to coherent quantum states, which include the phase information and can only be described through the second quantization, discussed in Sec.~\ref{sec:secondquantization}, is beyond the scope of the present work.

\subsection{\label{sec:probabilitycurrent}Photon probability current four-vector}

Next, we consider the photon probability current four-vector $\mathcal{I}_\mathrm{M}^\nu=(\rho_{\hat{n}}^\mathrm{(M)},\mathbf{j}_\mathrm{M}/c)$ for eigenstates of the Hamiltonian operator. Again, we effectively assume the normalizability of a narrow-frequency-band wavepacket state, given by $\frac{1}{2}\int\Psi^\dag\Psi d^3r=\hbar\omega$. The quantity $\mathbf{j}_\mathrm{M}$ is the conventional probability current three-vector. The time component $\rho_{\hat{n}}^\mathrm{(M)}$ of the photon probability current four-vector is equal to the photon probability density in Eq.~\eqref{eq:rhon2} as
\begin{equation}
 \rho_{\hat{n}}^\mathrm{(M)}=\frac{\Psi^\dag\Psi}{2\hbar\omega}=\frac{1}{4\hbar\omega}\Big(\varepsilon_0|\mathbf{E}|^2+\frac{|\mathbf{B}|^2}{\mu_0}\Big).
 \label{eq:I0}
\end{equation}
The fundamental relation between the photon probability density $\rho_{\hat{n}}^\mathrm{(M)}$ and the probability current $\mathbf{j}_\mathrm{M}$ is given by
\begin{equation}
 \frac{d}{dt}\int_V\rho_{\hat{n}}^\mathrm{(M)}d^3r=-\int_V\nabla\cdot\mathbf{j}_\mathrm{M}d^3r=-\int_{\partial V}\mathbf{j}_\mathrm{M}\cdot d\mathbf{S}.
\end{equation}
This equation is required to apply to an arbitrary volume $V$ and its boundary $\partial V$.

Next, we elaborate an expression for the photon probability current three-vector $\mathbf{j}_\mathrm{M}$ in terms of the photon spinor. First, we write the Hamiltonian operator in Eq.~\eqref{eq:Hamiltonian} as
\begin{equation}
 \hat{H}=c\boldsymbol{\alpha}_\mathrm{B}\cdot\hat{\mathbf{p}}=-i\hbar c\boldsymbol{\gamma}_\mathrm{B}^0\boldsymbol{\gamma}_\mathrm{B}^i\partial_i.
\end{equation}
The dynamical equations for the photon spinor and its conjugate transpose are then given by
\begin{equation}
 \hat{H}\Psi=i\hbar\frac{\partial\Psi}{\partial t}=-i\hbar c\boldsymbol{\gamma}_\mathrm{B}^0\boldsymbol{\gamma}_\mathrm{B}^i\partial_i\Psi,
\end{equation}
\vspace{0.01cm}
\begin{equation}
 \Psi^\dag\hat{H}=i\hbar\frac{\partial\Psi^\dag}{\partial t}=-i\hbar c\partial_i\Psi^\dag\boldsymbol{\gamma}_\mathrm{B}^0\boldsymbol{\gamma}_\mathrm{B}^i.
\end{equation}
Using these equations, the partial time derivative of $\Psi^\dag\Psi$ is written as
\begin{align}
 \frac{\partial}{\partial t}(\Psi^\dag\Psi) &=
 \frac{\partial\Psi^\dag}{\partial t}\Psi+\Psi^\dag\frac{\partial\Psi}{\partial t}\nonumber\\
 &=-c(\partial_i\Psi^\dag)\boldsymbol{\gamma}_\mathrm{B}^0\boldsymbol{\gamma}_\mathrm{B}^i\Psi
 -c\Psi^\dag\boldsymbol{\gamma}_\mathrm{B}^0\boldsymbol{\gamma}_\mathrm{B}^i(\partial_i\Psi)\nonumber\\
 &=-\partial_i(c\Psi^\dag\boldsymbol{\gamma}_\mathrm{B}^0\boldsymbol{\gamma}_\mathrm{B}^i\Psi)\nonumber\\
 &=-\partial_i(c\bar{\Psi}\boldsymbol{\gamma}_\mathrm{B}^i\Psi).
 \label{eq:probabilitydensityderivative}
\end{align}
For eigenstates of the Hamiltonian operator, the photon energy $\hbar\omega$ is constant. Therefore, we can place $2\hbar\omega$ as the numerator inside the derivatives of Eq.~\eqref{eq:probabilitydensityderivative} so that the left-hand side corresponds to the partial time derivative of the photon probability density in Eq.~\eqref{eq:I0}. Thus, we obtain
\begin{equation}
 \frac{\partial\rho_{\hat{n}}^\mathrm{(M)}}{\partial t}=\frac{\partial}{\partial t}\Big(\frac{\Psi^\dag\Psi}{2\hbar\omega}\Big)
 =-\partial_i\Big(\frac{c\bar{\Psi}\boldsymbol{\gamma}_\mathrm{B}^i\Psi}{2\hbar\omega}\Big)
 =-\nabla\cdot\mathbf{j}_\mathrm{M},
\end{equation}
where the components of the probability current three-vector $\mathbf{j}_\mathrm{M}$ are, thus, given by
\begin{equation}
 j_\mathrm{M}^i=\frac{c\bar{\Psi}\boldsymbol{\gamma}_\mathrm{B}^i\Psi}{2\hbar\omega}.
 \label{eq:j}
\end{equation}
Furthermore, in terms of the electric and magnetic-field amplitudes, we can write the photon probability current three-vector as
\begin{equation}
 \mathbf{j}_\mathrm{M}=\frac{\epsilon_0c^2}{2\hbar\omega}\mathrm{Re}(\mathbf{E}^*\times\mathbf{B}).
 \label{eq:probabilitycurrent3d}
\end{equation}
Thus, the spatial components of the probability current four-vector are equal to the components of the Poynting vector of a classical time-harmonic field averaged over the harmonic cycle and normalized by the photon energy.

Using Eqs.~\eqref{eq:I0} and \eqref{eq:j}, the photon probability current four-vector for photon spinors is then given by
\begin{equation}
 \mathcal{I}_\mathrm{M}^a=(\rho_{\hat{n}}^\mathrm{(M)},\mathbf{j}_\mathrm{M}/c)=\frac{\bar{\Psi}\boldsymbol{\gamma}_\mathrm{B}^a\Psi}{2\hbar\omega}.
 \label{eq:probabilitycurrent}
\end{equation}
The covariance of this representation can be verified for plane-wave states, discussed in Sec.~\ref{sec:spinorialwavefunctions}, by checking that the Lorentz transformation of spinors and the angular frequency gives the same result as the direct Lorentz transformation of the probability current four-vector as $\Lambda^a_{\;b}\mathcal{I}_\mathrm{M}^b=\frac{1}{2\hbar\omega'}\overline{\boldsymbol{\Lambda}_\mathrm{S}\Psi}\boldsymbol{\gamma}_\mathrm{B}^a\boldsymbol{\Lambda}_\mathrm{S}\Psi$, where $\omega'$ is the Doppler-shifted angular frequency. The Lorentz transformations are discussed in more detail in Sec.~\ref{sec:Lorentz}.

In the case of spherical photon states, discussed in Sec.~\ref{sec:spinorialwavefunctions}, it is important to note that these states when transformed to another inertial frame are no longer eigenstates of the energy operator in the transformed inertial frame. This is because the field components propagating in different directions experience different Doppler shifts. Thus, the angular frequency of a spherical state is not a global state parameter in the transformed inertial frame. Therefore, Eq.~\eqref{eq:probabilitycurrent} cannot be applied to spherical states in transformed inertial frames.

We find that the probability current three-vector in Eq.~\eqref{eq:probabilitycurrent3d} is related to the momentum and spin-density expectation values in Eqs.~\eqref{eq:rhop} and \eqref{eq:rhoS} as
\begin{align}
 \mathbf{j}_\mathrm{M} &=\frac{c^2}{\hbar\omega}\Big(\boldsymbol{\rho}_{\hat{\mathbf{p}}}^\mathrm{(M)}+\frac{1}{2}\nabla\times\boldsymbol{\rho}_{\hat{\mathbf{S}}}^\mathrm{(M)}\Big)\nonumber\\
 &=\frac{c^2}{2\hbar\omega}\frac{\Psi^\dag\hat{\mathbf{p}}\Psi-(\hat{\mathbf{p}}\Psi^\dag)\Psi+\nabla\times(\Psi^\dag\hat{\mathbf{S}}\Psi)}{\int\Psi^\dag\Psi d^3r}.
 \label{eq:probabilitycurrent3d2}
\end{align}
In nonrelativistic quantum mechanics of spinless particles with a rest mass $m_0$, a formula equivalent to Eq.~\eqref{eq:probabilitycurrent3d2} is well known to give the probability current of the particle. In this case, the photon energy $\hbar\omega$ is replaced by the rest mass energy $m_0c^2$ and the spin term is dropped out \cite{Shankar1994,Landau1977}. In the case of Dirac fermions, the momentum operator is replaced by the corresponding components of the gauge-covariant momentum operator containing a dependence on the electromagnetic four-potential, and the coefficient of the spin term is different by the electron-spin $g$ factor \cite{Peleg1998}. The four-vector generalization of the relation in Eq.~\eqref{eq:probabilitycurrent3d2} is given by
\begin{equation}
 \mathcal{I}_\mathrm{M}^a=\frac{c}{2\hbar\omega}\frac{\Psi^\dag\hat{P}^a\Psi-(\hat{P}^a\Psi^\dag)\Psi+\partial_b(\Psi^\dag\hat{S}^{ab}\Psi)}{\int\Psi^\dag\Psi d^3r}.
 \label{eq:probabilitycurrent2}
\end{equation}
Here $\hat{P}^a$ is the four-momentum operator in Eq.~\eqref{eq:fourmomentumoperator} and $\hat{S}^{ab}$ is the spin tensor operator in Eq.~\eqref{eq:spintensoroperator}. Equations above strongly indicate that, assuming the narrow-frequency-band wavepacket approximation, the photon spinor of the present work can be used as an \emph{equally consistent} wave-function-like concept as the electron wave function of the Dirac theory. The obvious difference comes from changing the number of particles in the system since any single-photon state overlaps in energy scale with an infinite number of many-photon states. No such overlapping takes place in the case of Dirac fermions at low energies. For Dirac fermions, one must create a particle-antiparticle pair, which has a large threshold energy of $2m_\mathrm{e}c^2$, where $m_\mathrm{e}$ is the electron rest mass. The photon spinor representation of the effective wave function for narrow-frequency-band wavepacket states and Eqs.~\eqref{eq:probabilitycurrent} and \eqref{eq:probabilitycurrent2} enable discovering photon-particle form similarities that are not equally obvious in the conventional four-potential-based formulation of QED.

\subsection{\label{sec:CPTsymmetry}CPT symmetry}

The CPT symmetry is the combined charge-conjugation (C), parity (P), and time-reversal (T) symmetry that is observed to be an exact symmetry of the laws of nature. It is formed by discrete Lorentz transformations \cite{Tully2011,Schwartz2014}. Electromagnetic interaction is known to satisfy each of these symmetries separately. Here we limit our study to the transformation properties of photon spinors. Another part of the CPT symmetry is the transformation of the derivatives in the equations of motion, which we do not consider here. We define the time-reversal $\hat{T}$, parity $\hat{P}$, and charge-conjugation $\hat{C}$ transformations of photon spinors as
\begin{equation}
 \hat{T}\Psi(t,\mathbf{r})=\Psi(-t,\mathbf{r}),
 \label{eq:Tsymmetry}
\end{equation}
\begin{equation}
 \hat{P}\Psi(t,\mathbf{r})=\Psi(t,-\mathbf{r}),
 \label{eq:Psymmetry}
\end{equation}
\begin{equation}
 \hat{C}\Psi(t,\mathbf{r})=\eta_c\mathbf{C}\bar{\Psi}^T(t,\mathbf{r}).
 \label{eq:Csymmetry}
\end{equation}
Here $\mathbf{C}$ is the charge-conjugation matrix and $\eta_c$ is a phase factor. Since photons are their own antiparticles, the charge-conjugation matrix does not change the particle type when it operates on photon spinors. Instead, for photon spinors, it is related to the inversion of the angular momentum. Accordingly, for plane-wave photon states, the charge-conjugation matrix operates on the spinor components so that the helicity is inverted as
\begin{equation}
 \hat{C}\Psi_{\mathbf{k},q}(t,\mathbf{r})=-\bar{\Psi}_{\mathbf{k},-q}^T(t,\mathbf{r}).
\end{equation}
For spherical photon states, the charge-conjugation matrix inverts the angular momentum $z$ component as
\begin{equation}
 \hat{C}\Psi_{\omega,J,M}^{(\eta)}(t,\mathbf{r})=(-1)^{J+M+\delta_{\eta,m}}\bar{\Psi}_{\omega,J,-M}^{(\eta)T}(t,\mathbf{r}).
\end{equation}
General photon spinors do not remain unchanged in each of the three symmetry transformations in Eqs.~\eqref{eq:Tsymmetry}--\eqref{eq:Csymmetry} separately, but the combination of the three symmetry transformations preserves photon spinors unchanged as
\begin{equation}
 \hat{C}\hat{P}\hat{T}\Psi(t,\mathbf{r})=\Psi(t,\mathbf{r}).
 \label{eq:CPTinvariance}
\end{equation}
This means that photon spinors satisfy the CPT symmetry. Equation \eqref{eq:CPTinvariance} is straightforward to verify by using the explicit expressions of the plane-wave and spherical photon state spinors, given in Appendix \ref{apx:spinorialwavefunctions}.

\section{\label{sec:Lagrangian}QED Lagrangian density and Euler-Lagrange equations}

In this section, we investigate the derivation of the present spinorial electromagnetic theory from the Lagrangian density. Thus, we follow the conventional approach of deriving field theories. There are formal differences in our spinorial representation compared with the conventional four-potential-based QED, but the resulting theory is equivalent to the conventional QED as discussed in more detail below.

\subsection{Lagrangian density of QED}
Here we present the conventional Lagrangian density of the coupled system of the electromagnetic field and the Dirac field. It will be given in terms of the potential spinor $\Theta_\Re$ in Eq.~\eqref{eq:Theta} and the Dirac electron-positron field $\psi$ since the interaction between the electromagnetic field and the Dirac field is known in terms of the four-potential and the direct representation in terms of the electric and magnetic fields is unknown. The conventional four-potential is a real-valued quantity and described by a Hermitian operator in QED \cite{Peskin2018,Landau1982}. Therefore, here we use the real-valued form $\Theta_\Re$ of the potential spinor. The definition of the Dirac field requires no quantization and the Lagrangian density is called the Lagrangian density of QED as a historical artifact. The second quantization is described in Sec.~\ref{sec:secondquantization} below.

Using our spinor notation, the derivation of the theory below differs from the conventional four-potential-based derivation, even though the resulting dynamical equations are equivalent. The conventional formulation of the complete QED Lagrangian density starting from the Lagrangian density of the free Dirac field is briefly described in Appendix \ref{apx:fourcurrent}.
The conventional QED Lagrangian density $\mathcal{L}_\mathrm{QED}$ is written in the eight-spinor notation as
\begin{align}
 \mathcal{L}_\mathrm{QED} &=\mathcal{L}_\mathrm{QED,D}+\mathcal{L}_\mathrm{QED,M}+\mathcal{L}_\mathrm{QED,DM}\nonumber\\
 &=\frac{i\hbar c}{2}\bar{\psi}(\bar{\boldsymbol{\gamma}}_\mathrm{F}\vec{D}-\cev{D}\boldsymbol{\gamma}_\mathrm{F})\psi-m_\mathrm{e}c^2\bar{\psi}\psi
 +\bar{\Psi}_\Re\Psi_\Re,
 \nonumber\\
 \mathcal{L}_\mathrm{QED,D} &=\frac{i\hbar c}{2}\bar{\psi}(\bar{\boldsymbol{\gamma}}_\mathrm{F}\vec{\partial}-\cev{\partial}\boldsymbol{\gamma}_\mathrm{F})\psi-m_\mathrm{e}c^2\bar{\psi}\psi\nonumber\\
 \mathcal{L}_\mathrm{QED,M} &=\bar{\Psi}_\Re\Psi_\Re=\bar{\Theta}_\Re\cev{\partial}_a\boldsymbol{\gamma}_\mathrm{B}^a\boldsymbol{\gamma}_\mathrm{B}^b\vec{\partial}_b\Theta_\Re
  =-\frac{1}{4\mu_0}F_{ab}F^{ab},\nonumber\\
 \mathcal{L}_\mathrm{QED,DM} &=\bar{\Phi}_\Re\Theta_\Re+\bar{\Theta}_\Re\Phi_\Re
 =\frac{g_\mathrm{e}}{2}\bar{\psi}(\bar{\boldsymbol{\gamma}}_\mathrm{F}\Theta_\Re+\bar{\Theta}_\Re\boldsymbol{\gamma}_\mathrm{F})\psi\nonumber\\
 &=-J_\mathrm{e\Re}^a A_{\Re a}.
 \label{eq:Lagrangiandensity1}
\end{align}
Here $\mathcal{L}_\mathrm{QED,D}$ is the Lagrangian density of the free Dirac field and $\mathcal{L}_\mathrm{QED,M}$ is the Lagrangian density of the free Maxwell electromagnetic field. The Lagrangian density $\mathcal{L}_\mathrm{QED,DM}$ describes the minimal-coupling interaction between the Dirac and electromagnetic fields. The quantity $g_\mathrm{e}=q_\mathrm{e}\sqrt{2/\varepsilon_0}$ in Eq.~\eqref{eq:Lagrangiandensity1} is the electric coupling constant of the present theory, where $q_\mathrm{e}=\pm e$ is the electric charge of the particle with $e$ being the elementary charge. The Dirac field is normalized so that $\psi^\dag\psi$ corresponds to the number density. The vector arrows above symbols in Eq.~\eqref{eq:Lagrangiandensity1} indicate the direction in which the differential operators operate. We have defined the eight-spinor electromagnetic-gauge-covariant derivative operator $\vec{D}$ and its adjoint $\cev{D}$ as
\begin{align}
 \vec{D} &=\vec{\partial}-\frac{ig_\mathrm{e}}{\hbar c}\Theta_\Re
 =[0,\vec{D}_x,\vec{D}_y,\vec{D}_z,-\vec{D}_0,0,0,0]^T,\nonumber\\
 \cev{D} &=\cev{\partial}+\frac{ig_\mathrm{e}}{\hbar c}\bar{\Theta}_\Re=[0,\cev{D}_x,\cev{D}_y,\cev{D}_z,\cev{D}_0,0,0,0].
\end{align}
Here the well-known electromagnetic-gauge-covariant derivative of the Dirac field is given by \cite{Peskin2018}
\begin{equation}
 \vec{D}_a=\vec{\partial}_a+i\frac{q_\mathrm{e}}{\hbar}A_{\Re a},
 \hspace{0.5cm}\cev{D}_a=\cev{\partial}_a-i\frac{q_\mathrm{e}}{\hbar}A_{\Re a}.
 \label{eq:covariantderivativeEM}
\end{equation}
The eight-spinor partial differential operator $\vec{\partial}$, its adjoint $\cev{\partial}$, the four-vector-type eight-spinor $\boldsymbol{\gamma}_\mathrm{F}$, made of the Dirac gamma matrices, its adjoint $\bar{\boldsymbol{\gamma}}_\mathrm{F}$,
and the eight-spinor timelike unit vector $\mathbf{e}_8$ and its adjoint spinor $\bar{\mathbf{e}}_8$, for use in Sec.~\ref{sec:Lagrangian2}, are given by
\begin{align}
 \vec{\partial} &=[0,\vec{\partial}_x,\vec{\partial}_y,\vec{\partial}_z,-\vec{\partial}_0,0,0,0]^T,\nonumber\\
 \cev{\partial} &=[0,\cev{\partial}_x,\cev{\partial}_y,\cev{\partial}_z,\cev{\partial}_0,0,0,0],\nonumber\\
 \boldsymbol{\gamma}_\mathrm{F} &=[\mathbf{0},\boldsymbol{\gamma}_\mathrm{F}^x,\boldsymbol{\gamma}_\mathrm{F}^y,\boldsymbol{\gamma}_\mathrm{F}^z,\boldsymbol{\gamma}_\mathrm{F}^0,\mathbf{0},\mathbf{0},\mathbf{0}]^T,\nonumber\\
 \bar{\boldsymbol{\gamma}}_\mathrm{F} &=[\mathbf{0},\bar{\boldsymbol{\gamma}}_\mathrm{F}^x,\bar{\boldsymbol{\gamma}}_\mathrm{F}^y,\bar{\boldsymbol{\gamma}}_\mathrm{F}^z,-\bar{\boldsymbol{\gamma}}_\mathrm{F}^0,\mathbf{0},\mathbf{0},\mathbf{0}],\nonumber\\
 \mathbf{e}_8 &=[0,0,0,0,1,0,0,0]^T,\nonumber\\
 \bar{\mathbf{e}}_8 &=[0,0,0,0,-1,0,0,0].
\end{align}
Here $\bar{\boldsymbol{\gamma}}_\mathrm{F}^a=\boldsymbol{\gamma}_\mathrm{F}^0\boldsymbol{\gamma}_\mathrm{F}^{a\dag}\boldsymbol{\gamma}_\mathrm{F}^0=\boldsymbol{\gamma}_\mathrm{F}^a$. In the equality between the first two forms of $\mathcal{L}_\mathrm{DM}$ in Eq.~\eqref{eq:Lagrangiandensity1}, we have used
\begin{equation}
 \Phi_\Re=\frac{1}{2}g_\mathrm{e}\bar{\psi}(\boldsymbol{\gamma}_\mathrm{F})\psi,\hspace{0.5cm}
 \bar{\Phi}_\Re=\Phi_\Re^\dag\boldsymbol{\gamma}_\mathrm{B}^0=\frac{1}{2}g_\mathrm{e}\bar{\psi}(\bar{\boldsymbol{\gamma}}_\mathrm{F})\psi.
 \label{eq:PhiR}
\end{equation}
These forms are obtained from the definition of the charge-current spinor in Eq.~\eqref{eq:wavefunctions} by using the well-known expression of the four-current density in terms of the Dirac spinors, given by $J_{\mathrm{e}\Re}^a=(c\rho_\mathrm{e},\mathbf{J}_{\mathrm{e}\Re})=q_\mathrm{e}c\bar{\psi}\boldsymbol{\gamma}_\mathrm{F}^a\psi$ \cite{Landau1982}. In the last form of $\textstyle\mathcal{L}_\mathrm{DM}$ in Eq.~\eqref{eq:Lagrangiandensity1}, we have used the covariant form of the electromagnetic four-potential, given by $A_{\Re a}=\eta_{ab}A_\Re^{b}=(\phi_\mathrm{e\Re}/c,-\mathbf{A}_\Re)$.

\subsection{Euler-Lagrange equations of the potential spinor field}

When writing the Euler-Lagrange equations, the spinors $\Theta_\Re$ and $\bar{\Theta}_\Re$ are treated as independent dynamical variables. The Euler-Lagrange equation of $\bar{\Theta}_\Re$ is given by
\begin{equation}
 \frac{\partial\mathcal{L}_\mathrm{QED}}{\partial\bar{\Theta}_\Re}-\partial_a\Big[\frac{\partial\mathcal{L}_\mathrm{QED}}{\partial(\partial_a\bar{\Theta}_\Re)}\Big]=0.
 \label{eq:EulerLagrangeTheta}
\end{equation}
Using the Lagrangian density in Eq.~\eqref{eq:Lagrangiandensity1}, we obtain $\partial\mathcal{L}_\mathrm{QED}/\partial\bar{\Theta}_\Re=\Phi_\Re$ and $\partial\mathcal{L}_\mathrm{QED}/\partial(\partial_a\bar{\Theta}_\Re)=\boldsymbol{\gamma}_\mathrm{B}^a\boldsymbol{\gamma}_\mathrm{B}^b\partial_b\Theta_\Re$. Substituting these derivatives into Eq.~\eqref{eq:EulerLagrangeTheta}, using $\partial_a\boldsymbol{\gamma}_\mathrm{B}^a\boldsymbol{\gamma}_\mathrm{B}^b\partial_b=\mathbf{I}_8\partial^a\partial_a$, and rearranging the terms, the Euler-Lagrange equations of motion become
\begin{equation}
 \partial^a\partial_a\Theta_\Re=\Phi_\Re.
 \label{eq:photonDirac2}
\end{equation}
Equation \eqref{eq:photonDirac2} is equal to the real part of the equation of the potential spinor in Eq.~\eqref{eq:photonDiracA1}. Using Eq.~\eqref{eq:photonDiracA2}, we can see that it is also equal to the spinorial Maxwell equation in Eq.~\eqref{eq:photonDirac}. Thus, the Lagrangian density in Eq.~\eqref{eq:Lagrangiandensity1} enables a compact derivation of the full set of Maxwell's equations. In contrast, in the conventional Lagrangian theory derivation of electrodynamics using the four-potential, one obtains only Gauss's law for electricity and the Amp\`ere-Maxwell law directly from the Euler-Lagrange equations \cite{Landau1989,Bliokh2013b,Partanen2019b}. Faraday's law of induction and Gauss's law for magnetism are obtained from the definition of the electromagnetic field tensor in terms of the four-potential by using the Bianchi identity.

As a further consistency check, we investigate the Euler-Lagrange equations for $\Theta_\Re$, which are identical to Eq.~\eqref{eq:EulerLagrangeTheta} with $\bar{\Theta}_\Re$ replaced by $\Theta_\Re$. In this case, we have $\partial\mathcal{L}_\mathrm{QED}/\partial\Theta_\Re=\bar{\Phi}_\Re$ and $\partial\mathcal{L}_\mathrm{QED}/\partial(\partial_a\Theta_\Re)=\partial_b\bar{\Theta}_\Re\boldsymbol{\gamma}_\mathrm{B}^b\boldsymbol{\gamma}_\mathrm{B}^a$. Substituting these derivatives into the Euler-Lagrange equation, using $\partial_a\partial_b\bar{\Theta}_\Re\boldsymbol{\gamma}_\mathrm{B}^b\boldsymbol{\gamma}_\mathrm{B}^a=\partial^a\partial_a\bar{\Theta}_\Re$, and rearranging the terms, we obtain $-\partial^a\partial_a\bar{\Theta}_\Re=-\bar{\Phi}_\Re$. Using $\bar{\Theta}_\Re=\Theta_\Re^\dag\boldsymbol{\gamma}_\mathrm{B}^0$ and $\bar{\Phi}_\Re=\Phi_\Re^\dag\boldsymbol{\gamma}_\mathrm{B}^0$, we obtain $-\partial^a\partial_a\Theta_\Re^\dag\boldsymbol{\gamma}_\mathrm{B}^0=-\Phi_\Re^\dag\boldsymbol{\gamma}_\mathrm{B}^0$. Taking the conjugate transpose of this equation, multiplying the resulting equation by $\boldsymbol{\gamma}_\mathrm{B}^0$ from the left, and rearranging the terms leads to $\partial^a\partial_a\Theta_\Re=\Phi_\Re$. This is equal to Eq.~\eqref{eq:photonDirac2}.

\subsection{Euler-Lagrange equations of the Dirac field}

Next, we consider dynamical equations of the Dirac field. As conventional, we treat $\psi$ and $\bar{\psi}=\psi^\dag\boldsymbol{\gamma}_\mathrm{F}^0$ as independent dynamical variables and write the Euler-Lagrange equations for $\bar{\psi}$ as \cite{Peskin2018}
\begin{equation}
 \frac{\partial\mathcal{L}_\mathrm{QED}}{\partial\bar{\psi}}-\partial_a\Big[\frac{\partial\mathcal{L}_\mathrm{QED}}{\partial(\partial_a\bar{\psi})}\Big]=0.
 \label{eq:EulerLagrangeDirac1}
\end{equation}
Using the Lagrangian density in Eq.~\eqref{eq:Lagrangiandensity1} with the identities $\bar{\boldsymbol{\gamma}}_\mathrm{F}\vec{\partial}=\boldsymbol{\gamma}_\mathrm{F}^a\vec{\partial}_a$ and $\cev{\partial}\boldsymbol{\gamma}_\mathrm{F}=\cev{\partial}_a\boldsymbol{\gamma}_\mathrm{F}^a$, we obtain $\partial\mathcal{L}_\mathrm{QED}/\partial\bar{\psi}=\frac{i}{2}\hbar c\boldsymbol{\gamma}_\mathrm{F}^a\partial_a\psi-m_\mathrm{e}c^2\psi+\frac{1}{2}g_\mathrm{e}(\bar{\boldsymbol{\gamma}}_\mathrm{F}\Theta_\Re+\bar{\Theta}_\Re\boldsymbol{\gamma}_\mathrm{F})\psi$ and $\partial\mathcal{L}_\mathrm{QED}/\partial(\partial_a\bar{\psi})=-\frac{i}{2}\hbar c\boldsymbol{\gamma}_\mathrm{F}^a\psi$. Substituting these derivatives into Eq.~\eqref{eq:EulerLagrangeDirac1}, the Euler-Lagrange equations of motion become
\begin{equation}
 i\hbar c\boldsymbol{\gamma}_\mathrm{F}^a\partial_a\psi-m_\mathrm{e}c^2\psi+\textstyle\frac{1}{2}g_\mathrm{e}(\bar{\boldsymbol{\gamma}}_\mathrm{F}\Theta_\Re+\bar{\Theta}_\Re\boldsymbol{\gamma}_\mathrm{F})\psi=0.
 \label{eq:Dirac1}
\end{equation}
Using the relation $\textstyle\frac{1}{2}g_\mathrm{e}(\bar{\boldsymbol{\gamma}}_\mathrm{F}\Theta_\Re+\bar{\Theta}_\Re\boldsymbol{\gamma}_\mathrm{F})=-q_\mathrm{e}c\boldsymbol{\gamma}_\mathrm{F}^a A_{\Re a}$, and the definition of the electromagnetic-gauge-covariant derivative of the Dirac field in Eq.~\eqref{eq:covariantderivativeEM}, we can write Eq.~\eqref{eq:Dirac1} compactly as
\begin{equation}
 i\hbar c\boldsymbol{\gamma}_\mathrm{F}^a\vec{D}_a\psi
 -m_\mathrm{e}c^2\psi=0.
 \label{eq:Dirac11}
\end{equation}
In terms of the components of the four-potential $A_{\Re\mu}=(\phi_\mathrm{e\Re}/c,-\mathbf{A}_\Re)$, momentum operator $\hat{\mathbf{p}}$, Dirac alpha and beta matrices $\boldsymbol{\alpha}_\mathrm{F}^i=\boldsymbol{\gamma}_\mathrm{F}^0\boldsymbol{\gamma}_\mathrm{F}^i$ and $\boldsymbol{\beta}_\mathrm{F}=\boldsymbol{\gamma}_\mathrm{F}^0$, and the associated vector $\boldsymbol{\alpha}_\mathrm{F}=(\boldsymbol{\alpha}_\mathrm{F}^x,\boldsymbol{\alpha}_\mathrm{F}^y,\boldsymbol{\alpha}_\mathrm{F}^z)$, Eq.~\eqref{eq:Dirac11} can be further rewritten as
\begin{equation}
 \hat{H}\psi=i\hbar\frac{\partial}{\partial t}\psi.
 \label{eq:Dirac2}
\end{equation}
Here the Hamiltonian operator $\hat{H}$ takes for the Dirac field the form
\begin{equation}
 \hat{H}=c\boldsymbol{\alpha}_\mathrm{F}\cdot(\hat{\mathbf{p}}-q_\mathrm{e}\mathbf{A}_\Re)+\boldsymbol{\beta}_\mathrm{F}m_\mathrm{e}c^2+q_\mathrm{e}\phi_\mathrm{e\Re}\mathbf{I}_4.
 \label{eq:DiracHamiltonian1}
\end{equation}
Equations \eqref{eq:Dirac11}--\eqref{eq:DiracHamiltonian1} show that the dynamical equation of the Dirac field is equivalent to the conventional minimally coupled Dirac equation in the presence of interaction with the electromagnetic field \cite{Landau1982}, as expected.

For checking the consistency, we derive the Dirac equation from the Euler-Lagrange equations for $\psi$, which is identical to Eq.~\eqref{eq:EulerLagrangeDirac1} with $\bar{\psi}$ replaced by $\psi$. In this case, we have $\partial\mathcal{L}_\mathrm{QED}/\partial\psi=-\frac{i}{2}\hbar c\partial_a\bar{\psi}\boldsymbol{\gamma}_\mathrm{F}^a-m_\mathrm{e}c^2\bar{\psi}+\frac{1}{2}g_\mathrm{e}\bar{\psi}(\bar{\boldsymbol{\gamma}}_\mathrm{F}\Theta_\Re+\bar{\Theta}_\Re\boldsymbol{\gamma}_\mathrm{F})$ and $\partial\mathcal{L}_\mathrm{QED}/\partial(\partial_a\psi)=\frac{i}{2}\hbar c\bar{\psi}\boldsymbol{\gamma}_\mathrm{F}^a$. Substituting these derivatives into the Euler-Lagrange equation, we obtain $-i\hbar c\partial_a\bar{\psi}\boldsymbol{\gamma}_\mathrm{F}^a-m_\mathrm{e}c^2\bar{\psi}+\frac{1}{2}g_\mathrm{e}\bar{\psi}(\bar{\boldsymbol{\gamma}}_\mathrm{F}\Theta_\Re+\bar{\Theta}_\Re\boldsymbol{\gamma}_\mathrm{F})$. Taking the conjugate transpose of this equation, using $\bar{\psi}=\psi^\dag\boldsymbol{\gamma}_\mathrm{F}^0$, and multiplying the resulting equation by $\boldsymbol{\gamma}_\mathrm{F}^0$ from the left leads to $i\hbar c\boldsymbol{\gamma}_\mathrm{F}^0\boldsymbol{\gamma}_\mathrm{F}^{a\dag}\boldsymbol{\gamma}_\mathrm{F}^0\partial_a\psi-\boldsymbol{\gamma}_\mathrm{F}^0\boldsymbol{\gamma}_\mathrm{F}^0m_\mathrm{e}c^2\psi+\frac{1}{2}g_\mathrm{e}\boldsymbol{\gamma}_\mathrm{F}^0(\bar{\boldsymbol{\gamma}}_\mathrm{F}\Theta_\Re+\bar{\Theta}_\Re\boldsymbol{\gamma}_\mathrm{F})^\dag\boldsymbol{\gamma}_\mathrm{F}^0\psi$. Using $\boldsymbol{\gamma}_\mathrm{F}^0\boldsymbol{\gamma}_\mathrm{F}^{a\dag}\boldsymbol{\gamma}_\mathrm{F}^0=\boldsymbol{\gamma}_\mathrm{F}^a$, $\boldsymbol{\gamma}_\mathrm{F}^0\boldsymbol{\gamma}_\mathrm{F}^0=\mathbf{I}_8$, and $\boldsymbol{\gamma}_\mathrm{F}^0(\bar{\boldsymbol{\gamma}}_\mathrm{F}\Theta_\Re+\bar{\Theta}_\Re\boldsymbol{\gamma}_\mathrm{F})^\dag\boldsymbol{\gamma}_\mathrm{F}^0=\bar{\boldsymbol{\gamma}}_\mathrm{F}\Theta_\Re+\bar{\Theta}_\Re\boldsymbol{\gamma}_\mathrm{F}$, we finally obtain the Dirac equation in the form given in Eq.~\eqref{eq:Dirac1}.

\section{\label{sec:secondquantization}Second quantization of spinorial electromagnetic field}

The goal of this section is to give the reader a concise idea of how the definitions of the key quantum operators and their matrix elements in the spinorial form transform into the pertinent quantities in the conventional QED. The quantization of the electromagnetic spinor field form $\Psi_\Re$, made of real-valued fields, follows trivially from the conventional QED \cite{Loudon2000}. It is obtained by substituting the conventional Hermitian electric and magnetic-field operators in places of the electric and magnetic fields in the electromagnetic spinor in Eq.~\eqref{eq:wavefunctions}. Similarly, the charge-current spinor $\Phi_\Re$ is quantized by substituting the conventional Dirac field operator into Eq.~\eqref{eq:PhiR} or the corresponding components of the electric four-current density into Eq.~\eqref{eq:wavefunctions}.

This section is devoted to the quantization of the complex-valued electromagnetic spinor field form $\Psi$. Since $\Psi$ is complex valued, the corresponding operator is non-Hermitian in analogy to the Dirac field operator. We demonstrate the usability of the photon-spinor-field operator in the calculation of the correct second-quantized operators of physical quantities in a way analogous to the Dirac theory. We also compare the single-photon states in the second quantization formalism to the first-quantized spinorial photon eigenstates presented in Sec.~\ref{sec:spinorialwavefunctions} and in Appendix \ref{apx:spinorialwavefunctions}.

\subsection{Photon-spinor-field operator}

Following the conventional quantization procedure of fields in the plane-wave basis \cite{Peskin2018,Loudon2000}, the photon-spinor-field operator is given in the Heisenberg picture by
\begin{equation}
 \hat{\Psi}=\sum_q\int\frac{V}{(2\pi)^3}\frac{1}{\sqrt{2\hbar\omega_\mathbf{k}}}\Psi_{\mathbf{k},q}(\hat{a}_{\mathbf{k},q}+\hat{a}_{\mathbf{k},q}^\dag)d^3k.
 \label{eq:fieldoperator}
\end{equation}
Here $\hat{a}_{\mathbf{k},q}$ and $\hat{a}_{\mathbf{k},q}^\dag$ are the conventional photon annihilation and creation operators, respectively, and $V$ is the quantization volume. The operators $\hat{a}_{\mathbf{k},q}$ and $\hat{a}_{\mathbf{k},q}^\dag$ satisfy the well-known bosonic commutation relations, given by
\begin{align}
 [\hat{a}_{\mathbf{k},q},\hat{a}_{\mathbf{k}',q'}^\dag] &=\frac{(2\pi)^3}{V}\delta_{q,q'}\delta(\mathbf{k}-\mathbf{k}'),\nonumber\\
 [\hat{a}_{\mathbf{k},q},\hat{a}_{\mathbf{k}',q'}] &=[\hat{a}_{\mathbf{k},q}^\dag,\hat{a}_{\mathbf{k}',q'}^\dag]=0.
 \label{eq:akqcommutation}
\end{align}
For the action of the photon annihilation and creation operators on photon states of the second quantization, see Sec.~\ref{sec:singlephotonsecondquantization} below.

We define the commutator of two field operators of the form in Eq.~\eqref{eq:fieldoperator} so that there is a scalar product between the spinor terms $\Psi_{\mathbf{k},q}$, whence the commutator applies to the annihilation and creation operators only. Consequently, since the sum $\hat{a}_{\mathbf{k},q}+\hat{a}_{\mathbf{k},q}^\dag$ is a Hermitian operator, the equal-time commutation relations for $\hat{\Psi}$ and $\hat{\Psi}^\dag$ are all zero as
\begin{equation}
 [\hat{\Psi}(\mathbf{r}),\hat{\Psi}^\dag(\mathbf{r}')]
 =[\hat{\Psi}(\mathbf{r}),\hat{\Psi}(\mathbf{r}')]
 =[\hat{\Psi}^\dag(\mathbf{r}),\hat{\Psi}^\dag(\mathbf{r}')]=0.\!\!
 \label{eq:psicommutation1}
\end{equation}
It also follows that the operators $\hat{\Psi}^{(+)}$ and $\hat{\Psi}^{(-)}$, which are the annihilation operator part of $\hat{\Psi}$ and the creation operator part of  $\hat{\Psi}^\dag$, respectively, satisfy the canonical equal-time commutation relations, given by
\begin{align}
 [\hat{\Psi}^{(+)}(\mathbf{r}),\hat{\Psi}^{(-)}(\mathbf{r}')] &=\delta(\mathbf{r}-\mathbf{r}'),\nonumber\\
 [\hat{\Psi}^{(+)}(\mathbf{r}),\hat{\Psi}^{(+)}(\mathbf{r}')] &=[\hat{\Psi}^{(-)}(\mathbf{r}),\hat{\Psi}^{(-)}(\mathbf{r}')]=0.
 \label{eq:psicommutation2}
\end{align}

The photon-spinor-field operator in Eq.~\eqref{eq:fieldoperator} can also be expressed in the spherical state basis, presented in Appendix \ref{apx:spinorialwavefunctions}. In this basis, it is written as
\begin{equation}
 \hat{\Psi}=\!\sum_{J,M,\eta}\int_0^\infty\!\!\!\!\frac{2V\omega^2}{\pi c^3\sqrt{2\hbar\omega}}\Psi_{\omega,J,M}^{(\eta)}(\hat{a}_{\omega,J,M}^{(\eta)}+\hat{a}_{\omega,J,M}^{(\eta)\dag})d\omega.
 \label{eq:fieldoperator2}
\end{equation}
Here the quantum number $J$ ranges from 1 to infinity, $M$ ranges from $-J$ to $J$, and the state parameter $\eta$ has values $\mathrm{e}$ and $\mathrm{m}$ for the states associated with the electric and magnetic multipoles of the multipole expansion of the electromagnetic field \cite{Landau1982,Jackson1999}.

The bosonic commutation relations of the photon annihilation operators $\hat{a}_{\omega,J,M}^{(\eta)}$ and creation operators $\hat{a}_{\omega,J,M}^{(\eta)\dag}$ in the spherical basis are given by
\begin{align}
 [\hat{a}_{\omega,J,M}^{(\eta)},\hat{a}_{\omega',J',M'}^{(\eta')\dag}] &=\frac{\pi c^3}{2V\omega^2}\delta_{\eta,\eta'}\delta_{J,J'}\delta_{M,M'}\delta(\omega-\omega'),\nonumber\\
 [\hat{a}_{\omega,J,M}^{(\eta)},\hat{a}_{\omega',J',M'}^{(\eta')}] &=[\hat{a}_{\omega,J,M}^{(\eta)\dag},\hat{a}_{\omega',J',M'}^{(\eta')\dag}]=0.
\end{align}

Another standard approach for the quantization of the electromagnetic field is to use wavepacket states and the corresponding wavepacket creation and annihilation operators \cite{Loudon2000}. This approach to the quantization of $\Psi$ is also possible.

\subsection{\label{sec:secondquantizedoperators}Second-quantized operators}

Next, we define quantum operators for the electromagnetic field in the second-quantized multiphoton picture. In distinction to the operators in the first quantization, we denote the second-quantized operators with an underline. Then, the second-quantized operator $\underline{\hat{O}}$, which corresponds to an arbitrary first-quantized operator $\hat{O}$, is given by
\begin{equation}
 \underline{\hat{O}}=\int:\!\hat{\Psi}^\dag\hat{O}\hat{\Psi}\!:d^3r.
 \label{eq:secondquantizedoperator}
\end{equation}
Here the colons denote the conventional normal ordering of operators, which removes the infinite constant term that otherwise arises from the commutator of the annihilation and creation operators \cite{Loudon2000}.

Second-quantized operators obtain simple forms in the eigenbasis of the pertinent first-quantized operators. Therefore, as examples of second-quantized operators, we present selected operators using the pertinent eigenbases. The operators are obtained by substituting the first-quantized pertinent operators from Sec.~\ref{sec:quantumoptics} and the photon-spinor-field operator from either Eq.~\eqref{eq:fieldoperator} or \eqref{eq:fieldoperator2} into Eq.~\eqref{eq:secondquantizedoperator}. Using the normal ordering relations of bosonic operators, given as an example in the plane-wave basis by $:\hat{a}_{\mathbf{k},q}^\dag\hat{a}_{\mathbf{k},q}:=\hat{a}_{\mathbf{k},q}^\dag\hat{a}_{\mathbf{k},q}=\hat{n}_{\mathbf{k},q}$ and $:\hat{a}_{\mathbf{k},q}\hat{a}_{\mathbf{k},q}^\dag:=\hat{a}_{\mathbf{k},q}^\dag\hat{a}_{\mathbf{k},q}=\hat{n}_{\mathbf{k},q}$, where $\hat{n}_{\mathbf{k},q}=\hat{a}_{\mathbf{k},q}^\dag\hat{a}_{\mathbf{k},q}$ is the single-mode number operator, after some algebra, we obtain the conventional results
\begin{equation}
 \underline{\hat{n}}
 =\sum_q\int\frac{V}{(2\pi)^3}\hat{n}_{\mathbf{k},q}d^3k
 =\sum_{J,M,\eta}\int_0^\infty\frac{2V\omega^2}{\pi c^3}\hat{n}_{\omega,J,M}^{(\eta)}d\omega,
 \label{eq:nop}
\end{equation}
\begin{align}
 \underline{\hat{H}}
 &=\sum_q\int\frac{V}{(2\pi)^3}\hbar\omega_\mathbf{k}\hat{n}_{\mathbf{k},q}d^3k\nonumber\\
 &=\sum_{J,M,\eta}\int_0^\infty\frac{2V\omega^2}{\pi c^3}\hbar\omega\hat{n}_{\omega,J,M}^{(\eta)}d\omega,
\end{align}
\begin{equation}
 \underline{\hat{\mathbf{p}}}
 =\sum_q\int\frac{V}{(2\pi)^3}\hbar\mathbf{k}\hat{n}_{\mathbf{k},q}d^3k,
\end{equation}
\begin{equation}
 \underline{\hat{\mathfrak{h}}}
 =\sum_q\int\frac{V}{(2\pi)^3}\hbar q\hat{n}_{\mathbf{k},q}d^3k,
\end{equation}
\begin{equation}
 \underline{\hat{J}^2}
 =\sum_{J,M,\eta}\int_0^\infty\frac{2V\omega^2}{\pi c^3}\hbar^2 J(J+1)\hat{n}_{\omega,J,M}^{(\eta)}d\omega,
\end{equation}
\begin{equation}
 \underline{\hat{J}_z}
 =\sum_{J,M,\eta}\int_0^\infty\frac{2V\omega^2}{\pi c^3}\hbar M\hat{n}_{\omega,J,M}^{(\eta)}d\omega.
 \label{eq:Jzop}
\end{equation}
The second-quantized square of the spin operator is trivially given by $\underline{\hat{S}^2}=\hbar^2S(S+1)\underline{\hat{n}}$. The conventional results in Eqs.~\eqref{eq:nop}--\eqref{eq:Jzop} strongly indicate the correctness of the present formalism since the results predicted by these conventional operators agree with ample experimental evidence.

Note that first-quantized density operators defined through Eq.~\eqref{eq:densityoperator} can also be used in place of $\hat{O}$ in Eq.~\eqref{eq:secondquantizedoperator}. Thus, Eq.~\eqref{eq:secondquantizedoperator} also defines the density operators in the second quantization picture. For example, using the first-quantized number-density operator $\hat{\rho}_{\hat{n}}(\mathbf{r},\mathbf{r}')=\delta(\mathbf{r}-\mathbf{r}')$, following from Eqs.~\eqref{eq:singlephotonnumber} and \eqref{eq:densityoperator}, we then obtain the second-quantized number density operator as
\begin{equation}
 \underline{\hat{\rho}_{\hat{n}}}^{\!\!(M)}=:\!\hat{\Psi}^\dag\hat{\Psi}\!:.
 \label{eq:rhonop}
\end{equation}
In analogy to the Dirac theory, $\underline{\hat{\rho}_{\hat{n}}}^{\!\!(M)}$ is equal to the time component of the number-density-current four-vector operator, which is given for photons by
\begin{equation}
 \underline{\hat{\mathcal{I}}_\mathrm{M}^\mu}=:\!\hat{\bar{\Psi}}\boldsymbol{\gamma}_\mathrm{B}^\mu\hat{\Psi}\!:.
 \label{eq:numberdensitycurrentoperator}
\end{equation}

Using Eq.~\eqref{eq:secondquantizedoperator} with the pertinent first-quantized density operators obtaned through Eq.~\eqref{eq:densityoperator},
we obtain the second-quantized Hamiltonian, momentum, orbital angular momentum, spin angular momentum, and helicity density operators of the electromagnetic field as
\begin{equation}
 \underline{\hat{\rho}_{\hat{H}}}^{\!\!(M)}=:\!\hat{\Psi}^\dag\hat{H}\hat{\Psi}\!:,
 \label{eq:rhoEop}
\end{equation}
\begin{equation}
 \textstyle\underline{\hat{\boldsymbol{\rho}}_{\hat{\mathbf{p}}}}^{\!\!(M)}=:\!\frac{1}{2}[\hat{\Psi}^\dag\hat{\mathbf{p}}\hat{\Psi}-(\hat{\mathbf{p}}\hat{\Psi}^\dag)\hat{\Psi}]\!:,
\end{equation}
\begin{equation}
 \textstyle\underline{\hat{\boldsymbol{\rho}}_{\hat{\mathbf{L}}}}^{\!\!(M)}=:\!\frac{1}{2}[\hat{\Psi}^\dag\hat{\mathbf{L}}\hat{\Psi}-(\hat{\mathbf{L}}\hat{\Psi}^\dag)\hat{\Psi}]\!:,
\end{equation}
\begin{equation}
 \textstyle\underline{\hat{\boldsymbol{\rho}}_{\hat{\mathbf{S}}}}^{\!\!(M)}=:\!\hat{\Psi}^\dag\hat{\mathbf{S}}\hat{\Psi}\!:,
\end{equation}
\begin{equation}
 \underline{\hat{\rho}_{\hat{\mathfrak{h}}}}^{\!\!(M)}=:\!\hat{\Psi}^\dag\hat{\mathfrak{h}}\hat{\Psi}\!:.
 \label{eq:rhohop}
\end{equation}
The expectation values of these operators for second-quantized single-photon states can be compared with the first-quantized density expectation values in Eqs.~\eqref{eq:rhoE}--\eqref{eq:rhoh}. This comparison is discussed in Sec.~\ref{sec:singlephotonsecondquantization} below.

\subsection{\label{sec:singlephotonsecondquantization}Photon states in the second quantization}

The single-photon plane-wave Fock state is the momentum eigenstate that is a continuum state written in the Heisenberg picture in terms of the plane-wave photon states $\Psi_{\mathbf{k},q}(t,\mathbf{r})$, presented in Sec.~\ref{sec:spinorialwavefunctions} and in Appendix \ref{apx:spinorialwavefunctions}, as
\begin{equation}
 |1_{\mathbf{k},q}\rangle
 =\int\Psi_{\mathbf{k},q}(0,\mathbf{r})|\mathbf{r}\rangle d^3r.
\label{eq:planewaveFockstate}
\end{equation}
Similarly, we can write the single-photon spherical Fock state in terms of the spherical photon states $\Psi_{\omega,J,M}^{(\eta)}(t,\mathbf{r})$, presented in Appendix \ref{apx:spinorialwavefunctions}, as
\begin{equation}
 |1_{\omega,J,M}^{(\eta)}\rangle
 =\int\Psi_{\omega,J,M}^{(\eta)}(0,\mathbf{r})|\mathbf{r}\rangle d^3r.
\label{eq:sphericalFockstate}
\end{equation}
As conventional, many-photon states are formed as symmetric
product states of the single-photon states. The direct sum of all product states of different numbers of photons then forms a Fock space.

In analogy to the standard QED, operating on a single-mode plane-wave vacuum state $|0_{\mathbf{k},q}\rangle$, the creation operator $\hat{a}_{\mathbf{k},q}^\dag$ creates a single-photon Fock state $|1_{\mathbf{k},q}\rangle$ with a wave vector $\mathbf{k}$ and the helicity quantum number $q$ as $|1_{\mathbf{k},q}\rangle=\sqrt{\hbar\omega_\mathbf{k}}\hat{a}_{\mathbf{k},q}^\dag|0_{\mathbf{k},q}\rangle$, and the annihilation operator $\hat{a}_{\mathbf{k},q}$ annihilates this state as $\hat{a}_{\mathbf{k},q}|1_{\mathbf{k},q}\rangle=\sqrt{\hbar\omega_\mathbf{k}}|0_{\mathbf{k},q}\rangle$. Correspondingly, the operator $\hat{\Psi}^{(-)}$ creates a single-photon state at position $\mathbf{r}$ as $\hat{\Psi}^{(-)}|0_\mathbf{r}\rangle=|1_\mathbf{r}\rangle$. The operator $\hat{\Psi}^{(+)}$ annihilates this state as $\hat{\Psi}^{(+)}|1_\mathbf{r}\rangle=|0_\mathbf{r}\rangle$. Similar relations apply to spherical photon states. The completeness relation for single-particle states is given in the plane-wave and spherical state bases by \cite{Peskin2018}
\begin{align}
 \hat{\mathbf{1}}_\mathrm{1} &=\sum_q\int\frac{V}{(2\pi)^3}\frac{1}{\hbar\omega_\mathbf{k}}|1_{\mathbf{k},q}\rangle\langle 1_{\mathbf{k},q}|d^3k\nonumber\\
 & =\sum_{J,M,\eta}\int_0^\infty\frac{2V\omega^2}{\pi c^3}\frac{1}{\hbar\omega}|1_{\omega,J,M}^{(\eta)}\rangle\langle 1_{\omega,J,M}^{(\eta)}|d\omega.
\end{align}
This operator is an identity operator within the subspace of single-particle states, and it is zero in the rest of the Hilbert space.

It is interesting to compare the single-photon expectation values of the second-quantized density operators in Eqs.~\eqref{eq:rhoEop}--\eqref{eq:rhohop} to the first-quantized density expectation values in Eqs.~\eqref{eq:rhoE}--\eqref{eq:rhoh}. For simple comparison, we do not form narrow-frequency-band wave-packet states here but use the $\delta$-function-normalized Fock states. As an example, for the Hamiltonian density operator expectation value for the single-photon plane-wave Fock state, we obtain
\begin{equation}
 \frac{\langle 1_{\mathbf{k},q}|\underline{\hat{\rho}_{\hat{H}}}|1_{\mathbf{k},q}\rangle}{\langle 1_{\mathbf{k},q}|1_{\mathbf{k},q}\rangle}=\frac{\Psi_{\mathbf{k},q}^\dag\hat{H}\Psi_{\mathbf{k},q}}{\int \Psi_{\mathbf{k},q}^\dag\Psi_{\mathbf{k},q}d^3r}
 =\hbar\omega_\mathbf{k}\frac{\Psi_{\mathbf{k},q}^\dag\Psi_{\mathbf{k},q}}{\int \Psi_{\mathbf{k},q}^\dag\Psi_{\mathbf{k},q}d^3r}.
 \label{eq:rhoE2nd}
\end{equation}
In the calculation, we have used $\langle 1_{\mathbf{k},q}|\hat{a}_{\mathbf{k}',q'}^\dag\hat{a}_{\mathbf{k}'',q''}|1_{\mathbf{k},q}\rangle=\hbar\omega_\mathbf{k}[(2\pi)^3/V]^2\delta_{q,q'}\delta_{q',q''}\delta(\mathbf{k}-\mathbf{k}')\delta(\mathbf{k}'-\mathbf{k}'')$.
The result of Eq.~\eqref{eq:rhoE2nd} corresponds to the first-quantized Hamiltonian density expectation value in Eq.~\eqref{eq:rhoE} if $\Psi_{\mathbf{k},q}$ is used in Eq.~\eqref{eq:rhoE}. Corresponding relations apply to other density operators. Thus, the second-quantized expectation value densities for single-photon states are equal to the first-quantized expectation value densities as expected. The comparison of the first-quantization and second-quantization density expectation values shows that the electromagnetic spinor $\Psi$ is a rigorous physical concept also in understanding single-photon states.

\section{\label{sec:Lagrangian2}Generating Lagrangian density of gravity, special unitary symmetry, and the SEM tensor}

In the sections above, we have presented how the conventional QED is expressed in the eight-spinor formalism. Next, we introduce \emph{the generating Lagrangian density of gravity} that plays, in the definition of the gauge theory of gravity \cite{Partanen2023c}, a similar role as the conventional Lagrangian density of the free Dirac field plays in the definition of the gauge theory of QED \cite{Peskin2018,Schwartz2014,Weinberg1996}. The generating Lagrangian density of gravity is associated with a special unitary symmetry of the quantum fields in the standard model, and it enables an elegant derivation of the symmetric SEM tensors as described in detail in the sections below. The Yang-Mills gauge theory of unified gravity that follows is presented in a separate work \cite{Partanen2023c}.

\subsection{Generating Lagrangian density of gravity}

As the starting point for defining the generating Lagrangian density of gravity, we use the analogy with standard gauge theory as it is applied in deriving the full QED from the Lagrangian density of the free Dirac field, $\mathcal{L}_\mathrm{QED,0}=\mathcal{L}_\mathrm{QED,D}$, in Eq.~\eqref{eq:Lagrangiandensity1}. We have presented a concise summary of the conventional gauge theory of QED in Appendix \ref{apx:fourcurrent}. As shown in Appendix \ref{apx:fourcurrent}, the unitary transformation $\psi\rightarrow e^{i\theta}\psi$ with parameter $\theta$ and the electromagnetic-gauge-covariant derivative $\vec{D}_a=\vec{\partial}_a+i\frac{q_\mathrm{e}}{\hbar}A_{\Re a}$ in Eq.~\eqref{eq:covariantderivativeEM} \emph{generate} from the Lagrangian density $\mathcal{L}_\mathrm{QED,0}$ the full QED. In Appendix \ref{apx:fourcurrent}, we also show that the variation of the generating Lagrangian density of QED, $\delta\mathcal{L}_\mathrm{QED,0}=-\frac{\hbar}{q_\mathrm{e}}J_\mathrm{e\Re}^a\partial_a\theta$, is proportional to the electric four-current density, which is the source term of the electromagnetic field.

Based on the known properties of gravitational interaction, we must set certain conditions that we require the generating Lagrangian density of gravity and the resulting full gauge theory of unified gravity to satisfy:
\begin{enumerate}
 \item The theory must satisfy the global Lorentz invariance and the general covariance, which means the form invariance of physical laws under general differentiable coordinate transformations. More strongly, we require diffeomorphism invariance.
 \item The SEM tensor must act as the source term of the gravitational field. It follows that the gravitational field is a tensor gauge field in contrast with the vector gauge fields of the standard model.
 \item Instead of the Abelian U(1) gauge theory of QED, we must use the non-Abelian Yang-Mills gauge theory analogous to the theories of weak and strong interactions \cite{Peskin2018,Schwartz2014,Weinberg1996}. This is because four symmetry generators are needed for the description of the tensor gauge field.
 \item To enable unification of gravity with the fundamental interactions of the standard model, the gauge theory of gravity must be based on an internal special unitary symmetry or subsymmetry of the quantum fields in the standard model.
 \item  The theory must contain a new coupling constant $g_\mathrm{g}$, called the coupling constant of unified gravity. The variation of the generating Lagrangian density of gravity with respect to the symmetry transformation parameters must be directly proportional to the SEM tensor divided by $g_\mathrm{g}$.
 \item The gauge theory of gravity must enable writing the dynamical equations for the gravitational field through the Euler-Lagrange equations. The gravitational field equations must reproduce the experimentally verified predictions of general relativity.
 \item Through the Euler-Lagrange equations, we must also obtain the generalized equations of motion containing gravitational coupling for all the fundamental interactions of the standard model \cite{Peskin2018,Schwartz2014,Weinberg1996}. In the Minkowski metric limit of the gravitational gauge field, these equations must reproduce the dynamical equations of the standard model.
\end{enumerate}

The generating Lagrangian density of gravity can be seen as the \emph{fundamental hypothesis} of the theory that unifies the standard model and gravity. All quantum fields of the standard model can be included on equal footing, but, for simplicity, here we present the theory using the Dirac electron-positron field and the electromagnetic field only. In a process of trial and error, we have heuristically ended up to the generating Lagrangian density of gravity, given by
\begin{align}
 \mathcal{L}_0 &=\Big[\frac{\hbar c}{4g_\mathrm{g}}\bar{\psi}(\cev{D}\bar{\mathbf{I}}_\mathrm{g}\boldsymbol{\gamma}_\mathrm{B}^5\boldsymbol{\gamma}_\mathrm{B}^\nu\vec{\partial}_\nu\mathbf{I}_\mathrm{g}\boldsymbol{\gamma}_\mathrm{F}-\bar{\boldsymbol{\gamma}}_\mathrm{F}\bar{\mathbf{I}}_\mathrm{g}\boldsymbol{\gamma}_\mathrm{B}^5\boldsymbol{\gamma}_\mathrm{B}^\nu\vec{\partial}_\nu\mathbf{I}_\mathrm{g}\vec{D})\psi\nonumber\\
 &\hspace{0.4cm}+\frac{im_\mathrm{e}c^2}{2g_\mathrm{g}}\bar{\psi}\bar{\mathbf{e}}_8\mathbf{I}_\mathrm{g}^\dag\boldsymbol{\gamma}_\mathrm{B}^5\boldsymbol{\gamma}_\mathrm{B}^\nu\vec{\partial}_\nu\bar{\mathbf{I}}_\mathrm{g}^\dag\mathbf{e}_8\psi+m_\mathrm{e}c^2\bar{\psi}\psi\nonumber\\
 &\hspace{0.4cm}+\frac{i}{g_\mathrm{g}}\bar{\Psi}_\Re\mathbf{I}_\mathrm{g}^\dag\boldsymbol{\gamma}_\mathrm{B}^5\boldsymbol{\gamma}_\mathrm{B}^\nu\vec{\partial}_\nu\bar{\mathbf{I}}_\mathrm{g}^\dag\Psi_\Re
 +\bar{\Psi}_\Re\Psi_\Re\Big]\sqrt{-g}.
 \label{eq:Lagrangiandensity2}
\end{align}
Here $g=\det(g_{\mu\nu})$ denotes the determinant of the metric tensor $g_{\mu\nu}$, whose definition in the present theory is discussed in Ref. [60]. The gamma matrices with Greek indices are given by $\boldsymbol{\gamma}_\mathrm{B}^\nu=g_a^{\;\,\nu}\boldsymbol{\gamma}_\mathrm{B}^a$, where $g_a^{\;\,\nu}$ is the tetrad field that maps the tangent space coordinates to general space-time coordinates \cite{Einstein1928,Sciama1964,Blagojevic2013}. The first three terms inside the square brackets of the generating Lagrangian density of gravity in Eq.~\eqref{eq:Lagrangiandensity2} describe the Dirac field. The last two terms describe the electromagnetic field. The coupling between the Dirac and electromagnetic fields is described through $\vec{D}$ and $\cev{D}$. The partial derivatives $\vec{\partial}_\nu$ in front of $\mathbf{I}_\mathrm{g}$ and $\bar{\mathbf{I}}_\mathrm{g}^\dag$ in Eq.~\eqref{eq:Lagrangiandensity2} act on $\mathbf{I}_\mathrm{g}$ and do not extend to the spinors $\Psi_\Re$ and $\psi$. In the present work, we use $\mathbf{I}_\mathrm{g}=\mathbf{I}_8$ and leave the study of possible space-time-dependent forms of $\mathbf{I}_\mathrm{g}$ as a topic of further work. Since $\mathbf{I}_\mathrm{g}=\mathbf{I}_8$ is a constant identity matrix, the terms $\vec{\partial}_\nu\mathbf{I}_\mathrm{g}$ and $\vec{\partial}_\nu\bar{\mathbf{I}}_\mathrm{g}^\dag$ are actually identically zero. These terms become nonzero after the partial derivatives $\vec{\partial}_\nu$ are replaced by gravitational-gauge-covariant derivatives as shown in the Yang-Mills gauge theory of unified gravity in Ref.~\cite{Partanen2023c}. The terms $\vec{\partial}_\nu\mathbf{I}_\mathrm{g}$ and $\vec{\partial}_\nu\bar{\mathbf{I}}_\mathrm{g}^\dag$ become nonzero also when the related space-time-dependent symmetry transformation is carried out for $\mathbf{I}_\mathrm{g}$. This symmetry transformation will be elaborated below.

\subsection{\label{sec:unitary}SU(8)$_\mathbf{4D}$ symmetry}

While the theories of the electromagnetic, weak, and strong interactions utilize U(1), SU(2), and SU(3) symmetries \cite{Peskin2018,Schwartz2014,Weinberg1996}, here we utilize a four-dimensional subsymmetry of SU(8), defined below and denoted by SU(8)$_\mathrm{4D}$. In analogy to the description of gauge symmeties in the conventional quantum field theory \cite{Peskin2018,Schwartz2014,Zee2010,Weinberg1996}, we consider the SU(8)$_\mathrm{4D}$ symmetry transformation under which the generating Lagrangian density in Eq.~\eqref{eq:Lagrangiandensity2} is globally invariant and whose space-time-dependent variation without introducing the gauge field is related to the source term of the gauge field. In our case, this source term is the SEM tensor, while in the case of the U(1) symmetry of QED it is the electric four-current density as briefly presented in Appendix \ref{apx:fourcurrent}.  The generating Lagrangian density of gravity is invariant with respect to any global, i.e., space-time-independent, symmetry transformation of $\mathbf{I}_\mathrm{g}=\mathbf{I}_8$, since such transformations do not make $\vec{\partial}_\nu\mathbf{I}_\mathrm{g}$ and $\vec{\partial}_\nu\bar{\mathbf{I}}^\dag_\mathrm{g}$ nonzero. Thus, the generating Lagrangian density of gravity is trivially globally invariant with respect to the SU(8)$_\mathrm{4D}$ symmetry.

The SU(8)$_\mathrm{4D}$ symmetry differs from the U(1) symmetry of the conventional QED: While the U(1) symmetry operates primarily on the Dirac field and generates the electromagnetic field as a gauge field, the SU(8)$_\mathrm{4D}$ symmetry operates primarily on both the Dirac and electromagnetic fields and generates the gravitational field as a gauge field. Such an operation is necessary since gravity is known to affect all fields and matter. Here we limit our study of the SU(8)$_\mathrm{4D}$ symmetry to its relation to the SEM tensors of the Dirac and electromagnetic fields without introducing the gravitational gauge field. The Yang-Mills gauge theory of unified gravity arising from the SU(8)$_\mathrm{4D}$ symmetry is studied in Ref.~\cite{Partanen2023c}. We define the SU(8)$_\mathrm{4D}$ symmetry transformation as
\begin{equation}
 \mathbf{I}_\mathrm{g}\rightarrow \mathbf{U}\mathbf{I}_\mathrm{g},
 \hspace{0.5cm}\text{where }\mathbf{U}=\exp(i\phi_a\mathbf{t}^a).
 \label{eq:transformationU}
\end{equation}
The symmetry transformation matrix $\mathbf{U}$ has determinant 1. Thus, it is an element of the special unitary group SU(8). The real-valued four-vector $\phi_a$ in Eq.~\eqref{eq:transformationU} defines the symmetry transformation parameters. The four transformation generators $\mathbf{t}^a$ are constant traceless Hermitian matrices, given in terms of the complex-conjugated bosonic gamma matrices as $\mathbf{t}^a=(\boldsymbol{\gamma}_\mathrm{B}^0\boldsymbol{\gamma}_\mathrm{B}^5\boldsymbol{\gamma}_\mathrm{B}^a)^*$. The commutation relation of the generators is given by $[\mathbf{t}^a,\mathbf{t}^b]=if_c^{\;\,ab}\mathbf{t}^c$, where $f_c^{\;\,ab}=2\varepsilon^{0cab}$. Thus, $\mathbf{t}^a$ generate a Lie algebra, which is a prerequisite for developing the Yang-Mills gauge theory in Ref.~\cite{Partanen2023c}. The generators are Lorentz invariant satisfying $\boldsymbol{\Lambda}_\mathrm{J}\mathbf{t}^a\boldsymbol{\Lambda}_\mathrm{J}^{-1}=\mathbf{t}^a$. The matrix $\mathbf{U}$ satisfies the commutation relations $[\mathbf{U},\boldsymbol{\gamma}_\mathrm{B}^5]=\mathbf{0}$, $[\mathbf{U},\boldsymbol{\gamma}_\mathrm{B}^a\boldsymbol{\gamma}_\mathrm{B}^b]=\mathbf{0}$, and $[\mathbf{U},\boldsymbol{\Lambda}_\mathrm{J}]=\mathbf{0}$. These relations are necessary for the gravitational gauge invariance of the Lagrangian density in the Yang-Mills gauge theory of unified gravity \cite{Partanen2023c}. The group of transformations defined by $\mathbf{U}$ corresponds to a four-dimensional subgroup of SU(8), which is isomorphic to SU(2)$\otimes$U(1), and which we denote SU(8)$_\mathrm{4D}$. For comparison with the present SU(8)$_\mathrm{4D}$ generators $\mathbf{t}^a$, in the case of the SU(2) symmetry of the electroweak interaction, the generators are the three Pauli matrices, and in the case of the SU(3) symmetry of quantum chromodynamics (QCD), the generators are the eight Gell-Mann matrices \cite{Schwartz2014}.

\subsection{\label{sec:SEMtensor}Symmetric SEM tensor}

In this section, we generalize the global SU(8)$_\mathrm{4D}$ symmetry transformation of Eq.~\eqref{eq:transformationU} into a local symmetry transformation by making the symmetry transformation parameters $\phi_a$ space-time dependent as $\phi_a=\phi(x^0,x^1,x^2,x^3)$. Would we follow the Yang-Mills gauge theory, we should simultaneously introduce the gauge-covariant derivative that makes the derivative terms of the generating Lagrangian density of gravity in Eq.~\eqref{eq:Lagrangiandensity2} invariant. The consequences of adding the gauge-covariant derivative are elaborated in the separate work on the Yang-Mills gauge theory of unified gravity in Ref.~\cite{Partanen2023c}. Here we consider the variation of the generating Lagrangian density of gravity without adding the gauge field. We recover the profound relationship between the SU(8)$_\mathrm{4D}$ symmetry and the symmetric SEM tensor source term of gravity. This relationship is completely analogous to the relation between the U(1) symmetry of QED and the electric four-current density source term of electromagnetism as briefly presented in Appendix \ref{apx:fourcurrent}.

The infinitesimal variation of $\mathbf{I}_\mathrm{g}=\mathbf{I}_8$ in the SU(8)$_\mathrm{4D}$ symmetry transformation of Eq.~\eqref{eq:transformationU} with respect to the transformation parameters $\phi_a$ is given by
\begin{equation}
 \delta\mathbf{I}_\mathrm{g}=i\mathbf{t}^a\delta\phi_a.
 \label{eq:infinitesimalvariation}
\end{equation}
Using this inifinitesimal variation, it is straightforward to calculate the variation of the generating Lagrangian density of gravity in Eq.~\eqref{eq:Lagrangiandensity2} with respect to $\phi_a$ as
\begin{align}
 \delta\mathcal{L}_0 &=\Big[\frac{\hbar c}{4g_\mathrm{g}}\bar{\psi}[\cev{D}\overline{(\delta\mathbf{I}_\mathrm{g})}\boldsymbol{\gamma}_\mathrm{B}^5\boldsymbol{\gamma}_\mathrm{B}^\nu\vec{\partial}_\nu\mathbf{I}_\mathrm{g}\boldsymbol{\gamma}_\mathrm{F}
 +\cev{D}\bar{\mathbf{I}}_\mathrm{g}\boldsymbol{\gamma}_\mathrm{B}^5\boldsymbol{\gamma}_\mathrm{B}^\nu\vec{\partial}_\nu(\delta\mathbf{I}_\mathrm{g})\boldsymbol{\gamma}_\mathrm{F}\nonumber\\
 &\hspace{0.4cm}-\bar{\boldsymbol{\gamma}}_\mathrm{F}\overline{(\delta\mathbf{I}_\mathrm{g})}\boldsymbol{\gamma}_\mathrm{B}^5\boldsymbol{\gamma}_\mathrm{B}^\nu\vec{\partial}_\nu\mathbf{I}_\mathrm{g}\vec{D}
 -\bar{\boldsymbol{\gamma}}_\mathrm{F}\bar{\mathbf{I}}_\mathrm{g}\boldsymbol{\gamma}_\mathrm{B}^5\boldsymbol{\gamma}_\mathrm{B}^\nu\vec{\partial}_\nu(\delta\mathbf{I}_\mathrm{g})\vec{D}]\psi\nonumber\\
 &\hspace{0.4cm}+\frac{im_\mathrm{e}c^2}{2g_\mathrm{g}}\bar{\psi}\bar{\mathbf{e}}_8(\delta\mathbf{I}_\mathrm{g})^\dag\boldsymbol{\gamma}_\mathrm{B}^5\boldsymbol{\gamma}_\mathrm{B}^\nu\vec{\partial}_\nu\bar{\mathbf{I}}_\mathrm{g}^\dag\mathbf{e}_8\psi\nonumber\\
 &\hspace{0.4cm}+\frac{im_\mathrm{e}c^2}{2g_\mathrm{g}}\bar{\psi}\bar{\mathbf{e}}_8\mathbf{I}_\mathrm{g}^\dag\boldsymbol{\gamma}_\mathrm{B}^5\boldsymbol{\gamma}_\mathrm{B}^\nu\vec{\partial}_\nu\overline{(\delta\mathbf{I}_\mathrm{g})}^\dag\mathbf{e}_8\psi\nonumber\\
 &\hspace{0.4cm}+\frac{i}{g_\mathrm{g}}\bar{\Psi}_\Re(\delta\mathbf{I}_\mathrm{g})^\dag\boldsymbol{\gamma}_\mathrm{B}^5\boldsymbol{\gamma}_\mathrm{B}^\nu\vec{\partial}_\nu\bar{\mathbf{I}}_\mathrm{g}^\dag\Psi_\Re\nonumber\\
 &\hspace{0.4cm}+\frac{i}{g_\mathrm{g}}\bar{\Psi}_\Re\mathbf{I}_\mathrm{g}^\dag\boldsymbol{\gamma}_\mathrm{B}^5\boldsymbol{\gamma}_\mathrm{B}^\nu\vec{\partial}_\nu\overline{(\delta\mathbf{I}_\mathrm{g})}^\dag\Psi_\Re\Big]\sqrt{-g}\nonumber\\
 &=\sqrt{-g}\Big[\frac{i\hbar c}{4g_\mathrm{g}}\bar{\psi}[\cev{D}\boldsymbol{\gamma}_\mathrm{B}^5\boldsymbol{\gamma}_\mathrm{B}^\nu\mathbf{t}^a\boldsymbol{\gamma}_\mathrm{F}
 -\bar{\boldsymbol{\gamma}}_\mathrm{F}\boldsymbol{\gamma}_\mathrm{B}^5\boldsymbol{\gamma}_\mathrm{B}^\nu\mathbf{t}^a\vec{D}]\psi\vec{\partial}_\nu\delta\phi_a\nonumber\\
 &\hspace{0.4cm}+\frac{m_\mathrm{e}c^2}{2g_\mathrm{g}}g^{a\nu}\bar{\psi}\psi\vec{\partial}_\nu\delta\phi_a
 +\frac{1}{g_\mathrm{g}}\bar{\Psi}_\Re\mathbf{t}^a\boldsymbol{\gamma}_\mathrm{B}^\nu\boldsymbol{\gamma}_\mathrm{B}^5\Psi_\Re\vec{\partial}_\nu\delta\phi_a\Big]\nonumber\\
 &=\frac{\sqrt{-g}}{g_\mathrm{g}}T^{a\nu}\vec{\partial}_\nu\delta\phi_a.
 \label{eq:Lagrangiandensityvariation}
\end{align}
In the second equality of Eq.~\eqref{eq:Lagrangiandensityvariation}, we have used an identity $\boldsymbol{\gamma}_\mathrm{B}^5\boldsymbol{\gamma}_\mathrm{B}^\nu\bar{\mathbf{t}}^a=-\mathbf{t}^a\boldsymbol{\gamma}_\mathrm{B}^\nu\boldsymbol{\gamma}_\mathrm{B}^5$ for the electromagnetic field term and the mass term of the Dirac field. For the mass term of the Dirac field, we have also used $\bar{\mathbf{e}}_8\mathbf{t}^a\boldsymbol{\gamma}_\mathrm{B}^\nu\boldsymbol{\gamma}_\mathrm{B}^5\mathbf{e}_8=g^{a\nu}=g_b^{\;\,\nu}\eta^{ab}$. The quantity $T^{a\nu}$, defined by the last equality of Eq.~\eqref{eq:Lagrangiandensityvariation}, is found to be related to the total symmetric SEM tensor $T^{\mu\nu}$ of the Dirac and electromagnetic fields by the inverse tetrad field as $T^{a\nu}=g_{\;\mu}^aT^{\mu\nu}$. Thus, the SEM tensor is given by
\begin{align}
 T^{\mu\nu} &=\frac{i\hbar c}{4}\bar{\psi}(\cev{D}\boldsymbol{\gamma}_\mathrm{B}^5\boldsymbol{\gamma}_\mathrm{B}^\nu\mathbf{t}^\mu\boldsymbol{\gamma}_\mathrm{F}-\bar{\boldsymbol{\gamma}}_\mathrm{F}\boldsymbol{\gamma}_\mathrm{B}^5\boldsymbol{\gamma}_\mathrm{B}^\nu\mathbf{t}^\mu\vec{D})\psi\nonumber\\
 &\hspace{0.4cm}+\frac{m_\mathrm{e}c^2}{2}g^{\mu\nu}\bar{\psi}\psi+\bar{\Psi}_\Re\mathbf{t}^\mu\boldsymbol{\gamma}_\mathrm{B}^\nu\boldsymbol{\gamma}_\mathrm{B}^5\Psi_\Re.
 \label{eq:semtensors}
\end{align}
The first two terms of Eq.~\eqref{eq:semtensors} are equal to the SEM tensor of the Dirac field in an external electromagnetic field. The last term is equal to the SEM tensor of the electromagnetic field. The conservation law for the SEM tensor is given by $\partial_\nu T^{\mu\nu}=0$. That Eq.~\eqref{eq:semtensors} leads to the well-known expressions of the symmetric SEM tensors of the Dirac and electromagnetic fields is revealed in the sections below.

\subsubsection{Stress-energy-momentum tensor of the Dirac field}

From the first two terms of Eq.~\eqref{eq:semtensors}, we obtain the SEM tensor of the Dirac field, given by
\begin{align}
 \!T_\mathrm{D}^{\mu\nu}
 &\!=\!\frac{i\hbar c}{4}\bar{\psi}(\cev{D}\boldsymbol{\gamma}_\mathrm{B}^5\boldsymbol{\gamma}_\mathrm{B}^\nu\mathbf{t}^\mu\boldsymbol{\gamma}_\mathrm{F}
 \!-\!\bar{\boldsymbol{\gamma}}_\mathrm{F}\boldsymbol{\gamma}_\mathrm{B}^5\boldsymbol{\gamma}_\mathrm{B}^\nu\mathbf{t}^\mu\vec{D})\psi
 \!+\!\frac{m_\mathrm{e}c^2}{2}g^{\mu\nu}\bar{\psi}\psi\nonumber\\
 &\!=\!\frac{i\hbar c}{4}\bar{\psi}(\boldsymbol{\gamma}_\mathrm{F}^\mu\vec{D}^\nu
 +\boldsymbol{\gamma}_\mathrm{F}^\nu\vec{D}^\mu
 -\cev{D}^\nu\boldsymbol{\gamma}_\mathrm{F}^\mu
 -\cev{D}^\mu\boldsymbol{\gamma}_\mathrm{F}^\nu\nonumber\\
 &\hspace{0.5cm}-g^{\mu\nu}\boldsymbol{\gamma}_\mathrm{F}^\rho\vec{D}_\rho+\cev{D}_\rho\boldsymbol{\gamma}_\mathrm{F}^\rho g^{\mu\nu})\psi
 +\frac{m_\mathrm{e}c^2}{2}g^{\mu\nu}\bar{\psi}\psi.
 \label{eq:SEMDrev}
\end{align}
In the last equality, we have used the mathematical identities $\bar{\boldsymbol{\gamma}}_\mathrm{F}\boldsymbol{\gamma}_\mathrm{B}^5\boldsymbol{\gamma}_\mathrm{B}^\nu\mathbf{t}^\mu\vec{D}=-\boldsymbol{\gamma}_\mathrm{F}^\mu\vec{D}^\nu-\boldsymbol{\gamma}_\mathrm{F}^\nu\vec{D}^\mu+g^{\mu\nu}\boldsymbol{\gamma}_\mathrm{F}^\rho\vec{D}_\rho$ and $\cev{D}\boldsymbol{\gamma}_\mathrm{B}^5\boldsymbol{\gamma}_\mathrm{B}^\nu\mathbf{t}^\mu\boldsymbol{\gamma}_\mathrm{F}=-\cev{D}^\mu\boldsymbol{\gamma}_\mathrm{F}^\nu-\cev{D}^\nu\boldsymbol{\gamma}_\mathrm{F}^\mu+\cev{D}_\rho\boldsymbol{\gamma}_\mathrm{F}^\rho g^{\mu\nu}$.
When the Dirac equation is satisfied, the terms proportional to the inverse metric tensor $g^{\mu\nu}$ cancel each other, and the SEM tensor of the Dirac field becomes
\begin{align}
 T_\mathrm{D}^{\mu\nu}
 &
 =\left[\begin{array}{cccc}
 W_\mathrm{D} & c\mathbf{G}_\mathrm{D}\\
 c\mathbf{G}_\mathrm{D} & \boldsymbol{\mathcal{T}}_\mathrm{D}
\end{array}\right]\nonumber\\
 &=\frac{i\hbar c}{4}\bar{\psi}(\boldsymbol{\gamma}_\mathrm{F}^\mu\vec{D}^\nu
 +\boldsymbol{\gamma}_\mathrm{F}^\nu\vec{D}^\mu
 -\cev{D}^\nu\boldsymbol{\gamma}_\mathrm{F}^\mu
 -\cev{D}^\mu\boldsymbol{\gamma}_\mathrm{F}^\nu)\psi.
 \label{eq:SEMD}
\end{align}
This is the well-known result for the symmetric SEM tensor of the Dirac field in an external electromagnetic field \cite{Peskin2018}. In general, all four terms on the last line of Eq.~\eqref{eq:SEMD} are necessary for the SEM tensor to be symmetric and its components real valued. In the Minkowski space-time, the energy density, momentum density, and stress tensor components of the Dirac field SEM tensor are given by
\begin{align}
 W_\mathrm{D} &=\rho_{\hat{H}}^\mathrm{(D)},\nonumber\\
 \mathbf{G}_\mathrm{D} &=\boldsymbol{\rho}_{\hat{\mathbf{p}}-q_\mathrm{e}\mathbf{A}_\Re}^\mathrm{(D)}
 +\frac{1}{2}\nabla\times\boldsymbol{\rho}_{\hat{\mathbf{S}}}^\mathrm{(D)},\nonumber\\
 \boldsymbol{\mathcal{T}}_\mathrm{D} &=\frac{1}{2}\big(\boldsymbol{\rho}_{\hat{\mathbf{v}}\otimes(\hat{\mathbf{p}}-q_\mathrm{e}\mathbf{A}_\Re)}^\mathrm{(D)}+\boldsymbol{\rho}_{(\hat{\mathbf{p}}-q_\mathrm{e}\mathbf{A}_\Re)\otimes\hat{\mathbf{v}}}^\mathrm{(D)}\big).
 \label{eq:DiracT}
\end{align}
Here $\hat{\mathbf{v}}=c\boldsymbol{\alpha}_\mathrm{F}$ is the effective velocity operator for the Dirac field, obtained in the Heisenberg picture from the time dependence of the position operator as $\hat{\mathbf{v}}=\frac{\partial\mathbf{r}}{\partial t}=\frac{i}{\hbar}[\hat{H},\mathbf{r}]=c\boldsymbol{\alpha}_\mathrm{F}$. The Dirac field Hamiltonian operator $\hat{H}$ used here is given in Eq.~\eqref{eq:DiracHamiltonian1}.

\subsubsection{Stress-energy-momentum tensor of the electromagnetic field}

From the last term of Eq.~\eqref{eq:semtensors}, we obtain the SEM tensor of the electromagnetic field, given by
\begin{align}
 T_\mathrm{M}^{\mu\nu} &=\bar{\Psi}_\Re\mathbf{t}^\mu\boldsymbol{\gamma}_\mathrm{B}^\nu\boldsymbol{\gamma}_\mathrm{B}^5\Psi_\Re
 =\left[\begin{array}{cccc}
 W_\mathrm{M} & c\mathbf{G}_\mathrm{M}\\
 c\mathbf{G}_\mathrm{M} & \boldsymbol{\mathcal{T}}_\mathrm{M}
\end{array}\right]\nonumber\\
 &=\frac{1}{\mu_0}\Big(F_{\;\;\rho}^{\mu}F^{\rho\nu}+\frac{1}{4}g^{\mu\nu}F_{\rho\sigma}F^{\rho\sigma}\Big).
 \label{eq:SEMtensorM}
\end{align}
In terms of the electric and magnetic fields, the components of the SEM tensor $T_\mathrm{M}^{\mu\nu}$ in Eq.~\eqref{eq:SEMtensorM} are given in the Minkowski space-time by
\begin{align}
 W_\mathrm{M} &=\frac{1}{2}\Big(\varepsilon_0\mathbf{E}_\Re^2+\frac{1}{\mu_0}\mathbf{B}_\Re^2\Big),\nonumber\\
 \mathbf{G}_\mathrm{M} &
 =\varepsilon_0\mathbf{E}_\Re\times\mathbf{B}_\Re,\nonumber\\
 \boldsymbol{\mathcal{T}}_\mathrm{M}
 &=W_\mathrm{M}\mathbf{I}_3-\varepsilon_0\mathbf{E}_\Re\otimes\mathbf{E}_\Re-\frac{1}{\mu_0}\mathbf{B}_\Re\otimes\mathbf{B}_\Re.
 \label{eq:SEMtensorcomponents}
\end{align}
The SEM tensor $T_\mathrm{M}^{\mu\nu}$ in Eq.~\eqref{eq:SEMtensorM} is equal to the well-known symmetric SEM tensor of the electromagnetic field \cite{Jackson1999,Landau1989}.

For a general classical time-harmonic field, the time averages of the energy and momentum density components $W_\mathrm{M}$ and $\mathbf{G}_\mathrm{M}$ in Eq.~\eqref{eq:SEMtensorcomponents} over the harmonic cycle, denoted by the brackets with a subscript $\omega$, are related to the expectation value densities in Eqs.~\eqref{eq:rhoE}, \eqref{eq:rhop}, and \eqref{eq:rhoS} as $\langle W_\mathrm{M}\rangle_\omega=\rho_{\hat{H}}^\mathrm{(M)}$ and $\langle\mathbf{G}_\mathrm{M}\rangle_\omega=\boldsymbol{\rho}_{\hat{\mathbf{p}}}^\mathrm{(M)}+\frac{1}{2}\nabla\times\boldsymbol{\rho}_{\hat{\mathbf{S}}}^\mathrm{(M)}$. The first one of these relations is widely known. The second relation is equivalent to Eq.~\eqref{eq:probabilitycurrent3d2} and it is also known from previous literature  \cite{Berry2009,Bliokh2017a}. Apart from the need of the harmonic cycle time average on the left-hand sides, these equations are analogous to those in the case of the Dirac field in Eq.~\eqref{eq:DiracT}.

\subsection{Summary of the generating Lagrangian density of gravity and its special unitary symmetry}

We expect that the eight-spinor formulation of the standard model is necessary for the unification of gravity into it. This conclusion is based on the surprising simplicity of how all the fermionic and bosonic fields of the standard model can be coupled to the gravity in the gauge theory \cite{Partanen2023c} arising from the SU(8)$_\mathrm{4D}$ symmetry presented in this work. The decades of work on this problem without an unambiguous solution also suggest that the formulation of the quantum field theory of gravity using the standard model without eight-spinor structures is extremely difficult if not impossible.

In the derivation of the SEM tensor using the generating Lagrangian density of gravity, the symmetric SEM tensor follows directly from the internal symmetry of the generating Lagrangian density of gravity. This is in accordance with Noether's theorem, which states that each generator of a continuous symmetry is associated with a conserved current \cite{Noether1918,Zee2010}. In the present case, the conserved currents of the four symmetry generators $\mathbf{t}^a$ combine to a single SEM tensor. Using the internal special unitary symmetry SU(8)$_\mathrm{4D}$ instead of the external space-time symmetry differs from the conventional Lagrangian derivation of the SEM tensor \cite{Landau1989,Bliokh2014b}. In the conventional derivation \cite{Landau1989}, the SEM tensor follows from the external space-time symmetry of the action and it is asymmetric if no additional symmetrization procedures are introduced, such as the Belinfante-Rosenfeld symmetrization \cite{Schroder1968,Ramos2015,Belinfante1940,Rosenfeld1940,Weinberg1995}. This observation is one of the foundations for the development of the Yang-Mills gauge theory of unified gravity in Ref.~\cite{Partanen2023c}.

We have shown how the SEM tensor acting as the source term in the Yang-Mills gauge theory of unified gravity arises from the special unitary symmetry, but have left the description of the tensor gauge field to a separate work \cite{Partanen2023c}. Here we briefly note that the tensor gauge field is a Lorentz-invariant tensor whose representation as a $4\times4$ matrix is invariant in the Lorentz transformation of second-rank tensors, and whose representation in terms of $8\times8$ matrices \cite{Partanen2023c} is invariant in the Lorentz transformation of spin-2 fields, given in Sec.~\ref{sec:Lorentztensor}. The different Lorentz transformation properties of vector and tensor gauge fields are one indication why the Yang-Mills gauge theory of unified gravity cannot be derived from the conventional Lagrangian density, but the generating Lagrangian density of gravity is needed. The conventional Lagrangian density follows from the gravitational-gauge-invariant form of the Lagrangian density of the Yang-Mills gauge theory of unified gravity in the Minkowski metric limit \cite{Partanen2023c}.


\section{\label{sec:conclusions}Conclusions}

In conclusion, we have presented QED based on the eight-dimensional spinorial Maxwell equation. The spinorial Maxwell equation is equivalent to the full set of Maxwell's equations. Consequently, it is equivalent to several formulations of electrodynamics, such as the most conventional three-vector-calculus \cite{Jackson1999}, electromagnetic field tensor \cite{Jackson1999}, exterior algebra of differential forms \cite{Flanders1963}, space-time algebra \cite{Dressel2015}, quaternions \cite{Majernik1999,Hong2019}, two-component spinors \cite{Sachs1961,Sachs1964}, and rank-two bi-spinors \cite{Kiessling2018}. It provides an elegant representation of classical electrodynamics and QED, but it does not produce new physics if no other elements are added to the theory.

In comparison to other known formulations of electrodynamics, the present formulation of the spinorial Maxwell equation is the most analogous to the Dirac equation. Here we interpret the analogy in such a way that, in our case of photons, the pertinent gamma matrices must reflect the properties of a spin-1 field instead of the spin-$\frac{1}{2}$ field of the Dirac theory. Instead of the conventional $4\times4$ Dirac gamma matrices, our spinorial Maxwell equation is given in terms of $8\times8$ bosonic gamma matrices satisfying the Dirac algebra. In contrast with $6\times6$ gamma matrices that have been studies in some previous works \cite{Mohr2010,Barnett2014}, the gamma matrices of the present theory allow the description the physics of all Maxwell's equations by a single equation, and accordingly, \emph{complete reformulation} of QED using eight-spinors. Many properties of the Dirac field are directly transferable to the eight-spinor electromagnetic theory. For example, we have formulated the Lorentz transformations of eight-spinors in analogy to the Lorentz transformation of Dirac spinors. As a result, the relativistic quantum spin operators of light emerge naturally through the generators of Lorentz transformations on eight-component electromagnetic spinors. Therefore, our work provides a well-defined electromagnetic spinor and field operator representations with a natural emergence of the relativistic quantum spin structure of light.

The spinorial Maxwell equation leads to the formulation of the Lagrangian density of QED using eight-spinors. It enables the generating Lagrangian density of gravity with internal special-unitary-symmetry-based coupling of the quantum fields of the standard model to gravity. The internal special unitary symmetry of the generating Lagrangian density of gravity was shown to enable an elegant derivation of the symmetric SEM tensors of the electromagnetic and Dirac fields. The possible space-time-dependent form of the quantity $\mathbf{I}_\mathrm{g}$ in the generating Lagrangian density of gravity was left as a topic of further work. Due to the analogous forms of the underlying theories, the present eight-spinor formalism can be extended to describe the other fundamental interactions and the related fermionic and bosonic fields of the standard model. The related fundamental consequence, the Yang-Mills gauge theory of unified gravity, is studied in a separate work \cite{Partanen2023c}.

\begin{acknowledgments}
This work has been funded by the Research Council of Finland under Contract No.~349971. Wolfram \emph{Mathematica} has been extensively used to verify the equations of the present work.
\end{acknowledgments}

\appendix

\section{\label{apx:spinorialwavefunctions}Photon spinors for plane-wave and spherical states}
\subsection{Plane-wave photon states}

The photon spinors for the plane-wave states corresponding to the wave vector $\mathbf{k}$ and the helicity quantum number $q$ are given by
\begin{equation}
 \Psi_{\mathbf{k},q}=\left[
 \begin{array}{c}
  0\\
  \boldsymbol{\mathcal{E}}_{\mathbf{k},q}\\
  0\\
  i\boldsymbol{\mathcal{B}}_{\mathbf{k},q}
 \end{array}\right].
 \label{eq:wavefunctions2}
\end{equation}
Here the electric and magnetic three-vector plane-wave state components $\boldsymbol{\mathcal{E}}_{\mathbf{k},q}$ and $\boldsymbol{\mathcal{B}}_{\mathbf{k},q}$ are defined to absorb the normalizing prefactors of Eq.~\eqref{eq:wavefunctions}, and they are written as
\begin{equation}
 \boldsymbol{\mathcal{E}}_{\mathbf{k},q}=\sqrt{\frac{\hbar\omega_\mathbf{k}}{2V}}\mathbf{e}_\mathbf{k}^{(q)}e^{i(\mathbf{k}\cdot\mathbf{r}-\omega_\mathbf{k}t)},
 \label{eq:planewavee}
\end{equation}
\begin{equation}
 \boldsymbol{\mathcal{B}}_{\mathbf{k},q}
 =\frac{\nabla\times\boldsymbol{\mathcal{E}}_{\mathbf{k},q}}{ik}
 =\frac{\mathbf{k}}{k}\times\boldsymbol{\mathcal{E}}_{\mathbf{k},q}.
 \label{eq:planewavem}
\end{equation}
Here the helicity quantum number $q=\pm1$ and the wave vector $\mathbf{k}$ 
appearing as indices in the spherical 
polarization vector $\mathbf{e}_\mathbf{k}^{(q)}$ 
correspond to a fixed 
polarization and wave vector state of a photon. We define the 
spherical polarization vectors as 
$\mathbf{e}_\mathbf{k}^{(0)}=\mathbf{R}_\mathbf{k}\cdot\hat{\mathbf{z}}$ and 
$\mathbf{e}_\mathbf{k}^{(\pm 
1)}=\mathbf{R}_\mathbf{k}\cdot\frac{1}{\sqrt{2}}(\hat{\mathbf{x } }
\pm i\hat{\mathbf{y}} )$, where $\hat{\mathbf{x}}$, $\hat{\mathbf{y}}$, and 
$\hat{\mathbf{z}}$ are unit vectors parallel to the $x$, $y$ and $z$ axes, 
respectively, and $\mathbf{R}_\mathbf{k}$ is a rotation matrix that rotates 
$\hat{\mathbf{z}}$ parallel to the wave vector $\mathbf{k}$. The rotation matrix $\mathbf{R}_\mathbf{k}$ is given by $\mathbf{R}_\mathbf{k}=\mathbf{I}_3\cos\theta_\mathbf{k}+(\mathbf{s}\cdot\hat{\mathbf{a}})\sin\theta_\mathbf{k}+(\hat{\mathbf{a}}\otimes\hat{\mathbf{a}})(1-\cos\theta_\mathbf{k})$, where $\theta_\mathbf{k}$ is the polar angle of the wave vector, $\hat{\mathbf{a}}=(\hat{\mathbf{z}}\times\mathbf{k})/|\hat{\mathbf{z}}\times\mathbf{k}|$, and $\mathbf{s}=(\mathbf{s}^x,\mathbf{s}^y,\mathbf{s}^z)$ is a vector made of the $3\times3$ rotation generator matrices, given by
\begin{align}
 &\mathbf{s}^x=\left[
 \begin{array}{ccc}
  0 & 0 & 0\\
  0 & 0 & -1\\
  0 & 1 & 0
 \end{array}\right]\!\!,\hspace{0.1cm}
 \mathbf{s}^y=\left[
 \begin{array}{ccc}
  0 & 0 & 1\\
  0 & 0 & 0\\
  -1 & 0 & 0
 \end{array}\right]\!\!,\nonumber\\
 &\hspace{1.5cm}\mathbf{s}^z=\left[
 \begin{array}{ccc}
  0 & -1 & 0\\
  1 & 0 & 0\\
  0 & 0 & 0
 \end{array}\right]\!\!.
 \label{eq:rotationgenerators3D}
\end{align}

The plane-wave states are eigenstates of the helicity operator satisfying $(\hat{\mathfrak{h}}/\hbar)[0,\boldsymbol{\mathcal{E}}_{\mathbf{k},q},0,i\boldsymbol{\mathcal{B}}_{\mathbf{k},q}]^T=q[0,\boldsymbol{\mathcal{E}}_{\mathbf{k},q},0,i\boldsymbol{\mathcal{B}}_{\mathbf{k},q}]^T=[0,i\boldsymbol{\mathcal{B}}_{\mathbf{k},q},0,\boldsymbol{\mathcal{E}}_{\mathbf{k},q}]^T$. In the last form of this equation, we have pointed out that the action of the helicity operator in Eq.~\eqref{eq:helicityoperator} and the division by $\hbar$ is equivalent to swapping of the upper and lower three-vector components in the photon spinor. This result is in agreement with the helicity-chirality equivalence for massless particles, $\hat{\mathfrak{h}}/\hbar\leftrightarrow\boldsymbol{\gamma}_\mathrm{B}^5$, discussed in Sec.~\ref{sec:fifthgamma}. From the point of view of the light source, the helicity quantum number $q=1$ corresponds to clockwise or right-handed circular polarization, while $q=-1$ corresponds to counter-clockwise or left-handed circular polarization.

The normalization condition for the plane-wave photon states is written as
\begin{equation}
 \int\Psi_{\mathbf{k},q}^\dag\Psi_{\mathbf{k}',q'}d^3r
 =\hbar\omega_\mathbf{k}\frac{(2\pi)^3}{V}\delta_{q,q'}\delta(\mathbf{k}-\mathbf{k}').
\end{equation}
This kind of a normalization of one-particle states is known to be Lorentz invariant \cite{Peskin2018}. The plane-wave photon states satisfy the space- and time-reversal symmetry relation, given by
\begin{equation}
 \Psi_{\mathbf{k},q}(-t,-\mathbf{r})=-\boldsymbol{\gamma}_\mathrm{B}^0\Psi_{\mathbf{k},-q}^*(t,\mathbf{r}).
\end{equation}
This relation is useful in checking that the CPT symmetry is satisfied as discussed in Sec.~\ref{sec:CPTsymmetry}.

\subsection{Spherical photon states}

The photon spinors for the spherical states are given by
\begin{equation}
 \Psi_{\omega,J,M}^\mathrm{(e)}=\left[
 \begin{array}{c}
  0\\
  \boldsymbol{\mathcal{E}}_{\omega,J,M}\\
  0\\
  i\boldsymbol{\mathcal{B}}_{\omega,J,M}
 \end{array}\right],\hspace{0.4cm}
 \Psi_{\omega,J,M}^\mathrm{(m)}=\left[
 \begin{array}{c}
  0\\
  i\boldsymbol{\mathcal{B}}_{\omega,J,M}\\
  0\\
  \boldsymbol{\mathcal{E}}_{\omega,J,M}
 \end{array}\right].
 \label{eq:wavefunctionsomegaJM}
\end{equation}
The spinors $\Psi_{\omega,J,M}^\mathrm{(e)}$ and $\Psi_{\omega,J,M}^\mathrm{(m)}$ are associated with the electric and magnetic multipoles of the multipole expansion of the electromagnetic field \cite{Landau1982,Jackson1999}. These spinors are found to be related to each other by the action of the helicity operator in Eq.~\eqref{eq:helicityoperator}  as $\Psi_{\omega,J,M}^\mathrm{(m)}=(\hat{\mathfrak{h}}/\hbar)\Psi_{\omega,J,M}^\mathrm{(e)}$ and $\Psi_{\omega,J,M}^\mathrm{(e)}=(\hat{\mathfrak{h}}/\hbar)\Psi_{\omega,J,M}^\mathrm{(m)}$. This corresponds to the helicity-chirality equivalence for massless particles, $\hat{\mathfrak{h}}/\hbar\leftrightarrow\boldsymbol{\gamma}_\mathrm{B}^5$, discussed in Sec.~\ref{sec:fifthgamma}. The normalized electric and magnetic three-vector spherical state components $\boldsymbol{\mathcal{E}}_{\omega,J,M}$ and $\boldsymbol{\mathcal{B}}_{\omega,J,M}$ in Eq.~\eqref{eq:wavefunctionsomegaJM} are written as
\begin{align}
 &\boldsymbol{\mathcal{E}}_{\omega,J,M}\nonumber\\
 &=\sqrt{\frac{\hbar\omega}{2V}}\,e^{-i\omega t}\Big[\sqrt{\frac{J}{2J+1}}j_{J+1}\Big(\frac{\omega r}{c}\Big)\boldsymbol{Y}_{J,J+1,M}(\theta_\mathbf{r},\phi_\mathbf{r})\nonumber\\
 &\hspace{0.4cm}-\sqrt{\frac{J+1}{2J+1}}j_{J-1}\Big(\frac{\omega r}{c}\Big)\boldsymbol{Y}_{J,J-1,M}(\theta_\mathbf{r},\phi_\mathbf{r})\Big],
\end{align}
\begin{align}
 \boldsymbol{\mathcal{B}}_{\omega,J,M}
 &=\frac{\nabla\times\boldsymbol{\mathcal{E}}_{\omega,J,M}}{i\omega/c}\nonumber\\
 &=\sqrt{\frac{\hbar\omega}{2V}}\,e^{-i\omega t}j_J\Big(\frac{\omega r}{c}\Big)\boldsymbol{Y}_{J,J,M}(\theta_\mathbf{r},\phi_\mathbf{r})
\end{align}
Here $j_J(kr)$ are the spherical Bessel functions of the first kind and 
$\mathbf{Y}_{J,L,M}(\theta_\mathbf{r},\phi_\mathbf{r})$ are the vector 
spherical harmonic functions, defined as \cite{Edmonds1974}
\begin{align}
 &\mathbf{Y}_{J,L,M}(\theta_\mathbf{r},\phi_\mathbf{r})\nonumber\\
 &=\sum_{n=-1}^1\langle L,M-n,1,n|J,M\rangle 
Y_{L,M-n}(\theta_\mathbf{r},\phi_\mathbf{r})\mathbf{u}^{(n)}.
\end{align}
The terms of this series are formed from the 
well-known Clebsch-Gordan coefficients $\langle L,M-n,1,n|J,M\rangle$ \cite{Rose1957}, 
the scalar spherical harmonic functions 
$Y_{L,M}(\theta_\mathbf{r},\phi_\mathbf{r})$, and the 
spherical unit vector $\mathbf{u}^{(n)}$. The spherical unit vectors 
$\mathbf{u}^{(n)}$ are defined as $\mathbf{u}^{(0)}=\hat{\mathbf{z}}$, 
$\mathbf{u}^{(-1)}=\frac{1}{\sqrt{2}}(\hat{\mathbf{x}}-i\hat{\mathbf{y}})$, and 
$\mathbf{u}^{(+1)}=-\frac{1}{\sqrt{2}}(\hat{\mathbf{x}}+i\hat{\mathbf{y}})$. 
For the spherical harmonic functions, we use the definition written in terms of 
the associated Legendre polynomials $P_{L,M}(x)$ as
\begin{equation} 
Y_{L,M}(\theta_\mathbf{r},\phi_\mathbf{r})=\sqrt{\frac{2L+1}{4\pi}\frac{
(L-M)!}{(L+M)!}}\,P_{L,M}(\cos\theta_\mathbf{r})e^{iM\phi_\mathbf{r}}.
\end{equation}
For the associated Legendre polynomials, we use the Condon-Shortley phase 
convention \cite{Condon1970}. Thus, the Condon-Shortley phase is also included in our definition of the spherical harmonic functions. The normalization condition for the spherical photon states is written as
\begin{align}
 &\int\Psi_{\omega,J,M}^{(\eta)\dag}\Psi_{\omega',J',M'}^{(\eta')}d^3r\nonumber\\
 &=\hbar\omega\frac{\pi c^3}{2V\omega^2}\delta_{\eta,\eta'}\delta_{J,J'}\delta_{M,M'}\delta(\omega-\omega').
\end{align}
The spherical photon states satisfy the space- and time-reversal symmetry relations, given by
\begin{equation}
 \Psi_{\omega,J,M}^{(\eta)}(t,-\mathbf{r})=(-1)^{J+1+\delta_{\eta,m}}\boldsymbol{\gamma}_\mathrm{B}^0\Psi_{\omega,J,M}^{(\eta)}(t,\mathbf{r}),
\end{equation}
\begin{equation}
 \Psi_{\omega,J,M}^{(\eta)}(-t,\mathbf{r})=(-1)^{M+1}\Psi_{\omega,J,-M}^{(\eta)*}(t,\mathbf{r}).
\end{equation}
These relations are useful in checking that the CPT symmetry is satisfied as discussed in Sec.~\ref{sec:CPTsymmetry}.

That the two linearly independent vectors in Eq.~\eqref{eq:wavefunctionsomegaJM} are needed for the spherical states is manifested by the fact that the plane-wave states $\Psi_{\mathbf{k},q}$ can be written as a sum over spherical photon states, where both $\Psi_{\omega,J,M}^\mathrm{(e)}$ and $\Psi_{\omega,J,M}^\mathrm{(m)}$ are needed, as \cite{Edmonds1974,Rose1957}
\begin{align}
\Psi_{\mathbf{k},q}
&=-i\sqrt{2\pi}\sum_ {J=1}^\infty\sum_{M=-J}^Ji^J\sqrt{2J+1}\nonumber\\
&\hspace{0.5cm}\times D_{M,q}^{(J)}(\phi_\mathbf{k},\theta_\mathbf{k},-\phi_\mathbf{k})(\Psi_{\omega,J,M}^{(m)}+q\Psi_{\omega,J,M}^\mathrm{(e)}).
\label{eq:sphericalstates}
\end{align}
Here $D_{M,q}^{(J)}(\phi_\mathbf{k},\theta_\mathbf{k},-\phi_\mathbf{k})$ are Wigner D-matrix elements corresponding to the three Eulerian angles that are in our case formed from the components of the wave vector $\mathbf{k}$ in the spherical coordinates, i.e., $\mathbf{k}=(k,\theta_\mathbf{k},\phi_{\mathbf{k}})$.

\section{\label{apx:fourcurrent}Derivation of the electric four-current density in QED}

Here we briefly present the derivation of the electromagnetic four-current density using the free Dirac field term $\mathcal{L}_\mathrm{QED,0}=\mathcal{L}_\mathrm{QED,D}$ of the conventional QED Lagrangian density in Eq.~\eqref{eq:Lagrangiandensity1}. The Lagrangian density $\mathcal{L}_\mathrm{QED,0}$ can be called the generating Lagrangian density of QED. The derivation of the electric four-current density using $\mathcal{L}_\mathrm{QED,0}$ highlights the complete analogy with the derivation of the symmetric SEM tensor using the generating Lagrangian density of gravity in Sec.~\ref{sec:Lagrangian2}. Thus, as the starting point, the generating Lagrangian density of QED is given by
\begin{equation}
 \mathcal{L}_\mathrm{QED,0}=\textstyle\frac{i}{2}\hbar c\bar{\psi}(\boldsymbol{\gamma}_\mathrm{F}^a\vec{\partial}_a-\cev{\partial}_a\boldsymbol{\gamma}_\mathrm{F}^a)\psi-m_\mathrm{e}c^2\bar{\psi}\psi.
 \label{eq:LD1}
\end{equation}
The generating Lagrangian density of QED satisfies the global unitary symmetry U(1). The unitary transformation associated with this symmetry is given by
\begin{equation}
 \psi\rightarrow U_\mathrm{e}\psi,
 \hspace{0.5cm}\text{where }U_\mathrm{e}=e^{i\theta}.
 \label{eq:U1transformation}
\end{equation}
The infinitesimal variations of the Dirac field $\psi$ and its adjoint $\bar{\psi}$ with respect to the symmetry transformation parameter $\theta$ are given by
\begin{equation}
 \delta\psi=i\psi\delta\theta,\hspace{0.5cm}\delta\bar{\psi}=-i\bar{\psi}\delta\theta.
\end{equation}
Using these infinitesimal variations, the variation of the generating Lagrangian density $\mathcal{L}_\mathrm{QED,0}$ in Eq.~\eqref{eq:LD1} is written as
\begin{align}
 \delta\mathcal{L}_\mathrm{QED,0} &=\textstyle\frac{i}{2}\hbar c(\delta\bar{\psi})(\boldsymbol{\gamma}_\mathrm{F}^a\vec{\partial}_a-\cev{\partial}_a\boldsymbol{\gamma}_\mathrm{F}^a)\psi
 -m_\mathrm{e}c^2(\delta\bar{\psi})\psi\nonumber\\
 &\hspace{0.5cm}+\frac{i}{2}\hbar c\bar{\psi}(\boldsymbol{\gamma}_\mathrm{F}^a\vec{\partial}_a-\cev{\partial}_a\boldsymbol{\gamma}_\mathrm{F}^a)(\delta\psi)
 -m_\mathrm{e}c^2\bar{\psi}(\delta\psi)\nonumber\\
 &=-\hbar c\bar{\psi}\boldsymbol{\gamma}_\mathrm{F}^a\psi\partial_a\delta\theta\nonumber\\
 &=-\frac{\hbar}{q_\mathrm{e}}J_\mathrm{e\Re}^a\partial_a\theta.
 \label{eq:LD1variation}
\end{align}
In the last equality of Eq.~\eqref{eq:LD1variation}, we have defined the electric four-current density $J_\mathrm{e\Re}^a$ as
\begin{equation}
 J_\mathrm{e\Re}^a=q_\mathrm{e}c\bar{\psi}\boldsymbol{\gamma}_\mathrm{F}^a\psi.
 \label{eq:Je}
\end{equation}
The derivation of the electric four-current density above using Eqs.~\eqref{eq:LD1}--\eqref{eq:LD1variation} can be seen to be completely analogous to the derivation of the symmetric SEM tensor using Eqs.~\eqref{eq:Lagrangiandensity2}--\eqref{eq:Lagrangiandensityvariation}.

For completeness, next we describe the derivation of the electromagnetic-gauge-invariant Lagrangian density of QED from the generating Lagrangian density of QED in Eq.~\eqref{eq:LD1} using the conventional gauge theory. In the case of the gravitational coupling of the present theory, analogous steps are not described in the present work, but they are left as a topic of the Yang-Mills gauge theory of unified gravity in Ref.~\cite{Partanen2023c}.

The variation of the generating Lagrangian density of QED in Eq.~\eqref{eq:LD1variation} is generally nonzero for a position- and time-dependent $\theta$, for which $\partial_a\theta\neq 0$. Therefore, the generating Lagrangian density of QED in Eq.~\eqref{eq:LD1} is not invariant with respect to the local form of the symmetry transformation in Eq.~\eqref{eq:U1transformation} with $\theta$ depending on the position and time. To promote the global symmetry of constant $\theta$ to a local symmetry of space-time dependent $\theta(t,\mathbf{r})$, the partial derivatives $\vec{\partial}_a$ and $\cev{\partial}_a$ in the generating Lagrangian density of QED in Eq.~\eqref{eq:LD1} are replaced by electromagnetic-gauge-covariant derivatives, given by Eq.~\eqref{eq:covariantderivativeEM} as $\vec{D}_a=\vec{\partial}_a+i\frac{q_\mathrm{e}}{\hbar}A_{\Re a}$ and $\cev{D}_a=\cev{\partial}_a-i\frac{q_\mathrm{e}}{\hbar}A_{\Re a}$. These derivatives bring the electromagnetic gauge potential $A_{\Re a}$ to the theory.

The electromagnetic-gauge-covariant derivative transforms by the same unitary transformation as the Dirac field itself in Eq.~\eqref{eq:U1transformation} as $\vec{D}_a\psi\rightarrow U_\mathrm{e}\vec{D}_a\psi$. The corresponding transformation for the Dirac adjoint is given by $\bar{\psi}\cev{D}_a\rightarrow\bar{\psi}\cev{D}_a U_\mathrm{e}^*$. That these transformations are satisfied requires that the gauge field $A_{\Re a}$ transforms as $A_{\Re a}\rightarrow(U_\mathrm{e}A_{\Re a}+\frac{i\hbar}{q_\mathrm{e}}\partial_a U_\mathrm{e})U_\mathrm{e}^*=A_{\Re a}-\frac{\hbar}{q_\mathrm{e}}\partial_a\theta$. Replacing the partial derivative operators $\vec{\partial}_a$ and $\cev{\partial}_a$ in the generating Lagrangian density of QED in Eq.~\eqref{eq:LD1} by the electromagnetic-gauge-covariant derivative operators $\vec{D}_a$ and $\cev{D}_a$ makes the Lagrangian density invariant with respect to the local space-time-dependent form of the U(1) symmetry transformation in Eq.~\eqref{eq:U1transformation}.

To construct the complete electromagnetic-gauge-invariant Lagrangian density, one must also add an electromagnetic-gauge-invariant term that depends on the gauge field $A_{\Re a}$ only. Utilizing the gauge theory, this is obtained from the commutator of the electromagnetic-gauge-covariant derivatives. The commutator $[\vec{D}_a,\vec{D}_b]=\frac{iq_\mathrm{e}}{\hbar}F_{ab}$ defines an antisymmetric field strength tensor $F_{ab}$, which corresponds to Eq.~\eqref{eq:Ftensor}, as
\begin{equation}
 F_{ab}=\partial_a A_{\Re b}-\partial_b A_{\Re a}.
\end{equation}
In the U(1) gauge symmetry transformation in Eq.~\eqref{eq:transformationU}, the field strength tensor $F_{ab}$ transforms as $F_{ab}\rightarrow U_\mathrm{e}F_{ab}U_\mathrm{e}^*=F_{ab}$. Using the gauge theory, we obtain an electromagnetic-gauge-invariant Lagrangian density term, which depends on the gauge field $A_{\Re a}$ only, as
\begin{equation}
 \mathcal{L}_\mathrm{QED,M}=-\frac{1}{4\mu_0}F_{ab}F^{ab},
\end{equation}
in agreement with Eq.~\eqref{eq:Lagrangiandensity1}. The prefactor of $\mathcal{L}_\mathrm{QED,M}$ is found from comparison of the resulting Euler-Lagrange equations of the electromagnetic gauge field to Maxwell's equations. The complete electromagnetic-gauge-invariant generalization of the generating Lagrangian density of QED in Eq.~\eqref{eq:LD1} is then given by
\begin{equation}
 \mathcal{L}_\mathrm{QED}=\frac{i\hbar c}{2}\bar{\psi}(\boldsymbol{\gamma}_\mathrm{F}^a\vec{D}_a-\cev{D}_a\boldsymbol{\gamma}_\mathrm{F}^a)\psi-m_\mathrm{e}c^2\bar{\psi}\psi
 -\frac{1}{4\mu_0}F_{ab}F^{ab}.
 \label{eq:L1}
\end{equation}
This equation is equivalent to Eq.~\eqref{eq:Lagrangiandensity1}.
The electromagnetic gauge invariance means that the Lagrangian density in Eq.~\eqref{eq:L1} satisfies $\delta\mathcal{L}_\mathrm{QED}=0$, when varied with respect to the symmetry transformation parameter $\theta$. Thus, the addition of the gauge field cancels the variation of the generating Lagrangian density of QED obtained without the gauge field in Eq.~\eqref{eq:LD1variation}.

\end{document}